\documentclass[sigconf,balance=false]{acmart}
\usepackage{popets}

\setcopyright{popets}
\copyrightyear{YYYY}

\acmYear{YYYY}
\acmVolume{YYYY}
\acmNumber{X}
\acmDOI{XXXXXXX.XXXXXXX}
\acmISBN{}
\acmConference{Proceedings on Privacy Enhancing Technologies}
\usepackage{hyperref}
\usepackage{algorithm}
\usepackage{algorithmic}
\def\Snospace~{\S{}}

\hypersetup{breaklinks,colorlinks,citecolor=blue,linkcolor=blue}

\makeatletter
\renewcommand{\ALG@name}{Protocol}
\makeatother

\usepackage{amsthm}
\newtheoremstyle{myDefinitionstyle}
{}
{}
{%
}
{}
{\bfseries}
{.}
{ }
{\thmname{#1}\thmnumber{ #2}\thmnote{ (#3)}}%

\usepackage{amsthm}
\newtheoremstyle{myTheoremstyle}
{0pt}
{0pt}
{%
	\itshape%
}
{}
{\bfseries}
{.}
{ }
{\thmname{#1}\thmnumber{ #2}\thmnote{ (#3)}}%

\theoremstyle{myDefinitionstyle}
\newtheorem{Definition}{Definition}

\theoremstyle{myTheoremstyle}
\newtheorem{Theorem}{Theorem}

\newcommand{\Sim}{\ensuremath{\mathcal{S}_{\textsf{\scheme}}}}

\newcommand{\Env}{\ensuremath{\mathcal{E}}}

\newcommand{\idealFunc}{\ensuremath{F}}

\newcommand{\idealFuncGenMask}{\ensuremath{\mathcal{F}_{\mathsf{GenMask}}}}

\newcommand{\idealFuncGenMaskShr}{\ensuremath{\mathcal{F}_{\mathsf{GenMaskShr}}}}

\newcommand{\idealFuncLTHReLU}{\ensuremath{\mathcal{F}_{\mathsf{LTHReLU}}}}

\newcommand{\idealFuncReLU}{\ensuremath{\mathcal{F}_{\mathsf{ReLU}}}}

\newcommand{\idealFuncLTHMMR}{\ensuremath{\mathcal{F}_{\mathsf{LTHMatMulReLU}}}}

\newcommand{\idealFuncMMR}{\ensuremath{\mathcal{F}_{\mathsf{MatMulReLU}}}}

\newcommand{\idealFuncLTHSoftmax}{\ensuremath{\mathcal{F}_{\mathsf{LTHSoftmax}}}}

\newcommand{\idealFuncSoftmax}{\ensuremath{\mathcal{F}_{\mathsf{Softmax}}}}

\newcommand{\blue}[1]{\textcolor{blue}{#1}}

\newcommand{\party}[1]{\ensuremath{#1}}

\newcommand{\Hybrid}{\ensuremath{\mathcal{H}}}

\usepackage{xcolor}
\usepackage{amsmath}
\usepackage{cleveref}
\usepackage{amsfonts}
\usepackage{stmaryrd}
\usepackage{marvosym}
\usepackage{lipsum}
\usepackage[normalem]{ulem}
\usepackage{enumitem}

\newcommand{\ph}[1]{}
\newcommand{\thang}[1]{}
\newcommand{\ed}[1]{}
\newcommand{\el}[1]{}
\newcommand{\de}[1]{}
\newcommand{\change}[1]{#1}

\usepackage{svg}
\usepackage{nccmath}
\usepackage{tikz}
\usepackage{adjustbox}
\usepackage{tabularx}
\usepackage{setspace}
\usepackage{svg}
\usepackage{tabulary}
\usepackage{multirow}
\usepackage{amsthm}
\usepackage{csquotes}
\usepackage[english]{babel}

\usepackage{pgf-pie}
\usepackage{float}
\usepackage{stfloats}
\usepackage{placeins}
\usepackage{subcaption}
\usepackage{pgfplots}
\pgfplotsset{compat=1.17}
\usepackage{soul}
\usepackage{textcomp}
\newcommand{\scheme}{\textsc{Stamp}\xspace}
\newcommand{\LTH}{LTH\xspace}
\usetikzlibrary{datavisualization}

\colorlet{shadecolor}{black!60}

\newcommand{\sbline}{\\[.5\normalbaselineskip]}

\newcolumntype{x}[1]{>{\centering\arraybackslash}p{#1}}

\newcounter{func}
\newenvironment{func}[1]
  {\par\addvspace{\topsep}
   \noindent
   \tabularx{\linewidth}{@{} X @{}}
    \hrulefill\\
    \multicolumn{1}{c}{#1} \\\\
    }
  { 
    \hrulefill
   \endtabularx
   \par\addvspace{\topsep}}

\usepackage{amsfonts}
\usepackage{mathrsfs}
\usepackage{xspace}
\usepackage{bm}
\usepackage{upgreek}

\newcommand{\safemath}[2]{\newcommand{#1}{\ensuremath{#2}\xspace}}

\safemath{\bma}{\mathbf{a}}
\safemath{\bmb}{\mathbf{b}}
\safemath{\bmc}{\mathbf{c}}
\safemath{\bmd}{\mathbf{d}}
\safemath{\bme}{\mathbf{e}}
\safemath{\bmf}{\mathbf{f}}
\safemath{\bmg}{\mathbf{g}}
\safemath{\bmh}{\mathbf{h}}
\safemath{\bmi}{\mathbf{i}}
\safemath{\bmj}{\mathbf{j}}
\safemath{\bmk}{\mathbf{k}}
\safemath{\bml}{\mathbf{l}}
\safemath{\bmm}{\mathbf{m}}
\safemath{\bmn}{\mathbf{n}}
\safemath{\bmo}{\mathbf{o}}
\safemath{\bmp}{\mathbf{p}}
\safemath{\bmq}{\mathbf{q}}
\safemath{\bmr}{\mathbf{r}}
\safemath{\bms}{\mathbf{s}}
\safemath{\bmt}{\mathbf{t}}
\safemath{\bmu}{\mathbf{u}}
\safemath{\bmv}{\mathbf{v}}
\safemath{\bmw}{\mathbf{w}}
\safemath{\bmx}{\mathbf{x}}
\safemath{\bmy}{\mathbf{y}}
\safemath{\bmz}{\mathbf{z}}
\safemath{\bmzero}{\mathbf{0}}
\safemath{\bmone}{\mathbf{1}}

\bmdefine{\biad}{a}
\bmdefine{\bibd}{b}
\bmdefine{\bicd}{c}
\bmdefine{\bidd}{d}
\bmdefine{\bied}{e}
\bmdefine{\bifd}{f}
\bmdefine{\bigd}{g}
\bmdefine{\bihd}{h}
\bmdefine{\biid}{i}
\bmdefine{\bijd}{j}
\bmdefine{\bikd}{k}
\bmdefine{\bild}{l}
\bmdefine{\bimd}{m}
\bmdefine{\bind}{n}
\bmdefine{\biod}{o}
\bmdefine{\bipd}{p}
\bmdefine{\biqd}{q}
\bmdefine{\bird}{r}
\bmdefine{\bisd}{s}
\bmdefine{\bitd}{t}
\bmdefine{\biud}{u}
\bmdefine{\bivd}{v}
\bmdefine{\biwd}{w}
\bmdefine{\bixd}{x}
\bmdefine{\biyd}{y}
\bmdefine{\bizd}{z}

\bmdefine{\bixid}{\xi}
\bmdefine{\bilambdad}{\lambda}
\bmdefine{\bimud}{\mu}
\bmdefine{\bithetad}{\theta}
\bmdefine{\biphid}{\phi}
\bmdefine{\bideltad}{\delta}

\safemath{\bmia}{\biad}
\safemath{\bmib}{\bibd}
\safemath{\bmic}{\bicd}
\safemath{\bmid}{\bidd}
\safemath{\bmie}{\bied}
\safemath{\bmif}{\bifd}
\safemath{\bmig}{\bigd}
\safemath{\bmih}{\bihd}
\safemath{\bmii}{\biid}
\safemath{\bmij}{\bijd}
\safemath{\bmik}{\bikd}
\safemath{\bmil}{\bild}
\safemath{\bmim}{\bimd}
\safemath{\bmin}{\bind}
\safemath{\bmio}{\biod}
\safemath{\bmip}{\bipd}
\safemath{\bmiq}{\biqd}
\safemath{\bmir}{\bird}
\safemath{\bmis}{\bisd}
\safemath{\bmit}{\bitd}
\safemath{\bmiu}{\biud}
\safemath{\bmiv}{\bivd}
\safemath{\bmiw}{\biwd}
\safemath{\bmix}{\bixd}
\safemath{\bmiy}{\biyd}
\safemath{\bmiz}{\bizd}

\safemath{\bmxi}{\bixid}
\safemath{\bmlambda}{\bilambdad}
\safemath{\bmmu}{\bimud}
\safemath{\bmtheta}{\bithetad}
\safemath{\bmphi}{\biphid}
\safemath{\bmdelta}{\bideltad}

\safemath{\bA}{\mathbf{A}}
\safemath{\bB}{\mathbf{B}}
\safemath{\bC}{\mathbf{C}}
\safemath{\bD}{\mathbf{D}}
\safemath{\bE}{\mathbf{E}}
\safemath{\bF}{\mathbf{F}}
\safemath{\bG}{\mathbf{G}}
\safemath{\bH}{\mathbf{H}}
\safemath{\bI}{\mathbf{I}}
\safemath{\bJ}{\mathbf{J}}
\safemath{\bK}{\mathbf{K}}
\safemath{\bL}{\mathbf{L}}
\safemath{\bM}{\mathbf{M}}
\safemath{\bN}{\mathbf{N}}
\safemath{\bO}{\mathbf{O}}
\safemath{\bP}{\mathbf{P}}
\safemath{\bQ}{\mathbf{Q}}
\safemath{\bR}{\mathbf{R}}
\safemath{\bS}{\mathbf{S}}
\safemath{\bT}{\mathbf{T}}
\safemath{\bU}{\mathbf{U}}
\safemath{\bV}{\mathbf{V}}
\safemath{\bW}{\mathbf{W}}
\safemath{\bX}{\mathbf{X}}
\safemath{\bY}{\mathbf{Y}}
\safemath{\bZ}{\mathbf{Z}}

\safemath{\bZero}{\mathbf{0}}
\safemath{\bOne}{\mathbf{1}}
\safemath{\bDelta}{\mathbf{\Delta}}
\safemath{\bLambda}{\mathbf{\UpLambda}}
\safemath{\bPhi}{\mathbf{\Phi}}
\safemath{\bPsi}{\mathbf{\Psi}}
\safemath{\bSigma}{\mathbf{\Upsigma}}
\safemath{\bOmega}{\mathbf{\Upomega}}
\safemath{\bTheta}{\mathbf{\Uptheta}}

\bmdefine{\biAd}{A}
\bmdefine{\biBd}{B}
\bmdefine{\biCd}{C}
\bmdefine{\biDd}{D}
\bmdefine{\biEd}{E}
\bmdefine{\biFd}{F}
\bmdefine{\biGd}{G}
\bmdefine{\biHd}{H}
\bmdefine{\biId}{I}
\bmdefine{\biJd}{J}
\bmdefine{\biKd}{K}
\bmdefine{\biLd}{L}
\bmdefine{\biMd}{M}
\bmdefine{\biOd}{N}
\bmdefine{\biPd}{O}
\bmdefine{\biQd}{P}
\bmdefine{\biRd}{R}
\bmdefine{\biSd}{S}
\bmdefine{\biTd}{T}
\bmdefine{\biUd}{U}
\bmdefine{\biVd}{V}
\bmdefine{\biWd}{W}
\bmdefine{\biXd}{X}
\bmdefine{\biYd}{Y}
\bmdefine{\biZd}{Z}

\bmdefine{\biDelta}{\Delta}
\bmdefine{\biLambda}{\Lambda}
\bmdefine{\biPhi}{\Phi}
\bmdefine{\biSigma}{\Sigma}
\bmdefine{\biOmega}{\Omega}
\bmdefine{\biTheta}{\Theta}

\safemath{\bimA}{\biAd}
\safemath{\bimB}{\biBd}
\safemath{\bimC}{\biCd}
\safemath{\bimD}{\biDd}
\safemath{\bimE}{\biEd}
\safemath{\bimF}{\biFd}
\safemath{\bimG}{\biGd}
\safemath{\bimH}{\biHd}
\safemath{\bimI}{\biId}
\safemath{\bimJ}{\biJd}
\safemath{\bimK}{\biKd}
\safemath{\bimL}{\biLd}
\safemath{\bimM}{\biMd}
\safemath{\bimN}{\biNd}
\safemath{\bimO}{\biOd}
\safemath{\bimP}{\biPd}
\safemath{\bimQ}{\biQd}
\safemath{\bimR}{\biRd}
\safemath{\bimS}{\biSd}
\safemath{\bimT}{\biTd}
\safemath{\bimU}{\biUd}
\safemath{\bimV}{\biVd}
\safemath{\bimW}{\biWd}
\safemath{\bimX}{\biXd}
\safemath{\bimY}{\biYd}
\safemath{\bimZ}{\biZd}

\safemath{\bimDelta}{\biDelta}
\safemath{\bimLambda}{\biLambda}
\safemath{\bimPhi}{\biPhi}
\safemath{\bimSigma}{\biSigma}
\safemath{\bimOmega}{\biOmega}
\safemath{\bimTheta}{\biTheta}

\safemath{\setA}{\mathcal{A}}
\safemath{\setB}{\mathcal{B}}
\safemath{\setC}{\mathcal{C}}
\safemath{\setD}{\mathcal{D}}
\safemath{\setE}{\mathcal{E}}
\safemath{\setF}{\mathcal{F}}
\safemath{\setG}{\mathcal{G}}
\safemath{\setH}{\mathcal{H}}
\safemath{\setI}{\mathcal{I}}
\safemath{\setJ}{\mathcal{J}}
\safemath{\setK}{\mathcal{K}}
\safemath{\setL}{\mathcal{L}}
\safemath{\setM}{\mathcal{M}}
\safemath{\setN}{\mathcal{N}}
\safemath{\setO}{\mathcal{O}}
\safemath{\setP}{\mathcal{P}}
\safemath{\setQ}{\mathcal{Q}}
\safemath{\setR}{\mathcal{R}}
\safemath{\setS}{\mathcal{S}}
\safemath{\setT}{\mathcal{T}}
\safemath{\setU}{\mathcal{U}}
\safemath{\setV}{\mathcal{V}}
\safemath{\setW}{\mathcal{W}}
\safemath{\setX}{\mathcal{X}}
\safemath{\setY}{\mathcal{Y}}
\safemath{\setZ}{\mathcal{Z}}
\safemath{\emptySet}{\varnothing}

\safemath{\colA}{\mathscr{A}}
\safemath{\colB}{\mathscr{B}}
\safemath{\colC}{\mathscr{C}}
\safemath{\colD}{\mathscr{D}}
\safemath{\colE}{\mathscr{E}}
\safemath{\colF}{\mathscr{F}}
\safemath{\colG}{\mathscr{G}}
\safemath{\colH}{\mathscr{H}}
\safemath{\colI}{\mathscr{I}}
\safemath{\colJ}{\mathscr{J}}
\safemath{\colK}{\mathscr{K}}
\safemath{\colL}{\mathscr{L}}
\safemath{\colM}{\mathscr{M}}
\safemath{\colN}{\mathscr{N}}
\safemath{\colO}{\mathscr{O}}
\safemath{\colP}{\mathscr{P}}
\safemath{\colQ}{\mathscr{Q}}
\safemath{\colR}{\mathscr{R}}
\safemath{\colS}{\mathscr{S}}
\safemath{\colT}{\mathscr{T}}
\safemath{\colU}{\mathscr{U}}
\safemath{\colV}{\mathscr{V}}
\safemath{\colW}{\mathscr{W}}
\safemath{\colX}{\mathscr{X}}
\safemath{\colY}{\mathscr{Y}}
\safemath{\colZ}{\mathscr{Z}}

\safemath{\opA}{\mathbb{A}}
\safemath{\opB}{\mathbb{B}}
\safemath{\opC}{\mathbb{C}}
\safemath{\opD}{\mathbb{D}}
\safemath{\opE}{\mathbb{E}}
\safemath{\opF}{\mathbb{F}}
\safemath{\opG}{\mathbb{G}}
\safemath{\opH}{\mathbb{H}}
\safemath{\opI}{\mathbb{I}}
\safemath{\opJ}{\mathbb{J}}
\safemath{\opK}{\mathbb{K}}
\safemath{\opL}{\mathbb{L}}
\safemath{\opM}{\mathbb{M}}
\safemath{\opN}{\mathbb{N}}
\safemath{\opO}{\mathbb{O}}
\safemath{\opP}{\mathbb{P}}
\safemath{\opQ}{\mathbb{Q}}
\safemath{\opR}{\mathbb{R}}
\safemath{\opS}{\mathbb{S}}
\safemath{\opT}{\mathbb{T}}
\safemath{\opU}{\mathbb{U}}
\safemath{\opV}{\mathbb{V}}
\safemath{\opW}{\mathbb{W}}
\safemath{\opX}{\mathbb{X}}
\safemath{\opY}{\mathbb{Y}}
\safemath{\opZ}{\mathbb{Z}}
\safemath{\opZero}{\mathbb{O}}
\safemath{\identityop}{\opI}

\safemath{\veca}{\bma}
\safemath{\vecb}{\bmb}
\safemath{\vecc}{\bmc}
\safemath{\vecd}{\bmd}
\safemath{\vece}{\bme}
\safemath{\vecf}{\bmf}
\safemath{\vecg}{\bmg}
\safemath{\vech}{\bmh}
\safemath{\veci}{\bmi}
\safemath{\vecj}{\bmj}
\safemath{\veck}{\bmk}
\safemath{\vecl}{\bml}
\safemath{\vecm}{\bmm}
\safemath{\vecn}{\bmn}
\safemath{\veco}{\bmo}
\safemath{\vecp}{\bmp}
\safemath{\vecq}{\bmq}
\safemath{\vecr}{\bmr}
\safemath{\vecs}{\bms}
\safemath{\vect}{\bmt}
\safemath{\vecu}{\bmu}
\safemath{\vecv}{\bmv}
\safemath{\vecw}{\bmw}
\safemath{\vecx}{\bmx}
\safemath{\vecy}{\bmy}
\safemath{\vecz}{\bmz}

\safemath{\veczero}{\bmzero}
\safemath{\vecone}{\bmone}
\safemath{\vecxi}{\bmxi}
\safemath{\veclambda}{\bmlambda}
\safemath{\vecmu}{\bmmu}
\safemath{\vectheta}{\bmtheta}
\safemath{\vecphi}{\bmphi}
\safemath{\vecdelta}{\bmdelta}

\safemath{\matA}{\bA}
\safemath{\matB}{\bB}
\safemath{\matC}{\bC}
\safemath{\matD}{\bD}
\safemath{\matE}{\bE}
\safemath{\matF}{\bF}
\safemath{\matG}{\bG}
\safemath{\matH}{\bH}
\safemath{\matI}{\bI}
\safemath{\matJ}{\bJ}
\safemath{\matK}{\bK}
\safemath{\matL}{\bL}
\safemath{\matM}{\bM}
\safemath{\matN}{\bN}
\safemath{\matO}{\bO}
\safemath{\matP}{\bP}
\safemath{\matQ}{\bQ}
\safemath{\matR}{\bR}
\safemath{\matS}{\bS}
\safemath{\matT}{\bT}
\safemath{\matU}{\bU}
\safemath{\matV}{\bV}
\safemath{\matW}{\bW}
\safemath{\matX}{\bX}
\safemath{\matY}{\bY}
\safemath{\matZ}{\bZ}
\safemath{\matzero}{\bmzero}

\safemath{\matDelta}{\bDelta}
\safemath{\matLambda}{\bLambda}
\safemath{\matPhi}{\bPhi}
\safemath{\matSigma}{\bSigma}
\safemath{\matOmega}{\bOmega}
\safemath{\matTheta}{\bTheta}

\safemath{\matidentity}{\matI}
\safemath{\matone}{\matO}

\safemath{\rnda}{A}
\safemath{\rndb}{B}
\safemath{\rndc}{C}
\safemath{\rndd}{D}
\safemath{\rnde}{E}
\safemath{\rndf}{F}
\safemath{\rndg}{G}
\safemath{\rndh}{H}
\safemath{\rndi}{I}
\safemath{\rndj}{J}
\safemath{\rndk}{K}
\safemath{\rndl}{L}
\safemath{\rndm}{M}
\safemath{\rndn}{N}
\safemath{\rndo}{O}
\safemath{\rndp}{P}
\safemath{\rndq}{Q}
\safemath{\rndr}{R}
\safemath{\rnds}{S}
\safemath{\rndt}{T}
\safemath{\rndu}{U}
\safemath{\rndv}{V}
\safemath{\rndw}{W}
\safemath{\rndx}{X}
\safemath{\rndy}{Y}
\safemath{\rndz}{Z}

\safemath{\rveca}{\bimA}
\safemath{\rvecb}{\bimB}
\safemath{\rvecc}{\bimC}
\safemath{\rvecd}{\bimD}
\safemath{\rvece}{\bimE}
\safemath{\rvecf}{\bimF}
\safemath{\rvecg}{\bimG}
\safemath{\rvech}{\bimH}
\safemath{\rveci}{\bimI}
\safemath{\rvecj}{\bimJ}
\safemath{\rveck}{\bimK}
\safemath{\rvecl}{\bimL}
\safemath{\rvecm}{\bimM}
\safemath{\rvecn}{\bimN}
\safemath{\rveco}{\bomO}
\safemath{\rvecp}{\bimP}
\safemath{\rvecq}{\bimQ}
\safemath{\rvecr}{\bimR}
\safemath{\rvecs}{\bimS}
\safemath{\rvect}{\bimT}
\safemath{\rvecu}{\bimU}
\safemath{\rvecv}{\bimV}
\safemath{\rvecw}{\bimW}
\safemath{\rvecx}{\bimX}
\safemath{\rvecy}{\bimY}
\safemath{\rvecz}{\bimZ}

\safemath{\rvecxi}{\bmxi}
\safemath{\rveclambda}{\bmlambda}
\safemath{\rvecmu}{\bmmu}
\safemath{\rvectheta}{\bmtheta}
\safemath{\rvecphi}{\bmphi}

\safemath{\rmatA}{\bimA}
\safemath{\rmatB}{\bimB}
\safemath{\rmatC}{\bimC}
\safemath{\rmatD}{\bimD}
\safemath{\rmatE}{\bimE}
\safemath{\rmatF}{\bimF}
\safemath{\rmatG}{\bimG}
\safemath{\rmatH}{\bimH}
\safemath{\rmatI}{\bimI}
\safemath{\rmatJ}{\bimJ}
\safemath{\rmatK}{\bimK}
\safemath{\rmatL}{\bimL}
\safemath{\rmatM}{\bimM}
\safemath{\rmatN}{\bimN}
\safemath{\rmatO}{\bimO}
\safemath{\rmatP}{\bimP}
\safemath{\rmatQ}{\bimQ}
\safemath{\rmatR}{\bimR}
\safemath{\rmatS}{\bimS}
\safemath{\rmatT}{\bimT}
\safemath{\rmatU}{\bimU}
\safemath{\rmatV}{\bimV}
\safemath{\rmatW}{\bimW}
\safemath{\rmatX}{\bimX}
\safemath{\rmatY}{\bimY}
\safemath{\rmatZ}{\bimZ}

\safemath{\rmatDelta}{\bimDelta}
\safemath{\rmatLambda}{\bimLambda}
\safemath{\rmatPhi}{\bimPhi}
\safemath{\rmatSigma}{\bimSigma}
\safemath{\rmatOmega}{\bimOmega}
\safemath{\rmatTheta}{\bimTheta}

\safemath{\llb}{\llbracket}
\safemath{\rrb}{\rrbracket}

\begin{document}

\title{Efficient Privacy-Preserving Machine Learning with Lightweight Trusted Hardware}




\author{Pengzhi Huang}
\affiliation{%
  \institution{Cornell University}
  \city{}
  \state{}
  \country{}}
\email{ph448@cornell.edu}

\author{Thang Hoang}
\affiliation{%
  \institution{Virginia Tech}
  \city{}
  \state{}
  \country{}}
\email{thanghoang@vt.edu}

\author{Yueying Li}
\affiliation{%
  \institution{Cornell University}
  \city{}
  \state{}
  \country{}}
\email{yl3469@cornell.edu}

\author{Elaine Shi}
\affiliation{%
  \institution{Carnegie Mellon University}
  \city{}
  \state{}
  \country{}}
\email{runting@gmail.com}

\author{G. Edward Suh}
\affiliation{%
  \institution{NVIDIA\textsuperscript{1} / Cornell University}
  \city{}
  \state{}
  \country{}}
\email{edward.suh@cornell.edu}


\renewcommand{\shortauthors}{Huang et al.}

\begin{abstract}
In this paper, we propose a new secure machine learning inference platform assisted by a small dedicated security processor, which will be easier to protect and deploy compared to today's TEEs integrated into high-performance processors.
Our platform provides three main advantages over the state-of-the-art:
(i) We achieve significant performance improvements compared to state-of-the-art distributed Privacy-Preserving Machine Learning (PPML) protocols, with only a small security processor that is comparable to a discrete security chip such as the Trusted Platform Module (TPM) or on-chip security subsystems in SoCs similar to the Apple enclave processor. 
In the semi-honest setting with WAN/GPU, our scheme is 4$\times$-63$\times$ faster than Falcon (PoPETs'21) and AriaNN (PoPETs'22) and 3.8$\times$-12$\times$ more communication efficient. 
We achieve even higher performance improvements in the malicious setting.
(ii) Our platform guarantees security with abort against malicious adversaries under honest majority assumption. 
(iii) Our technique is not limited by the size of secure memory in a TEE and can support high-capacity modern neural networks like ResNet18 and Transformer. 
While previous work investigated the use of high-performance TEEs in PPML, this work represents the first to show that even tiny secure hardware with very limited performance can be leveraged to significantly speed-up distributed PPML protocols
if the protocol can be carefully designed for lightweight trusted hardware.
 
\end{abstract}

\keywords{Multi-party computation, Secure hardware, Machine learning}

\maketitle

\footnotetext[1]{This work was done while the author was at Meta.}

\section{Introduction}

As the world increasingly relies on machine learning (ML) for everyday tasks, a large amount of potentially sensitive or private data need to be processed by ML learning algorithms.
For example, ML models for medical applications may need to use private datasets distributed in multiple nations as inputs \cite{kaissis2021end}. A cloud-based ML services process private data from users with pre-trained models to provide predictions \cite{Azure, GoogleAI}. 
The data to be shared in these applications are often private and sensitive and must be protected from the risk of leakage. 
Government regulations may play an essential role as a policy, but cannot guarantee actual protection. 
We need technical protection for privacy-preserving machine learning (PPML) for strong confidentiality and privacy guarantees.

In this paper, we propose a new PPML framework, named {\scheme} (Small Trusted hardware Assisted MPc), which enables far more efficient secure multiparty computation (MPC) for machine learning through a novel use of small lightweight trusted hardware (\LTH).
MPC refers to a protocol that allows multiple participants to jointly evaluate a particular function while preventing their inputs from being revealed to each other. Ever since Yao's initial studies (later called Garbled Circuit) \cite{yao1982protocols,yao1986generate} which gave such a secure protocol in the case of two semi-honest parties, many studies have been conducted to improve the efficiency\de{ further}, to expand to more than two parties, and to ensure the feasibility against malicious behaviors. Recently, there has been significant interest in using and optimizing MPC for secure machine learning computation \cite{mohassel2018aby3, wagh2019securenn, wagh2020falcon, ryffel2020ariann, crypten2020}. However, the overhead for MPC-based PPML is still significant.

For low-overhead secure computation, trusted execution environments (TEEs) in modern microprocessors such as Intel SGX \cite{costan2016intel} AMD SEV \cite{amdsev} aim to provide hardware-based protection for the confidentiality and integrity of data and code inside.
If the TEE protection and the software inside can be trusted, secure machine learning computation can be performed directly inside a TEE with relatively low overhead \cite{kim2020vessels}.
The TEE can also be used to improve cryptographic protocols by accelerating bootstrapping \cite{katz2007universally,lu2021correlated} or simplifying protocols \cite{choi2019hybrid,bahmani2017secure, katz2007universally,felsen2019secure}.  
However, it is challenging to build a secure environment inside a high-performance processor due to its large trusted computing base (TCB) and complex performance optimizations such as out-of-order execution, speculation, and caching.
For example, multiple attacks have been shown for SGX \cite{van2018foreshadow,van2017sgx, gotzfried2017cache}. 
Moreover, the TEE requires adding hardware protection to each type of computing engines (CPU, GPU, and accelerators), and significant changes to the software stack.
As a result, developing and deploying a TEE for a new piece of hardware requires significant effort and time. 

In this paper, \de{instead of relying on a TEE on a complex high-performance processor, }we propose to leverage a small dedicated security processor, another type of trusted hardware that is widely deployed today, to reduce the MPC overhead.
For example, small discrete security chips such as trusted platform module (TPM), Google Titan, and Apple T1 are widely used as a platform root-of-trust.
For system-on-chip (SoC) designs, on-chip security subsystems like \de{such as}the Apple enclave processor perform security-critical operations such as secure booting, attestation, and key management.

While the high-level idea to combine trusted hardware and MPC has been explored before, we believe this work represents the first to investigate MPC acceleration using a small security processor.
Clearly, such lightweight trusted hardware can only provide relatively low performance. The main question is if a low-performance trusted hardware
can still be leveraged to provide meaningful speed-ups for MPC.
In the following discussion, we refer to such small security processors as lightweight trusted hardware (\LTH).

The key insight we leverage in {\scheme} is that non-linear operations, which can be performed very efficiently in plaintext, account for the major part of the overhead in MPC.
MPC-based deep learning inference is not particularly expensive in computation but introduces large communication overhead due to multi-round data exchanges, especially when the network latency is high. This overhead leads to a very different cost distribution for MPC compared to plaintext computation.
Profiling an inference task of AlexNet \cite{krizhevsky2012imagenet}, which represents a classical deep learning model, shows that 85\% of total plaintext execution time comes from linear operations such as convolution and fully-connected layers, while for MPC, this portion drops to only 5\% with the remaining 95\% coming from non-linear operations.
Most of those non-linear operations are simple and cheap in plaintext (e.g., \textsf{ReLU}, \textsf{MaxPooling}, which are generally comparisons) with some exceptions (e.g., \textsf{Softmax})\de{ which have higher complexity}. 
This observation implies that even a lightweight trusted hardware can potentially speed up MPC-based PPML significantly if we can efficiently offload non-linear operations.




{\scheme} combines the advantages of MPC and trusted hardware by performing linear operations in MPC\de{ for its security and ease of deployment,} while leveraging {\LTH} for non-linear operations.
To realize this approach, we introduce new MPC protocols that efficiently offload non-linear operations while minimizing communications among multiple parties and between the {\LTH} and an untrusted CPU/GPU.
Although simple nonlinear operations can be performed inside the small {\LTH} with sufficiently high performance, expensive operations such as \textsf{Softmax} require higher performance.
To address the challenge, {\scheme} securely offloads parts of the expensive exponentiation operations \change{to an untrusted CPU/GPU}.
The following describes the main technical contributions and advantages of {\scheme}.







\ph{redundancy in subsequent paragraphs}
\textbf{Insight and performance improvement.}
\change{{\scheme} represents the first work to investigate if tiny low-performance security processors can still be leveraged to meaningfully speed-up MPC protocols.}
Our results demonstrate that even with trust in a tiny piece of discrete secure hardware similar to a TPM, significant speedups can be achieved for privacy-preserving neural network inference \change{when the MPC protocol is carefully redesigned for efficient offloading of non-linear operations}. 
\change{We compared \scheme with three state-of-the-art MPC protocols (Falcon~\cite{wagh2020falcon}, AriaNN~\cite{ryffel2020ariann}, and CryptGPU~\cite{tan2021cryptgpu}).} 
The results show that {\scheme} achieves significantly lower inference overhead compared to the state-of-the-art MPC protocols on either CPUs or GPUs, under either a WAN or LAN setting, and 
\change{using either a discrete security chip (\LTH-chip) or a security processor on an SoC (\LTH-SoC)}. 
\change{
{\scheme} is $4\times$ to $63\times$ faster in the semi-honest WAN/GPU setting, even with the tiny {\LTH-chip} with a low-bandwidth interconnection, and 
reduces the inter-party communication by $7\times$ to $10\times$.
\scheme can also improve the performance of the MPC-based secure inference in malicious settings. 
Interestingly, the experimental results show that \scheme (\LTH-SoC) can even outperform a protocol that leverages a high-performance TEE (Intel SGX) with secure GPU outsourcing (Goten~\cite{ng2021goten}) thanks to its ability to significantly reduce the inter-party communication.
While {\scheme} can also be used with a high-performance TEE to further improve performance, this result suggests that tiny low-performance secure hardware can indeed be sufficient if it is primarily used for non-linear operations.
}
\change{\scheme provides the most significant performance improvements under GPU/WAN settings when WAN communication represents more of a performance bottleneck compared to GPU-based computation.}

\textbf{Malicious security.} \scheme provides security guarantees under the honest-majority setting similar to previous schemes \cite{wagh2020falcon,mohassel2018aby3}, assuming that the majority (2 out of the 3 participants) are behaving honestly. If the corrupted party behaves semi-honestly, the protocol ensures that no information is obtained by any party without reconstructing a value. If a party is actively corrupted and behaves maliciously, we guarantee detection of such a behavior and output ``abort'' while still keeping the confidentiality of the data with extra steps. We show the security of {\scheme} using the standard simulation-based paradigm in \autoref{sec:proof}. We implement both semi-honest and malicious protocols in our end-to-end framework\de{, providing an option to choose either lower overhead or higher security guarantees based on the knowledge of the participants}.

\textbf{Prototype implementation.} 
We implemented a functional prototype of both semi-honest and malicious protocols of \scheme in \verb!C++!.
The compilation framework and a small number of pure MPC-based operations (see \autoref{sec:mpc} and \autoref{sec:MPCbasic}) are based on \cite{wagh2020falcon}. 
The baseline framework was significantly modified to incorporate new non-linear operation protocols, GPUs support, new networks and datasets, and a better socket library. 
The prototype implementation supports both CPU-only and GPU-assisted settings, and adds the same GPU support to our baseline for a fair comparison.

\textbf{Evaluation and analysis.} 
We demonstrate \scheme by supporting the secure inference of various networks including AlexNet \cite{krizhevsky2012imagenet}, VGG16 \cite{simonyan2014very}, ResNet18 \cite{he2016deep} and Transformer \cite{wolf-etal-2020-transformers}, over multiple datasets including MNIST \cite{deng2012mnist}, CIFAR-10 \cite{krizhevsky2014cifar}, ImageNet \cite{ILSVRC15} and Wikitext-2 \cite{merity2016pointer}, under both WAN and LAN, and semi-honest and malicious settings. We provide theoretical analysis of the overhead \change{and scalability analysis.} We perform detailed experimental studies \change{against state-of-the-art MPC protocols, and also protocols leveraging high-performance TEEs for a balanced discussion on the trade-off.}
We show that even a very small trusted hardware reduces the overhead of MPC protocols significantly while supporting various high-capacity networks, 
\change{and \scheme can support larger models without extra overhead in most cases}. 

\section{Model}

\begin{figure}[t]
	\centering

         \includegraphics{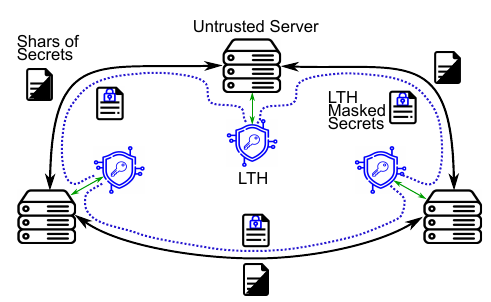}
         
         \caption{{\scheme} system and threat model. The black local machines owned by three parties, green local buses, and black inter-party communication channels are untrusted. The blue {\LTH}s are trusted and contain secret keys shared among {\LTH}s.}
         \label{fig:STAMP1}
\end{figure}

\noindent \textbf{System Model.}
{In our system, there are three parties who want to run a common ML model together using inputs from individuals. 
	We assume that the model structure is publicly known.
	%
	%
	We assume that each party consists of two components: an untrusted machine (CPU/GPU) and an \LTH module whose computational power is limited.
	%
	%
	%
	\LTH in each party communicates with each other through its host 
	by establishing pairwise secure communication channels. 
	%
	
}

\begin{figure}[t]
     \centering
     \includegraphics{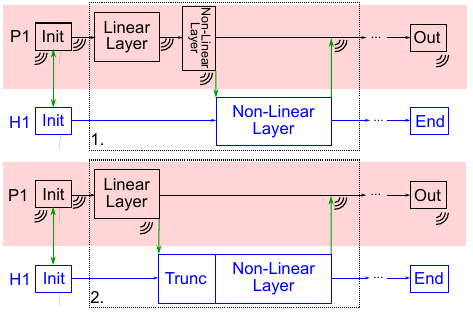}
     \caption{\change{The {\scheme} execution flow on one of the parties. Inter-party communication (wave symbol) and the local communication with the \LTH (green arrows) happen during initialization and execution, with (1) or without (2) the optimization in \autoref{sec:Merge}. An adversary has complete control over the data and operations in the red zone.}}
     \label{fig:STAMP2}
\end{figure}

\noindent \textbf{Threat Model.}
The goal of \scheme is to protect the confidentiality and the integrity of ML model inference in the presence of a malicious adversary. We capture such confidentiality and integrity through simulation-based security \cite{goldreich2019play, canetti2000security, canetti2001universally}: 

\begin{Definition}[Simulation-based security: privacy and verifiability] \label{def:security} A protocol $\pi_\mathcal{F}$ is said to securely realize the ideal functionality $\mathcal{F}$ if for any probabilistic polynomial time (PPT) real-world adversary $\mathcal{A}$, there exists an ideal-world adversary $\mathcal{S}$ such that for any PPT environment $\mathcal{Z}$,  there exists a negligible function $\mathsf{negl}$ such that
	$$
	|\Pr[\mathsf{Real}_{\Pi_{\mathcal{F},\mathcal{A},\mathcal{Z}(\lambda)}}=1] - \Pr[\mathsf{Ideal}_{\mathcal{F},\mathcal{S},\mathcal{Z}(\lambda)}=1]| \le \mathsf{negl}(\lambda)
	$$
\end{Definition}

We consider honest majority, meaning that at most 
one party (except its \LTH) can be malicious.
The other two parties can be semi-honest, in which they may try to learn secrets (e.g., inputs or weights provided by other parties) while still following the protocol faithfully.
The malicious adversary can deviate arbitrarily from the honest protocol, and its goal can be breaking the integrity of the evaluation by providing incorrect results without being noticed, or breaking the confidentiality of the data by learning the secrets. 
We assume that there is no collusion between any of the parties.

\change{\autoref{fig:STAMP1} provides an overview of the \scheme system. 
We assume that a party or its server is untrusted except for its \LTH. In other words, an adversary may control any part of the server, including a virtual machine monitor, an operating system, drivers, storage, and others except for \LTH.
We assume that the confidentiality and integrity of \LTH are protected and an adversary cannot obtain data on an \LTH or alter its execution. 
To ensure that only valid secure hardware can participate in the protocol, \LTH contains a unique private key and is authenticated
through a Certificate Authority (CA) during the initialization step.
As shown in \autoref{fig:STAMP1}, the three {\LTH}s act as three trusted third parties with established correlations (secret keys). 
\autoref{fig:STAMP2} shows that the data flow between an untrusted server (red) and an \LTH (blue) during the {\scheme} execution. Data should be encrypted before being sent to the red zone, and any data from or operations conducted in the red zone should be verified assuming the presence of a malicious adversary.}

%

\change{
Note that {\scheme}, similar to other secure computation techniques based on TEE, MPC and / or homomorphic encryption (HE), does not prevent attacks that poison the model through malicious inputs or extract information from the trained parameters or model outputs~\cite{tramer2022truth,carlini2023extracting}.
To defend against such algorithmic attacks, a secure computing framework such as \scheme needs to be combined with other orthogonal defense techniques (e.g., Differential Privacy \cite{abadi2016deep}, out-of-distribution points removal \cite{yang2021generalized}).
Additionally, \scheme primarily targets private inference, not training, so training data poisoning attacks are not its main concern.
}

\change{The security model and its detailed analysis are presented in \autoref{sec:proof}.} Although we assume that \LTH is secure, we discuss how \LTH provides security benefits over TEEs in \autoref{sec:TEE} under a hybrid MPC+trusted hardware threat model.

\section{Background}

In this section, we \de{discuss the background materials for our method. We first }describe our notation, and then provide some basics of MPC and trusted hardware.

\subsection{Notation}\label{sec:notation}

We define $L$ as the finite field size, and $\mathbb{Z}_L$ to be the finite field we generally consider in this work. \textsf{fp} is the fix-point precision. We use the bold font $\bma$ or $\bA$ to represent a vector or a matrix. We use $a_i$, $(\bma)_i$ or $A_{i,j}$ to represent the $i^{th}$ element of the vector $\bma$ or the element of the matrix $\bA$ in the $i^{th}$ row and in the $j^{th}$ column. 
This is different from the bold $\bA_i$, which still represents a matrix.
Throughout the paper, if not specifically mentioned, all operations are carried out within the finite field $\mathbb{Z}_L$. When needed, we use $(a+b)_{L}$ to represent the modulo $L$ operation for the output of the integer operations in brackets. We add a bar to a variable or operation, as $\bar a,\overline{\exp(a)}$, to represent that a number or an output is a real number. The right-shift operation is indicated as $\gg $ (e.g., $a\gg b=a/2^b$). We will often use two signed integers $m,q$ to represent a positive real number $\bar a$ as $\bar a = 2^q\cdot (m\gg 52)$ where $m$ represents the mantissa part of 52 bits with $m\gg 52\in[0,1)$, and $q_L$ is the exponent part. This is actually the format in which floating point numbers are represented following the IEEE Standard for Binary Floating-Point Arithmetic (IEEE
754-1985) \cite{kahan1996ieee}, but without sign on the mantissa part. We use $\lfloor \bar a\rfloor$ to round a real number $\bar a$ down to an integer.

\subsection{Multiparty Computation}\label{sec:mpc}

\textbf{Notation.} The sharing scheme used in this work is the 2-out-of-3 replicated secret sharing scheme (RSS) modulo $L$. Let $P_1,P_2,P_3$ be the three parties participating in the evaluation. For convenience, we use $P_{i-1},P_{i+1}$ to refer to the previous and next party of one party (e.g., the previous and the next party of $P_1$ are $P_3$ and $P_2$).
The RSS of an integer secret $x\in\mathbb{Z}_L$ is denoted as $\llbracket x\rrbracket^L=(\llb x\rrb^L_1,\llb x\rrb^L_2,\llb x\rrb^L_3)$,
where $L$ is the size of the finite field to which the shares belong and $x= \llb x\rrb^L_1+\llb x\rrb^L_2+\llb x\rrb^L_3$. When a secret $x$ is shared as $\llbracket x\rrbracket^L$, party $P_i$ holds $(\llb x\rrb^L_i,\llb x\rrb^L_{i+1})$ for $i=1,2,3$. To generate the integer representation $x$ based on the real value $\bar x$, we use two's complement fixed-point encoding with \textsf{fp} bits of precision. For a positive $\bar x$ we have $x = \lfloor\bar x \cdot 2^{\textsf{fp}}\rfloor$, while for a negative $\bar x$, $x = \lfloor\bar x \cdot 2^{\textsf{fp}}\rfloor + L$, assuming that $x$ is within the bound $[-L/2^{\textsf{fp}},L/2^{\textsf{fp}})$. 

In our experiments, we mainly use the cases of $l=32$, with $\textsf{fp}=13$ and $L=2^l$ (which supports inputs from -262144 to $262144-2^{-13}$), to match the bit-width used in the baseline MPC schemes.
The security of a $l=32$ setting naturally comes from the random masking creating the shares. Multiple existing MPC schemes \cite{wagh2020falcon,ryffel2020ariann,riazi2018chameleon,chaudhari2019astra} use the 32-bit secret sharing setting and already prove its security.
Our protocol can also use a larger field such as a 64-bit setting if a wider range of values need to be supported.

Here, we explain how multiplications are performed under 2-out-of-3 RSS. These operations follow the protocols defined in \cite{mohassel2018aby3, wagh2019securenn, wagh2020falcon, demmler2015aby}.
We describe the rest of the baseline MPC operations including share creation, reconstruction, and aggregation in \autoref{sec:MPCbasic}. 

\textbf{Multiplications.} $\llbracket x \cdot  y\rrbracket^L\gets \Pi_{\textsf{Mul}}(\llbracket x \rrbracket^L, \llbracket y\rrbracket^L)$ : To get $\llbracket z\rrbracket^L=\llbracket x \cdot  y\rrbracket^L$, $P_i$ first computes $\hat z_i = \llbracket x_i\rrbracket^L\llbracket y_i\rrbracket^L+\llbracket x_{i+1}\rrbracket^L \llbracket y_i\rrbracket^L+\llbracket x_i\rrbracket^L \llbracket y_{i+1}\rrbracket^L$, then $(\hat z_1,\hat z_2, \hat z_3)$ is already a valid 3-out-of-3 secret sharing of $xy$ since $z_1+z_2+z_3 = xy $. A reshare is needed to maintain the consistency of the 2-out-of-3 sharing scheme. To avoid any possible leakage of information, $P_i$ uses the 3-out-of-3 randomness $\{\alpha_i\}$ to mask $\hat z_i$ as $z_i = \hat z_i + \alpha_i$, then share it with $P_{i-1}$. Therefore, the parties obtain the necessary shares and build $\llbracket z\rrbracket^L=\llbracket xy\rrbracket^L=( z_1, z_2,  z_3)$.

\textbf{Matrix Multiplications.} $(\llbracket \mathbf{A} \mathbf{B} \rrbracket^L)\gets\Pi_{\textsf{MatMul}} (\llbracket \mathbf{A}\rrbracket^L, \llbracket \mathbf{B}\rrbracket^L )$: To perform matrix multiplication $\llbracket \mathbf{C}_{a\times c} \rrbracket^L=\llbracket \mathbf{A}_{a\times b}\mathbf{B}_{b\times c}\rrbracket^L$\de{ in the RSS scheme, a similar protocol can be used as in $\Pi_{\textsf{Mul}}$.}, simply applying $\Pi_{\textsf{Mul}}$ for each multiplication leads to $\mathcal{O}(abc)$ shares to be sent. The parties can instead perform part of the addition of shares locally (i.e., $\llbracket\hat\bC\rrbracket^L_i =\llbracket \bA \rrbracket^L_i\llbracket \bB \rrbracket^L_i+\llbracket \bA \rrbracket^L_i\llbracket \bB \rrbracket^L_{i+1}+\llbracket \bA \rrbracket^L_{i+1}\llbracket \bB \rrbracket^L_i$) and then share $\llbracket\hat\bC\rrbracket^L_i$ at once. This strategy yields only $\mathcal{O}(ac)$ transmission overhead\de{, saving a factor $b$ of the number of shares to be sent}. 
As stated in \cite{wagh2019securenn}, convolutions can be expanded into overall larger matrix multiplications. 

The above multiplication protocols work well for integer representations, but will cause errors with fixed-point representations. A truncation protocol (right-shift the results by \textsf{fp} bits) must follow after a multiplication to correct the fixed-point precision in 3-party MPC. We refer the readers to ABY3 \cite{mohassel2018aby3} for more details on the 3-party truncation protocol, to prior work \cite{mohassel2018aby3,furukawa2017high} for the malicious variant of $\Pi_{\textsf{MatMul}}$, and to \autoref{sec:MPCbasic} for other basic operations.

\subsection{Trusted/Secure Hardware}\label{sec:TEE}

Dedicated security hardware has a long history of being successfully used in many high-security use cases, starting as (co-)processors specializing in crypto operations. 
For example, smart cards~\cite{rankl2004smart} are widely used in financial transactions.
Similarly, hardware security modules (HSMs) such as IBM 4758~\cite{ibm4758} have also been used to protect critical secret keys.
Discrete security chips such as TPM~\cite{pearson2003trusted}, Google Titan~\cite{googletitan}, and Apple T1 provide hardware root-of-trust on many platforms. 
Modern System-on-Chip (SoC) designs also typically include a dedicated security processor with crypto engines for secure booting and other high-security operations: 
Synopsys tRoot hardware security module~\cite{Synopsys}, Rambus RT-630 programmable root-of-trust (RoT)~\cite{Rambus}, Apple secure enclave~\cite{AppleS}, Qualcomm secure processing unit~\cite{QSPU}, etc.
Even though their performance is limited and their implementations may still have security vulnerabilities~\cite{moghimi2020tpm, han2018bad, butterworth2013bios}, the small dedicated security processors are considered to be far more secure compared to high-performance processors. 
The dedicated security processors are also relatively easy to deploy\de{because they can be integrated into a system} as a separate chip or as an IP block\de{ without affecting the rest of the system}.  

For high-performance processors, the idea of trusted hardware developed into a trusted execution environment (TEE), which adds hardware-based security protection on a shared general-purpose processor running a full software stack \cite{van2017sgx,amdsev,hua2020guardnn}.
A TEE aims to protect the integrity and confidentiality of the code and data inside, even when low-level software and/or the environment cannot be trusted. 
TEEs on modern processors can typically provide much higher performance compared to the dedicated security hardware, but also introduce new security challenges due to the large TCB and complex optimizations in high-performance processors
\cite{fei2021security,xu2015controlled,shinde2015preventing, gyselinck2018off,lee2017inferring,gotzfried2017cache,brasser2017software,lee2017hacking, wang2018interface}. 
The high-performance TEEs also require new hardware protection for each computing engine and significant changes to a complex software stack, making their deployment for new hardware challenging.

\begin{figure}
\centering
\begin{subfigure}{0.25\textwidth}
  \centering
    \includegraphics[]{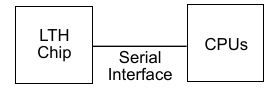}
      \vspace{-4mm}
  \caption{\LTH-Chip}
  \label{fig:LTH1}
\end{subfigure}%
\begin{subfigure}{0.25\textwidth}
  \centering
  \includegraphics[]{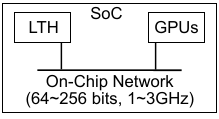}
  \caption{\LTH-SoC}
  \label{fig:LTH2}
\end{subfigure}
\vspace{-5mm}
\caption{Two types of {\LTH}s that \scheme considers.}
\label{fig:LTH}
\vspace{2mm}
\end{figure}

In this work, we consider lightweight trusted hardware ({\LTH}) with performance and complexity similar to a traditional security chip or on-chip security subsystems in modern SoCs: a dedicated low-performance security processor that supports remote attestation to validate its identity and shared key exchanges (\autoref{sec:init}), has hardware crypto engines, and includes a programmable processor that can run code. 
We consider two types of {\LTH} designs as shown in \autoref{fig:LTH}: 1) a discrete security chip similar to a TPM (\LTH-chip) running at a low clock frequency (tens of MHz), and connected to a CPU through a low-bandwidth interface; and 2) a security subsystem on an SoC (\LTH-SoC) running at a much higher SoC clock frequency (1-3GHz), and connected to other processing engines (CPUs, GPUs, NPUs, etc.) on the same SoC through high-bandwidth on-chip networks.
Our study suggests that even the {\LTH-chip} can significantly improve the performance of the MPC-based PPML.

\textbf{Notation.} Each party $P_i$ is equipped with a \LTH $H_i$, which has a built-in PRF unit $F$ (e.g., an AES engine) for pseudo-random number generation. We assume that even malicious participants cannot break the integrity and confidentiality guarantees that \LTH provides. The protocols executed in $H_i$ will be introduced in \autoref{sec:protocol}.


\subsubsection{\LTH Benefits} 
\label{sec:whyLTH}
While it is difficult to quantify the security, we believe that LTH provides strong security and deployment benefits over high-performance TEEs. For example, the previous survey~\cite{fei2021security} provides an overview of the vulnerabilities in Intel SGX (high-performance TEE) and countermeasures. Most vulnerability categories (address translation, CPU cache, DRAM, branch prediction, rowhammer) in the survey do not apply to LTH due to the following reasons\de{ that \LTH has}. 
Here, we provide a more detailed discussion of the security of \LTH and list some {\it attacks that \LTH is more robust to}.


{\bf Physical isolation}: \LTH is dedicated to a small set of security tasks, and physically separate from main processing cores with potentially malicious software. \LTH tightly controls its software using secure booting and typically does not allow user software. Because hardware is not shared with potential attack software, there is much less concern for {\it timing-channel attacks} - a major challenge in today's TEEs.
    
{\bf Smaller TCB/attack surface, lower complexity}: LTH uses a simple (in-order) processor with limited interfaces/commands for a small set of security tasks. Both hardware and software are much smaller and simpler compared to the main processors. 
For example, the dedicated security processors usually occupy less than $1mm^2$ of the silicon area. On the other hand, a high-performance CPU takes hundreds of $mm^2$ and contains millions of lines of code (LoCs) \cite{lee2020keystone}, and is shared with many software components.
Because there is no speculation or out-of-order execution, {\it transient-execution attacks} such as {\it Meltdown/Spectre} that can run commands or read memory without permission are not a concern for LTH.
LTH does not have external memory (DRAM), and is not exposed to {\it attacks on external memory} such as {\it DRAM probing and rowhammer attacks}. 


{\bf Side-channel protection}: \LTH such as smartcards, TPM, and others are usually equipped with dedicated crypto engines and countermeasures (e.g., tamper-resistant circuits \cite{sparks2007security} for TPM, randomized block design \cite{moein2017hardware} for smart cards, etc.) against physical side channels such as {\it power side channels}, and without off-chip memory. In that sense, \LTH is more robust against {\it physical attacks}.

\vspace{-0.1in}
\subsubsection{\change{\LTH Limitations}}\label{sec:lthlimitation} 
%
\change{The main limitation of \LTH comes from its performance. \LTH is typically not designed for high-performance computation. Both computation and communication on \LTH are much slower compared to high-performance TEEs.
As a result, the use of \LTH comes with the additional challenges to support sufficient end-to-end performance. Traditionally, \LTH is only used for small security-critical operations such as key management and infrequent signing. In order to leverage \LTH for larger applications such as ML inference, we need to divide the workload and only offload small parts to \LTH in a way that \LTH does not become the performance bottleneck. 
In fact, {\scheme} had to be carefully designed to leverage low-performance \LTH and our experimental results show that the overall performance still depends on the performance of \LTH (\LTH-chip vs. \LTH-SoC). 
On the other hand, TEEs can closely match the performance of the underlying high-performance processors and can often be used to run the entire task such as ML inference inside, with minimal changes to the workload.
If a high-performance TEE can be fully trusted, a TEE can replace the \LTH in our scheme to provide higher performance or be used to run the entire ML inference without MPC. 
}

\change{
While we believe that \LTH provides stronger security compared to high-performance TEEs, we note that \LTH can still have security vulnerabilities, similar to how secure cryptographic algorithms can be broken due to implementation-level vulnerabilities. 
}
For example, 
timing side channels and power interrupts may make TPM private key recovery possible \cite{moghimi2020tpm,han2018bad}. 
Smart cards, although practically considered secure enough and widely developed, have faced challenges including reverse engineering \cite{quadir2016survey}, micro probing \cite{skorobogatov2017microprobing}, optical fault induction attacks \cite{skorobogatov2003optical}, and others. 


Compared to complex high-performance TEEs, \LTH has far fewer vulnerabilities, making countermeasures easier to apply in terms of cost and design complexity.
In practice, the main security concerns for today’s TEE come from software-exploitable vulnerabilities.
In that sense, \LTH provides a major security benefit by removing most timing-channel or transient-execution vulnerabilities.
While physical attacks are not considered a major threat in data-center environments, \LTH can also provide strong physical security. \LTH has no off-chip memory to protect, and often has anti-tamper/DPA countermeasures. 
In contrast, recent TEEs target weaker threat models against physical attacks. Intel removed the integrity tree for replay protection in Icelake/TDX. AMD SEV has no replay protection against physical attacks. NVIDIA GPU TEE (H100) does not even encrypt its high-bandwidth memory (HBM).


\vspace{-0.05in}
\section{The {\scheme} Protocol} \label{sec:protocol}

This section introduces the details of the \scheme protocols for both semi-honest and malicious settings. We refer the reader to \autoref{sec:proof} for detailed security analysis.



\vspace{-0.05in}
\subsection{Initialization phase}\label{sec:init}






The initialization phase $\Pi_{\textsf{Init}}$ is a part of the offline phase (which needs no input data or model weights) of the protocol where the {\LTH}s will have shared keys and initial values established in them if their identities are proven. 
Although $\Pi_{\textsf{Init}}$ plays an important role in our scheme, it is not where our main contribution lies, since mature remote attestation protocols already exist \cite{banks2021remote}.
A simplified description of $\Pi_{\textsf{Init}}$ is shown in ~\autoref{p:init} .

The communication out of $H_i$ has to go through $P_i$, which provides a corrupted party with a natural way to observe or even alter the communication among the {\LTH}s. For semi-honest adversaries, the Diffie–Hellman key exchange protocol already prevents them from obtaining the key with bounded computational resources. If the corrupted party behaves maliciously, $\Pi_{\textsf{Init}}$ does not have to take extra steps to detect such actions. If a malicious $P_i$ modifies the remote attestation, a CA will not provide a certificate and $P_i$ cannot create a certificate on its own, causing an abort. If a malicious $P_i$\de{ changes the initial parameters or} alters the transmission during key exchange, there will be no correct initialization established, and the protocol will abort later when data inconsistency is detected.

\begin{algorithm}
\small 
\caption{$\Pi_{\textsf{Init}}$ Initialization}\label{p:init}
\setstretch{0}
\raggedright
\textbf{Input.} Security parameter $\lambda$.

\textbf{Result.} Output (\textsf{Success},$L$) if the remote attestation succeeds and aborts if failed. After the initialization, LTHs ($H_i$) obtain shared keys and initial parameters.
\begin{enumerate}[topsep=0.5pt,itemsep=0.5ex,partopsep=0ex,parsep=0ex]
    \item Parties first agree to a $L$ for the finite 
    field $\mathbb{Z}_L$, size $l=\log L$ bits, prime $p$, and their order to define the previous and next party.
    
    \item $P_i$s perform remote attestation on each $H_i$ to obtain a certificate from the CA and publicly share them to validate $H_i$. Abort if validation fails.
    
    \item $H_i$ performs Diffie–Hellman key exchange with signature through the secure channel between $P_i$s to obtain $\mathcal{O}(\lambda)$-bit PRF keys $k_{i,i+1},k_{i-1,i}$, then use $F_{k_{i,i+1}}$ to mask one key $k'_{i-1,i} \equiv k_{i-1,i} + F_{k_{i,i+1}}(0)\mod{p}$ and send $k'_{i-1,i}$ to $H_{i+1}$ through $P_i$. $H_i$ would receive $k'_{i+1,i-1}$ from $H_{i-1} $ and can recover   $k_{i+1,i-1}=k'_{i+1,i-1}-F_{k_{i+1,i-1}}(0)$.

\end{enumerate}

\end{algorithm}

The shared keys and the PRF in the {\LTH}s can support the pseudorandom number generation 
and are kept only known to the {\LTH}, unlike the correlated randomness introduced in \autoref{sec:MPCbasic}.
With the shared keys in \autoref{sec:init} and a built-in PRF $F$,
we can now construct $\Pi_{\textsf{LTH.GenMask}}$ and $\Pi_{\textsf{LTH.GenMaskShare}}$ in the \LTH as \autoref{p:LTH.GenMask} and \autoref{p:LTH.GenMaskShare}. 
They are very similar with only a minor difference that $\Pi_{\textsf{LTH.GenMaskShare}}$ always generates shares of $\mathbf{0}$. 
Four counters $\{\mathsf{ctr}^i_1,\mathsf{ctr}^i_2,\mathsf{ctr}^i_3,\mathsf{ctr}^i_s\}$ 
are used in each $H_i$ to maintain consistency among $H_i$s in a semi-honest setting, and additional four $\{\hat {\mathsf{ctr}}^i_1,\hat {\mathsf{ctr}}^i_2,\hat {\mathsf{ctr}}^i_3,\hat {\mathsf{ctr}}^i_s\}$ are needed in a malicious setting for reduplicate execution for the detection of inconsistency.
Notice that \autoref{p:LTH.GenMask} and \autoref{p:LTH.GenMaskShare} give the outputs to $H_i$, not $P_i$, and $H_i$ may be set to give partial outputs in some protocols.

\begin{algorithm}[t]
\small
\caption{$\bmm \gets  \Pi_{\textsf{LTH.GenMask}}(n, L ,j ; i ,\mathsf{ctr}^i_j,k_{j,j+1})$} 
\label{p:LTH.GenMask}
\setstretch{0}
\raggedright
\textbf{Input. }The number of masks to be generated $n$, the index $j$ for which counter, and which key to choose. The size of the finite field $L$, the counter $\mathsf{ctr}^i_j$ and the key $k_{j,j+1}$ are stored in the {\LTH}.
\sbline
\textbf{Output.} pseudo-random masks $\bmm\in\mathbf{Z}_L$ and updated counter $\mathsf{ctr}^i_j$.
\sbline
$\bmm=(F_{k_{j,j+1}}(\mathsf{ctr}^i_j),
F_{k_{j,j+1}}(\mathsf{ctr}^i_j+1),...,F_{k_{j,j+1}}(\mathsf{ctr}^i_j+n-1))$.

Update $\mathsf{ctr}^i_j\leftarrow \mathsf{ctr}^i_j+n$.

\end{algorithm}

\begin{algorithm}[t]
\small 
\caption{$(\llb \bmm_j\rrb^L_i,\llb \bmm_j\rrb^L_{i+1}) \gets\Pi_{\textsf{LTH.GenMaskShare}}(n, L; \\ i, \mathsf{ctr}^i_s, k_{i,i+1},k_{i+1,i-1},k_{i-1,i})$}
\label{p:LTH.GenMaskShare}
\setstretch{0}
\raggedright
\textbf{Input. }The number of masks to be generated $n$. The size of the finite field $L$ and the counter and keys are stored in the {\LTH}.
\sbline
\textbf{Output.} pseudorandom masks $\llb\bmm\rrb^L_i,\llb\bmm\rrb^L_{i+1}\in\mathbb{Z}_L$
\sbline
$\llb m_j\rrb^L_i= F_{k_{i,i+1}}(\mathsf{ctr}^i_s+j)-F_{k_{i+1,i-1}}(\mathsf{ctr}^i_s+j)$ for $j=0,...,n-1$

$\llb m_j\rrb^L_{i+1}= F_{k_{i+1,i-1}}(\mathsf{ctr}^i_s+j)-F_{k_{i-1,i}}(\mathsf{ctr}^i_s+j)$ for $j=0,...,n-1$

Update $\mathsf{ctr}^i_s\leftarrow \mathsf{ctr}^i_s+n$.



\end{algorithm}

A proper remote attestation protocol is commonly supported on secure hardware, such as a TPM \cite{pearson2003trusted}, and validates the {\LTH}'s identity and its state.
This process can involve the acquisition of the certificate of a {\LTH} from a trusted CA/Verifiers, which is usually the manufacturer of it. $H_i$ after being verified, can perform pairwise Diffie-Hellman key exchanges with a signature 
to obtain the shared key $k_{i-1,i}$ with $H_{i-1}$, $k_{i,i+1}$ with $H_{i+1}$, and then also $k_{i+1,i-1}$ through sharing masks. 
The shared keys can support the pseudorandom number generation with PRF 
and are kept only known to the {\LTH}.



\vspace{-0.05in}
\subsection{Optimized ReLU with Matrix Multiplication}\label{sec:Merge}

Non-linear layers used in a machine learning model are computationally light under plaintext. \textsf{ReLU}, for example, takes only one comparison and multiplexing. However, its complexity gets amplified significantly under the RSS scheme with more local computation steps and significant communication overhead\de{, both in communication rounds and the amount of transmitted data}. 

Using each party's {\LTH} and the common randomness established in \autoref{sec:init}, we can significantly reduce the overhead by ``offloading'' the non-linear operations to the {\LTH}. 
For example, \textsf{ReLU} can be performed by: invoking $\Pi_{\textsf{LTH.GenMask}}$ to get the pseudo-random masks, transmitting the masked shares, recovering the plaintext to compute inside {\LTH}s, and then generate and distribute the psuedorandom shares of the results. We show the details of this protocol ($\Pi_{\textsf{ReLU}}$) in \autoref{sec:reluapp}.

In {\scheme}, we further optimize \textsf{ReLU} by combining it with truncation.
For typical ML models \cite{krizhevsky2012imagenet, wolf-etal-2020-transformers, simonyan2014very, he2016deep}, \textsf{ReLU} is applied after matrix multiplications in convolution (Conv) or fully-connected (FC) layers.
As introduced in \autoref{sec:mpc}, in a fixed-point setting, truncation is required after each multiplication to keep the consistency of the precision.\de{ The classic MPC protocol \cite{mohassel2018aby3} requires pregeneration and a round of communication.} 
If we apply $\Pi_{\textsf{ReLU}}$ directly after the completion of multiplications, the communication overhead will be the multiplication / truncation overhead and the $\Pi_{\textsf{ReLU}}$ overhead summed, which is not optimal. Since the truncation itself is also a simple non-linear function in plaintext (which is just right-shift), we can exploit this common structure in deep learning models and merge the truncation with the following non-linear operations to be simply computed together in plaintext inside the trusted {\LTH}. 

The protocol $\Pi_{\textsf{MatMulReLU}}$, detailed in \autoref{p:matmulrelu}, demonstrates how ReLU can be combined with truncation after matrix multiplication. The steps for a semi-honest setting are colored black, with additional steps for a malicious adversary marked blue. We use this notation in other protocols as well. $\Pi_{\textsf{MatMulReLU}}$ reduces the total communication rounds of matrix multiplication and ReLU combined to 2 from at least 3, by merging the transmission needed for truncation and sharing shares masked by psuedorandom masks generated by {\LTH}s in step 2) and 3).
The malicious version generally adds replicate parallel operations and requires replicate sharing of the same values to validate the integrity.
Parties compare the copies of intermediate results and final outputs from different sources to achieve malicious security with abort. We also use $\Pi_{\textsf{mal-arith-mult}}$ of \cite{mohassel2018aby3} to ensure correct 2-out-of-3 shares after local multiplication.

\begin{algorithm*}[!ht]
\small
\caption{$\llbracket\textsf{ReLU}(\bA\times\bB)\rrbracket^L\gets\Pi_{\textsf{MatMulReLU}}(\llb\bA\rrb^L,\llb\bB\rrb)$: Multiply $\bA$ and $\bB$, then output the shares of the \textsf{ReLU} of the results.}
\label{p:matmulrelu}
\setstretch{0}
\raggedright
\textbf{Input. }$\{P_i\}$ have shares of $ \bA \in\mathbf{Z}_L^{a\times b}$ and $ \bB \in\mathbf{Z}_L^{b\times c}$.
\sbline
\textbf{Output. }$\{P_i\}$ get shares of $\llbracket\bZ\rrbracket^L=\llbracket\textsf{ReLU}(\bA\times\bB)\rrbracket^L$.
\begin{enumerate}[topsep=0.5pt,itemsep=0.5ex,partopsep=0ex,parsep=0ex]
    \item $P_1,P_2$, and $P_3$ locally computes $\llbracket\hat\bC\rrbracket^L_i = \llbracket \bA \rrbracket^L_i\times\llbracket \bB \rrbracket^L_i+\llbracket \bA \rrbracket^L_i\times\llbracket \bB \rrbracket^L_{i+1}+\llbracket \bA \rrbracket^L_{i+1}\times\llbracket \bB \rrbracket^L_i$.
    
    \textcolor{blue}{\underline{\textit{Malicious:}} Parties instead perform $\Pi_{\textsf{mal-arith-mult}}$ of \cite{mohassel2018aby3} to ensure that the multiplications (before truncation) were performed faithfully by parties. In the end, the 2-out-of-3 sharing $\llbracket\hat\bC\rrbracket$ is distributed.}
    
    \item $P_i$ calls $H_i$ to execute
    $\Pi_{\textsf{LTH.GenMaskShare}}(a\times c,L)$ to obtain the masks $\llb\bM\rrb_i\in\mathbf{Z}_L^{a\times c}$, then compute $\hat \bC_i'=\llb\hat\bC\rrb^L_i+\llb\bM\rrb_i$. $P_{i-1}$ also calls $H_{i-1}$ to perform $\Pi_{\textsf{LTH.GenMaskShare}}(a\times c,L)$ to obtain $\llb\bM\rrb_{i-1}\in\mathbf{Z}_L^{a\times c}$ and $\hat \bC_{i-1}'=\llb\hat\bC\rrb^L_{i-1}+\llb\bM\rrb_{i-1}$.
    
    \textcolor{blue}{\underline{\textit{Malicious:}} Instead of doing the semi-honest protocol, $P_{i+1},P_{i-1}$ generates $\llb\bM\rrb_{i+1}\gets\Pi_{\textsf{LTH.GenMask}}(a\times c,L,i+1)$ and $\hat \bC_{i+1}'=\llb\hat \bC'\rrb^L_{i+1}+\llb\bM\rrb_{i+1}$. $P_{i-1}$, $P_{i}$ also generates $\llb\bM\rrb_{i-1}$ invoking $\Pi_{\textsf{LTH.GenMask}}'(a\times c,L,i-1)$, $\hat \bC'_{i-1}=\llb\hat \bC'\rrb^L_{i-1}+\llb\bM\rrb_{i-1}$.}
    
    \item $P_i$ and $P_{i-1}$ send $\hat \bC_i'$ and $\hat \bC_{i-1}'$to $P_{i+1}$.
    
    \textcolor{blue}{\underline{\textit{Malicious:}} Instead of doing the semi-honest protocol, $P_{i+1},P_{i-1}$ send $\hat \bC_{i+1}'$ to $P_{i}$; $P_{i-1}$, $P_{i}$ send $\bC'_{i-1}$ to $P_{i+1}$.}
    
    \item $P_{i+1}$ computes $\hat \bC'=\hat \bC_i'+\hat \bC_{i-1}'+\llb\hat \bC'\rrb^L_{i+1}$ and passes it to $H_{i+1}$.
    
    \textcolor{blue}{\underline{\textit{Malicious:}} Instead of doing the semi-honest protocol, $P_{i}$, $P_{i+1}$ compare the two received copies and abort if inconsistency is found. $P_{i}$ computes $\hat \bC'=\hat \bC_{i+1}'+\llb\hat \bC\rrb^L_{i-1}+\llb\hat \bC\rrb^L_{i}$ and passes it to $H_{i}$, $P_{i+1}$ computes $\hat \bC'=\hat \bC_{i-1}'+\llb\hat \bC\rrb^L_{i}+\llb\hat \bC\rrb^L_{i+1}$ and passes it to $H_{i+1}$.}
    
    \item \textbf{{\LTH} Only} 
    : $H_{i+1}$ generates the masks $\llb\bM\rrb_{i+1}$ by invoking $\Pi_{\textsf{LTH.GenMaskShare}}(a\times c,L)$ and recovers the plaintext through truncation: $\hat\bC=\lfloor(\hat \bC'+\llb\bM\rrb_{i+1})\rfloor\gg\textsf{fp}$.
    Note that $\llb\bM\rrb_{i+1}+\llb\bM\rrb_i+\llb\bM\rrb_{i-1}=\mathbf{0}$.
    
    Set $\bD=(\hat\bC>\mathbf{0})$.
    Then $H_{i+1}$ invokes $\Pi_{\textsf{LTH.GenMaskShare}}(a\times c,L)$
    to get 
    $\llb \bZ^*\rrb^L_{i}\in\mathbb{Z}_L^{a\times c}$, 
    $\llb \bZ^*\rrb^L_{i+1}\in\mathbb{Z}_L^{a\times c}$, 
    and compute: $(\llb Z_{j,k}\rrb^L _i,\llb Z_{j,k}\rrb^L _{i+1})=((D_{j,k}?C_{j,k}:0)+\llb Z^*_{j,k}\rrb^L _i,\llb Z^*_{j,k}\rrb^L _{i+1})$. 
    Return them to $P_{i+1}$.
    
    $P_i$ and $P_{i-1}$ call $H_i$ and $H_{i-1}$ to invoke $\Pi_{\textsf{LTH.GenMaskShare}}({a\times c},L)$ to get $\llb \bZ\rrb^L_{i-1}\in\mathbb{Z}_L^{a\times c}$.
    
    \textcolor{blue}{\underline{\textit{Malicious:}} Instead, $H_{i+1}$ generates masks with $\Pi_{\textsf{LTH.GenMask}}(a\times c,L,i-1)$ to recover the plaintext 
    $\hat\bC=\lfloor(\hat \bC'-\llb\bM\rrb_{i-1})\rfloor\gg\textsf{fp}$, with the remaining being the same. 
    $H_{i}$ does as above respectively with index $i$ replacing index $i+1$, replacing 
    $\Pi_{\textsf{LTH.GenMask}}$ with $\Pi_{\textsf{LTH.GenMask}}'$, and fixes that the plaintext results are 
    added to mask share $i$.
    $\Pi_{\textsf{LTH.GenMaskShare}}({a\times c},L)$ is invoked at last to generate the shares.
    $H_{i-1}$ also generates the remaining share for $P_{i-1}$.
    }
    
    \item $P_{i+1}$ send $\llb \bZ\rrb^L_i,\llb \bD\rrb^L_i$ to $P_i$, send $\llb \bZ\rrb^L_{i+1},\llb \bD\rrb^L_{i+1}$ to $P_{i-1}$. Now, $\llb \bZ\rrb^L$ and $\llb \bD\rrb^L$ are calculated and shared with each party.
    
    \textcolor{blue}{\underline{\textit{Malicious:}} $P_{i}$ shares $\llb \bZ\rrb^L_{i-1},\llb \bD\rrb^L_{i-1}$ and $\llb \bZ\rrb^L_i,\llb \bD\rrb^L_i$ to $P_{i-1}$ and $P_{i+1}$ respectively. Each party checks the results from the two parallel computations and aborts if an inconsistency is found.}
\end{enumerate}

\end{algorithm*}

\begin{algorithm*}[!ht]
\small  
\caption{$\Pi_{\textsf{Softmax}} (\mathbf{x})$ compute softmax}
\label{p:softmax}
\setstretch{0}
\raggedright
\textbf{Input. }$\{P_i\}$ have replicative shares of $\mathbf{x}\in\mathbb{Z}_L^{n}$.
\sbline

\textbf{Output. }$\{P_i\}$ get $\llb\exp(\bmx)\rrb^L$ 
\sbline
\textbf{Initial values. } The {\LTH}s save $(q_L ,m_L)$ for later use, setting
$(m_L\gg52)=\lfloor\overline{\exp(L\gg\textsf{fp})\cdot 2^{-q_L}}\rfloor$ with $\textsf{fp}=13$ is the fixed-point precision under $L=2^{32}$, where we save
$q_L\in\mathbb{Z}_{2^{32}}$, $m_L\in\mathbb{Z}_{2^{52}}$, so $(m_L\gg52) \in[0,1)$ and $q_L$ will not overflow.
We note $\bar \bmx$ to be the real values that $\bmx$ represents.
    \begin{enumerate}[topsep=0.5pt,itemsep=0.5ex,partopsep=0ex,parsep=0ex]

        \item For each $\llb x_j\rrb^L$, $P_i$ computes $\bar r = \overline{ \exp(\llb x_j\rrb^L_{i-1}\gg\textsf{fp})}$. Let $\bar r = \overline{ \exp(\llb x_j\rrb^L_{i-1}\gg\textsf{fp}) }= 2^{q_j}\cdot (m_j\gg 52)$ where $m_j$ has no sign, since it is always positive.
        (Notice that $|q_j|=|\lfloor\overline{\log_2( \exp(\llb x_j\rrb^L_{i-1}\gg\textsf{fp}))}\rfloor|=\lfloor\overline{|\llb x_j\rrb^L_{i-1}\gg\textsf{fp}|\cdot \log_2(e))}\rfloor<2^{30}$, so 32 bits are enough to store $q_j$ and support additions without overflow).
         
        Invoke $\Pi_{\textsf{LTH.GenMask}}$ from $H_i$ to generate two masks $\alpha_j\in \mathbb{Z}_{2^{52}}$ and $\beta_j\in\mathbb{Z}_{2^{32}}$ for $j=1,...,n$ with the corresponding dimensions, and send $\{m^*_j= (m_j+\alpha_j)_{2^{52}}, q^{*}_j=(q_j+\beta_j)_{2^{32}}\}$ for $i=1,...,n$ to $P_{i+1}$.
        
        \textcolor{blue}{\underline{\textit{Malicious:}} $P_{i-1}$ follows the same computation to get $\{m^*_j, q^{*}_j\}$ to $P_{i+1}$. $P_{i-1}$, $P_{i+1}$ additionally compute $ '\bar r=\overline{ \exp(\llb x_j\rrb^L_{i+1}\gg\textsf{fp})}$ and obtain $\{'m^*_j, 'q^{*}_j\}$ by masking the mantissa and exponent part with masks from $\Pi_{\textsf{LTH.GenMask}}'$, send them to $P_i$.}
        
        \item $P_{i+1} $ receives $\{q^*_j,m^*_j\}$ and computes $ 2^{\hat q_j}\cdot \hat m_j:=\overline{ \exp((\llb x_j \rrb^L_{i+1}+\llb x_j \rrb^L_{i})_L\gg\textsf{fp}) }$. Send $\{\bmq^*+\hat \bmq,\bmm^*,\hat \bmm\}$ to $H_{i+1}$. 
        
        \textcolor{blue}{\underline{\textit{Malicious:}} $P_{i+1}$ and $P_{i}$ compare the two received copies and abort if an inconsistency is found. $P_i$ do the same computation as above with index $i$ replacing index $i-1$, and send the obtained $\{'\bmq^*+'\hat \bmq,'\bmm^*,'\hat \bmm\}$ to $H_i$.}
        
        \item \textbf{{\LTH} Only}: $H_{i+1}$ Generate $\alpha_j$ and $\beta_j$, Compute $q'_j=(q^*_j+\hat q_j)-\beta_j=q_j+\hat q_j$, $m'_j=(m^*_j-\alpha_j)\cdot \hat m_j=m_j\cdot \hat m_j$. Define the results $ 2^{\hat q'_j}\cdot m'_j:=\overline{ \exp(\llb x_j \rrb^L_{i}\gg\textsf{fp}) }\cdot\overline{ \exp((\llb x_j \rrb^L_{i-1}+\llb x_j \rrb^L_{i+1})_L\gg\textsf{fp})}=\overline{ \exp(((\llb x_j \rrb^L_{i-1}+\llb x_j \rrb^L_{i+1})_L+\llb x_j \rrb^L_{i})\gg\textsf{fp}) }$
        
        For the fixed-point representation $x_j\in[0,L)$, the real value it represents $\bar x_j\in[-L/2\gg\textsf{fp},L/2\gg\textsf{fp})$, $\lceil\log_2(\exp(x))\rceil\in[-(L/2)\gg\textsf{fp}\cdot \log_2(e),(L/2)\gg\textsf{fp}\cdot \log_2(e)]$. Set the $q$-bound: $\textsf{qb} = ((L/2)\gg\textsf{fp})\cdot \log_2(e)$. Define $\overline{ \exp(\bar x_j)}:=2^{q''_j}\cdot (m''_j\gg 52)$, then $q''_j$ should be in $[-\textsf{qb},\textsf{qb}]$.
        
        $q'_j$ and $m'_j$ may alter from the correct $q''_j$ and $m''_j$ for two possible reasons: We are missing an $L$ to be subtracted if $(\llb x_j \rrb^L_{i-1}+\llb x_j \rrb^L_{i+1})_L+\llb x_j \rrb^L_{i} \neq (\llb x_j \rrb^L_i+\llb x_j \rrb^L_{i+1}+\llb x_j \rrb^L_{i-1})_L = x_j$; or $\bar x_j$, or the real value $x_j$ represents is actually negative, so $\bar x_j=((\llb x_j \rrb^L_i+\llb x_j \rrb^L_{i+1}+\llb x_j \rrb^L_{i-1})_L-L)\gg\textsf{fp}$. $H_{i+1}$ runs:
        \begin{enumerate}
        \item If $q'_j\in(0,\textsf{qb}]$, $q''_j=q'_j$, $m''_j=m'_j$. 
        
        \item If $q'_j\in(\textsf{qb},3*\textsf{qb}]$,
        we are missing one $\exp(-L\gg\textsf{fp})$ to be multiplied for either reason mentioned above. Compute $q''_j=q'_j-q_L$, $m''_j=(m^*_j-\alpha_j)\cdot \hat m_j/ m_L$. 
        
         \item If $q'_j\in(3*\textsf{qb},5*\textsf{qb}]$,
        we are missing $\exp(-2L\gg\textsf{fp})$ to be multiplied for both reasons. Compute $q''_j=q'_j-2*q_L$, $m''_j=(m^*_j-\alpha_j)\cdot \hat m_j/ (m_L)^2$. 
        \end{enumerate}
        
        Now $H_{i+1}$ obtains the corrected $\exp(\bar x_j)=2^{q''_j}(m''_j\gg52)$. $H_{i+1}$.
        
        $\bullet$Then compute softmax of the real values directly by $\textsf{Softmax}(\bar \bmx)= \exp(\bar \bmx) /\sum(\exp(\bar \bmx))$ which involves only $\mathcal{O}(n)$ additions, $\mathcal{O}(1)$ multiplications and divisions. Then convert the results to fixed-point representations, and invoke $\Pi_{\textsf{LTH.GenMaskShare}}$ for masks $\llb\bmm\rrb^L_{i+1},\llb\bmm\rrb^L_{i}$ to output $(\llb\mathbf{y}\rrb^L_{i},\llb\mathbf{y}\rrb^L_{i+1})=(\textsf{Softmax}(\mathbf{x})+\llb\bmm\rrb^L_{i}, \llb\bmm\rrb^L_{i+1})$ to $P_{i+1}$. 
        
        $P_i$ and $P_{i-1}$ call $H_i$ and $H_{i-1}$ to generate $\llb\bmm\rrb^L_i= \Pi_{\textsf{LTH.GenMaskShare}}(n)$ as $\llb\mathbf{y}\rrb^L_{i-1}$.
        
        \textcolor{blue}{\underline{\textit{Malicious:}} $H_i$ does the computation accordingly, while also masking the results with the index masks $i$. $P_{i-1}$ calls $H_{i-1}$ to generate $(\llb\bmm\rrb^L_{i+1},\llb\bmm\rrb^L_{i-1})= \Pi_{\textsf{LTH.GenMaskShare}}(n,L)$ as $(\llb\mathbf{y}\rrb^L_{i+1},\llb\mathbf{y}\rrb^L_{i-1})$.}

        \item $P_{i+1}$ shares $\llb\mathbf{y}\rrb^L_{i}$ with $P_i$ and $\llb\mathbf{y}\rrb^L_{i+1}$ with $P_{i-1}$
        
        \textcolor{blue}{\underline{\textit{Malicious:}} $P_{i}$ shares $\llb\mathbf{y}\rrb^L_{i-1}$ to $P_{i-1}$ and $\llb\mathbf{y}\rrb^L_{i}$ to $P_{i+1}$ respectively. Each party checks the results of the two parallel computations and aborts if an inconsistency is found.}

    \end{enumerate}
\end{algorithm*}

One may notice that the workload is not balanced among the three parties if we fix $i$. 
In the protocol, the party index $i$ can be any of $\{1,2,3\}$, which means that the three parties can start the protocol simultaneously with a disjoint dataset. 
Therefore, when provided with a batch $B$ of inputs for evaluation, each party can work on the $B/3$ data and start the corresponding protocol 
simultaneously, balancing resource usage and reducing overall latency. 

\vspace{-0.1in}
\subsection{Extensions to Other Operations}\label{sec:expanse}

$\Pi_{\textsf{ReLU}}$ can be extended to $\Pi_{\textsf{MaxPooling}},\Pi_{\textsf{BatchNorm}}$,$\Pi_{\textsf{LayerNorm}}$ that are common non-linear operations needed in deep learning networks. $\Pi_{\textsf{MaxPooling}}$ needs comparisons and multiplexing. $\Pi_{\textsf{BatchNorm}}$ need about two and $\Pi_{\textsf{LayerNorm}}$ needs about three multiplications for each element on average. Their low complexity allows them to be offloaded to the {\LTH} in a similar way as $\Pi_{\textsf{ReLU}}$ by changing the exact plaintext function executed inside.
$\Pi_{\textsf{MatMulReLU}}$ can be extended to other operations in a similar way by
changing step $5)$ of it\de{ for other plaintext operations to be performed in the {\LTH}}. To optimize neural networks in our experiments, we mainly also use $\Pi_{\textsf{MatMulMaxPoolReLU}}$, $\Pi_{\textsf{MatMulBatchNormReLU}}$ which merge the truncation with different joint non-linear layers.

\de{However, $\Pi_{\textsf{Softmax}}$, which is also one of the most frequently used protocols in deep learning, cannot be adapted in the same way due to the costly exponentiation operations.}


\vspace{-0.05in}
\subsection{Softmax}\label{sec:softmax}

Exponentiation is crucial in modern deep learning models, such as logistic and softmax functions.
In this work, we focus on softmax, which is extensively used in modern models such as Transformers \cite{wolf-etal-2020-transformers}. Classical MPC softmax implementations  \cite{crypten2020,ramachandran2021s++} leads to a large overhead due to two main reasons: 
the complex protocol for approximating exponentiation and the max function applied before softmax.
A recent study \cite{wang2021characterizing} shows that softmax is the main source of overhead when running a Transformer network with an MPC protocol and also introduces a numerical stability problem.

The Softmax on a vector $\bmx$ is defined as follows:

\useshortskip
\begin{equation}
    \textsf{Softmax}(\bmx):=\exp{(\bmx)}/\sum_{i=1}^n\exp{(\bmx_i)}
\end{equation}
In a regular ML setting, exponentiation can easily lead to overflow, a problem exacerbated in fixed-point representations used by MPC protocols. The traditional solution is to subtract the maximum value of the input vector $\bmx$ from every element before applying the softmax function, ensuring the maximum input value is $0$ and preventing overflow. However, this additional $\max$ operation introduces significant MPC overhead as shown in a recent study~\cite{wang2021characterizing}.

A na\"ive extension of the previous protocol for $\exp$ would be to move $\exp$ to the {\LTH}, similar to the other non-linear operations in $\Pi_{\textsf{ReLU}}$. This would not work due to the low computational power of the {\LTH} and the large amount of computation required for $\exp$ compared to other operations. Under our assumption on the trusted hardware (details in \autoref{sec:experiment}), tests show that 1 million 32-bit multiplications take less than a second, while double-precision exponentiation takes over a minute. Unlike simple non-linear operations, $\exp$ needs to be done with floating point arithmetic for high precision, involving tens of multiplications per $\exp$ and creating significant overhead and a new bottleneck for our scheme on a small \LTH.

Our solution is to split and ``offload'' the computation to the untrusted local machine.
The most complex part of the $\exp$ operation is performed by the powerful but untrusted CPU/GPU, and then the results are assembled within the {\LTH}. The protocol is based on the property $\exp(a+b+c) = \exp(a) \exp(b) \exp(c)$, allowing untrusted machines to compute $\exp$ on individual shares so that only simple multiplications are needed on \LTH.
However, the conversion between fixed-point representations and real-number arithmetic is non-trivial under MPC. In \autoref{p:softmax}, we expand the exponent part (see \autoref{sec:notation}) to contain all possible results of $\exp(\llb x\rrb^L_i)$, specifically for $L=2^{32}$. Overflow would not occur after this adjustment, even without invoking the $\max$ function before \textsf{Softmax}.

\subsection{\change{Integrating \scheme into Real Systems}}\label{sec:extension}

\change{
A full implementation of \scheme requires four main functions to be performed by \LTH: attestation during the initialization phase, pseudorandom number generation for masking, communication between \LTH and a host CPU, and the rest of the protocol mainly for in-\LTH computation.
From the functionality point of view, all these operations can be implemented in software on any security processor if it is equipped with a unique device secret key in hardware that can be used for attestation.
Fortunately, most security hardware today supports attestation and meets this requirement.
From the performance point of view, our prototype and experimental evaluation assume that \LTH has hardware AES engines for pseudorandom number generation to match the \LTH-CPU communication bandwidth, while assuming that all other \LTH operations are performed in software. 
More specifically, the performance evaluation is based on software run-time on a tiny microcontroller (Arduino Due) with an ARM Cortext-M3 that is also used in TPM, which represents today's low-end security processor. 
}

\change{
Consider Apple's Secure Enclave~\cite{appleenclave} as another example, which already includes a dedicated nonvolatile storage and a unique ID root (UID) cryptographic key to protect device-specific secrets for remote attestation, . It also includes a true random number generator (TRNG), an AES engine that may be used for pseudo-random number generation, a general-purpose CPU (Secure Enclave Processor), and a communication channel with the main CPU. 
While Apple does not disclose the throughput of the AES engine or the performance of the Secure Enclave Processor, they are likely sufficient for \scheme, as the AES engine is designed to encrypt NAND flash storage, and the processor runs at a high SoC clock frequency. 
Other SoC security subsystems, such as the Synopsys tRoot hardware security module~\cite{Synopsys}, the Rambus RT-630 programmable root-of-trust (RoT)~\cite{Rambus}, and the Qualcomm secure processing unit~\cite{QSPU} also support comparable hardware features, including a device-specific secret key, a hardware AES engine, and a general-purpose processor.
Thus, we believe that \scheme can be realized on today's lightweight security hardware with minimal changes.
}

\change{
\scheme is designed to be used even with a tiny low-performance security processor, but it can also run on a high-performance TEE such as Intel SGX and AMD SEV 
implemented in software inside. Although performance should be better than the low-end \LTH implementation, we believe that \LTH can provide stronger security protection compared to the traditional TEEs (see \autoref{sec:TEE}).
}



\section{Evaluation}
\label{sec:experiment}

\subsection{Experimental Setup}\label{sec:setup}



\textbf{Implementation and baselines.} We implemented {\scheme} in \verb!C++!, building on Falcon \cite{wagh2020falcon}. We introduced new protocols, GPU support for linear layers, and we switched to ZeroMQ for networking. Falcon is the main framework we compare to, but the open-source project was not implemented to support GPUs and does not address a key protocol for Transformers: \textsf{Softmax}. Falcon+ introduces three main improvements: (1) GPU support for linear layers, (2) a new MPC protocol for $\Pi_{\textsf{softmax}}$ using the exponentiation protocol from \cite{crypten2020} combined with Falcon's $\Pi_{\textsf{Div}}$ and $\Pi_{\textsf{Max}}$, and (3) ZeroMQ for networking to ensure better performance and a fair comparison with {\scheme}. These changes do not alter the threat model.

We also compare our scheme with AriaNN \cite{ryffel2020ariann} and CryptGPU \cite{tan2021cryptgpu} as additional pure MPC baselines. 
\autoref{sec:analytical} compares the theoretical complexities for Falcon+, AriaNN, and {\scheme}.
AriaNN and CryptGPU show both advantages and disadvantages relative to Falcon in different settings prior to our optimizations. We discover in our experiment that AriaNN and CryptGPU consume a significant amount of memory: a server with 64GB DRAM can only process ResNet18 inference with a batch size of 8 using AriaNN, whereas 32GB suffices for a batch size of 128 with {\scheme}. Similarly, CryptGPU supports batch sizes of up to 8 for ResNet and 32 for VGG16. We adjusted batch sizes accordingly for these experiments, noting results with smaller batch sizes explicitly. Additionally, we compare {\scheme} with two high-performance TEE-based schemes: a full SGX solution, running entire inferences inside an SGX enclave; and Goten \cite{ng2021goten}, which accelerates TEE-based private inference by offloading linear operations to untrusted GPUs using a secret multiplication protocol with Beaver triples. Both CryptGPU and Goten support only a semi-honest GPU setting.

\textbf{Hardware and network.} We conducted our experiments on Cloudlab c240g5 machines with Ubuntu 20.04 LTS, equipped with an Intel Xeon Silver 4114 10-core CPU (2.20 GHz) and an NVIDIA 12GB P100 GPU. The network setup mirrors previous studies \cite{wagh2020falcon, wagh2019securenn, ryffel2020ariann, mohassel2017secureml}, with a LAN bandwidth of 625 MBps and a ping time of 0.2 ms, and a WAN bandwidth of 40 MBps and a ping time of 70 ms. Both semi-honest and malicious settings were tested. For the {\LTH}-chip, we used an Arduino Due with an Atmel SAM3X8E ARM Cortex-M3 CPU (84MHz, 512 KB of Flash, and up to 96 KB of SRAM), which is used for a commercial implementation of TPM \cite{TPMSPI}, to evaluate the {\LTH} runtime. The {\LTH} assumes a low-pin-count (LPC) bus, resulting in a 15MBps bandwidth limit. 
A maximum of 3MBps of random number generation can be achieved in a TPM~\cite{suciu2010benchmarking}, which is enough for its original use case, but not for our scheme. We assume an additional low-cost hardware AES engine \cite{dong201945nm} achieving a throughput of 14 GBps, making the {\LTH}'s pseudo-random number generation time negligible compared to data transmission time. Other details of the hardware can be seen in \autoref{sec:area}.
For the {\LTH}-SoC, we assume the same Cortex-M3 processor running at 1GHz and the 128-bit on-chip network (16GBps). The performance is estimated by scaling the execution time of the discrete {\LTH}. 
We additionally provide a memory usage analysis of \LTH in \autoref{sec:memory}, showing that \scheme can handle most models with our current \LTH setting, and can be modified to handle even larger models with increased \LTH local communication cost.


\textbf{Neural networks and dataset.}
We use 8 neural networks: a small 3-layer fully-connected network with ReLU activations (Network-A, as in SecureML \cite{mohassel2017secureml}), a small convolutional network with ReLU activation (Network-B, as in \cite{riazi2018chameleon}), a small convolutional network with ReLU activation and maxpooling (Network-C, as in \cite{liu2017oblivious}),  AlexNet \cite{krizhevsky2012imagenet}, VGG16 \cite{simonyan2014very}, ResNet18 \cite{he2016deep}, a small Transformer \cite{vaswani2017attention} \change{and a small Word2Vec \cite{mikolov2013efficient}}. We use a small Transformer and reduce the size of the last layer in Word2Vec to manage the computational expense of \textsf{Softmax} in pure MPC, especially in a WAN setting. The datasets used are MNIST \cite{deng2012mnist} for the first four networks, CIFAR-10 \cite{krizhevsky2014cifar} for AlexNet and VGG16, ImageNet \cite{ILSVRC15} for ResNet18, and Wikitext-2 \cite{merity2016pointer} for the Transformer and Word2Vec.

\textbf{Parameter choice.} As mentioned in \autoref{sec:softmax}, we choose $L=2^{32}$ and $\textsf{fp}=13$ in our implementation. We pick the group size to be 2048 bits in Diffie–Hellman key exchange and use AES-128 for the pseudo-random number generation.

\subsection{Performance}


\autoref{tab:time_semi} and \autoref{tab:time_mal} show the end-to-end latency (in seconds) of the inference on inputs of batch size 128 in semi-honest and malicious settings, respectively. \autoref{tab:comm_sem} and \autoref{tab:comm_mal} report the amount of data transmitted compared to baselines with traffic analysis tools or the data reported in the papers. `-' in the cells indicates that the implementation is missing or the network is too large for CPU evaluation. The brackets in the tables indicate an altered batch size for the cases when a large batch size did not fit into our machine. Only Falcon implemented its work in a malicious setting. 

\begin{table*}[ht]
\small 
\renewcommand\arraystretch{0.9}
    \centering
    \caption{\change{Inference time (s) of the entire batch of size 128 in a semi-honest setting. AriaNN has a reduced batch size of 64 and 8 for VGG16 and ResNet18 due to memory consumption, which also applies to other tables. Brackets indicate an altered batch size.}}\vspace{-.9em}
    \begin{tabular}{x{1.8cm}|x{0.86cm}x{0.86cm}x{0.86cm}x{0.86cm}|x{0.86cm}x{0.86cm}x{0.86cm}x{0.86cm}|x{0.86cm}x{0.86cm}x{0.86cm}x{0.86cm}}
    \hline\hline
        &  \multicolumn{4}{c|}{Network-A} &\multicolumn{4}{c|}{Network-B}
        &\multicolumn{4}{c}{Network-C}\\
        &  LAN & LAN &WAN &WAN & LAN & LAN &WAN &WAN & LAN & LAN &WAN &WAN \\
        Framework&  GPU & CPU &GPU &CPU & GPU & CPU &GPU &CPU & GPU & CPU &GPU &CPU \\\hline
        
        Falcon+ &0.082&0.112&1.478&1.393&0.203&0.2618&1.464&1.292&1.866&2.191&7.321&7.288\\
        AriaNN &0.256 &0.512 &-     &5.504 &-     &-     &-     &-  &3.072 &5.248 &-     &17.02   \\
        \change{CryptGPU }&\change{0.449}&\change{ - }&\change{ 13.49}&\change{ - }&\change{ 0.333 }&\change{ - }&\change{ 9.592 }&\change{ - }&\change{ 0.752 }&\change{ - }&\change{ 19.45 }&\change{ -} \\\hline
        \scheme-chip    &\textbf{0.073}&\textbf{0.077}&\textbf{0.251}&\textbf{0.294}&\textbf{0.117}&\textbf{0.114}&\textbf{0.278}&\textbf{0.293}&\textbf{0.680}&\textbf{1.298}&\textbf{1.024}&\textbf{1.494}\\
       speed-up & 1.11$\times$ &{ 1.43$\times$ }&{ 5.87$\times$ }&{ 4.72$\times$ }&{ 1.73$\times$ }&{ 2.29$\times$ }&{5.25$\times$ }&{ 4.39$\times$ }&{ 1.10$\times$ }&{ 1.68$\times$ }&{ 7.14$\times$ }& {4.87$\times$}\\
        \scheme-SoC    &0.0432	&0.0472	&0.2211	&0.2641	&0.0643	&0.0613	&0.1994	&0.2408 &0.1990	&0.8163	&0.5424	&1.012\\
         speed-up &1.91$\times$	& 2.37$\times$	& 6.69$\times$	& 5.28$\times$	&    3.17$\times$	& 4.28$\times$	& 7.34$\times$	& 5.37$\times$	&    3.78$\times$	& 3.10$\times$	& 13.5$\times$	& 7.20$\times$	   \\\hline
         \change{Goten  }&\change{0.261}&\change{ - }&\change{ 3.304}&\change{ - }&\change{ 0.376}&\change{ - }&\change{ 4.097}&\change{ - }&\change{ 0.602}&\change{ - }&\change{ 5.589}&\change{ - }\\
         \change{Full SGX }&  \multicolumn{4}{c|}{\change{0.462}}&  \multicolumn{4}{c|}{\change{0.461}}&  \multicolumn{4}{c}{\change{0.461}}\\\hline
    \hline
        &\multicolumn{4}{c|}{LeNet}&\multicolumn{4}{c|}{AlexNet}&\multicolumn{4}{c}{Transformer}\\
        &  LAN & LAN &WAN &WAN & LAN & LAN &WAN &WAN & LAN & LAN &WAN &WAN \\
        Framework&  GPU & CPU &GPU &CPU & GPU & CPU &GPU &CPU & GPU & CPU &GPU &CPU \\\hline
        Falcon+ &2.592&4.603&8.867&9.563&4.276&11.78&38.56&43.06&4.026&16.20&321.0&334.7
        \\
        AriaNN &4.480 &7.040 &-     &18.30&9.984&19.20&-&43.52  &-&-&-& -     \\
        \change{CryptGPU }&\change{1.337 }&\change{ - }&\change{ 19.12  }&\change{ - }&\change{ 1.918 }&\change{ - }&\change{ 35.90 }&\change{ - }&\change{ - }&\change{ - }&\change{ - }&\change{ -} \\\hline
        \scheme-chip    &\textbf{0.969}&\textbf{3.075}&\textbf{1.315}&\textbf{3.255}&\textbf{1.564}&\textbf{9.263}&\textbf{2.463}&\textbf{9.449}&\textbf{0.5130}&\textbf{12.18}&\textbf{5.024}&\textbf{16.66}\\
        
        speed-up    & 1.38$\times$ & 1.49$\times$ & 6.74$\times$ & 2.93$\times$ & 1.22$\times$ & 1.27$\times$ & 14.6$\times$ & 4.55$\times$ & 7.84$\times$ & 1.32$\times$& 63.8$\times$ &20.08$\times$\\
        \scheme-SoC &0.2869	&2.392	&0.6328	&2.573 &0.305	&8.029	&1.229	&8.215 &0.3618	&12.03	&4.873	&16.51\\
        speed-up   & 4.66$\times$	& 2.02$\times$	& 6.27$\times$	& 3.72$\times$	&    14.0$\times$	& 1.52$\times$	& 29.2$\times$	& 5.24$\times$	&   11.1$\times$	&1.34$\times$	& 65.8$\times$	&20.26$\times$
        \\\hline
        \change{Goten  }&\change{0.944}&\change{ - }&\change{ 6.233}&\change{ - }&\change{ 0.778}&\change{ - }&\change{ 9.127}&\change{ - }&\change{ -}&\change{ - }&\change{ -}&\change{ - }\\
        \change{ Full SGX } &  \multicolumn{4}{c|}{\change{0.507}}&\multicolumn{4}{c|}{\change{5.031}}&  \multicolumn{4}{c}{\change{0.563}}\\\hline
    \hline
        &\multicolumn{4}{c|}{VGG16}&\multicolumn{4}{c|}{ResNet18}&\multicolumn{4}{c}{\change{Word2Vec}}\\
        &  LAN & LAN &WAN &WAN & LAN & LAN &WAN &WAN & \change{LAN }&\change{ LAN }&\change{WAN }&\change{WAN}\\
        Framework&  GPU & CPU &GPU &CPU & GPU & CPU &GPU &CPU  &\change{GPU }&\change{ CPU }&\change{GPU }&\change{CPU}\\\hline
        Falcon+ &49.36&-&122.1&-&545.9&-&1439&   &\change{1.631}&\change{8.687}&\change{81.93}&\change{90.82 }       \\
        AriaNN  &198.4&-&-&-&1779(8)&-&-&-          &-&-&-&-
        \\
        \change{CryptGPU }&\change{11.31(32)}&\change{ - }&\change{ 55.43(32)}&\change{ - }&\change{ 45.20(8)}&\change{ - }&\change{ 600.3(8)}&\change{ - }&\change{ 1.469}&\change{ - }&\change{ 47.19 }&\change{ - }\\\hline
        \scheme-chip    &\textbf{26.55}&-&\textbf{30.05}&-&\textbf{309.7}&-&\textbf{350.7}&- &\change{\textbf{0.628}}&\change{\textbf{7.556}}&\change{\textbf{1.727}}&\change{\textbf{8.869}}\\
        speed-up    &1.85$\times$&-&4.06$\times$&-&1.76$\times$&-&4.10$\times$&-&\change{2.33$\times$}&\change{1.14$\times$}&\change{47.4$\times$}&\change{10.2$\times$}\\
    \scheme-SoC & 13.06	&-	&16.56	&- &106.9	&-	&148.0	&- &\change{0.593}&\change{7.521}&\change{1.692}&\change{8.835}\\
    speed-up   &3.78$\times$	&-	&7.37$\times$	&-    &5.10$\times$	&-	&9.72$\times$	&- &\change{2.74$\times$}&\change{1.15$\times$}&\change{27.32$\times$}&\change{10.2$\times$}
        \\\hline
        \change{Goten  }&\change{6.208}&\change{ - }&\change{ 25.71}&\change{ - }&\change{ -}&\change{ - }&\change{ -}&\change{ - }&\change{ -}&\change{ - }&\change{ -}&\change{ - }\\
        \change{Full SGX  }&  \multicolumn{4}{c|}{\change{32.422}}&\multicolumn{4}{c|}{\change{8.156}}&  \multicolumn{4}{c}{\change{2.23}}\\\hline\hline
    \end{tabular}

    \label{tab:time_semi}
\end{table*}

In the tables, we compare {\scheme} with Falcon \cite{wagh2020falcon}, AriaNN \cite{ryffel2020ariann}, and \change{CryptGPU \cite{tan2021cryptgpu}, three of the state-of-the-art MPC frameworks implementing different optimizations for non-linear layer inference.} AriaNN does not implement the execution of different parties on separate machines, but instead uses the local simulation of the network for performance evaluation. 
AriaNN does not support Transformer and Word2Vec \change{because it does not support \textsf{Softmax}, and we could not use the open-sourced CryptGPU to run Transformer, because the provided version raised error when generating the secret sharing model due to compatibility issues}. 
Depending on the structure of the machine learning network, {\scheme} with {\LTH}-chip is $4\times $ to $63\times $ or $6\times $ to $59\times $ faster than the state-of-the-art MPC results with semi-honest or malicious settings in WAN / GPU environments. The advantage that we obtain under a LAN or CPU environment is smaller compared to a WAN/GPU environment. In the LAN, communication overhead is significantly reduced. With a CPU, the computation accounts for a larger portion of the execution time. 
These factors reduce the speedup, which is mainly accomplished by reducing the communication overhead of non-linear functions. {\scheme} with {\LTH}-SoC achieves even higher speedup because {\LTH}-SoC has higher performance compared to {\LTH}-chip due to its higher clock frequency and the data movement between an \LTH and a CPU/GPU is also faster on an SoC.
The performance gap between {\LTH}-SoC and {\LTH}-chip is the largest in LAN / GPU environments where the local communication overhead is large. Also, for smaller networks, a GPU can be slower than a CPU. For a small amount of data and a small model, initialization and data movement may take more time than operating directly on a CPU.\de{ For larger models, the improvement with a GPU is more consistent.}

\change{
We also compare {\scheme} with other schemes that rely on a high-performance TEE (Intel SGX): the full SGX solution and Goten \cite{ng2021goten}. 
The full SGX solution assumes that semi-honest parties can securely share their data with one party's SGX for evaluation. The experiments are run on SGX V1 with 16GB enclave memory on Azure Standard DC4s v3. 
For smaller networks such as Network-B, the full SGX solution is slightly slower (0.46 seconds) compared to \scheme (0.12), mainly due to initialization overhead. However, for larger networks such as ResNet18, the full SGX solution only takes 8.15 seconds, while \scheme (\LTH-SoC) takes 148 seconds. 
This result is expected as the performance overhead of MPC-based secure computation is known to be substantially higher compared to the performance overhead of a TEE.
On the other hand, MPC is generally considered to be more secure compared to a high-performance TEE such as Intel SGX.
As discussed in \autoref{sec:TEE}, we believe that \LTH is easier to protect and deploy compared to high-performance TEEs.
}

\begin{figure}
\centering

\includegraphics{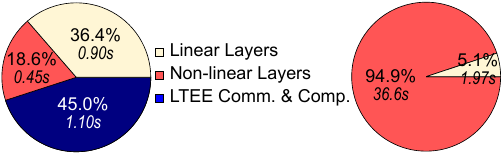}
\vspace{-3mm}
\caption{The breakdown of local machine execution time: linear layers, non-linear layers, and {\LTH}-Chip bus communication \& computation time. {\scheme} (left) and Falcon+ (right) on semi-honest inference over AlexNet under WAN/GPU.
}
\label{fig:chartlinear}
\end{figure}

\change{
When compared to Goten, which also relies on MPC for secure outsourcing of linear layers, the experimental results suggest that {\scheme} is faster in most cases even though \LTH has much lower performance compared to Intel SGX.
Goten is relatively slow for small networks, again partialy due to the SGX initialization overhead. For VGG16, Goten outperforms \scheme with a discrete security chip (\LTH-chip), mainly due to the large performance gap between the low-end \LTH and a high-performance TEE (SGX) used by Goten. 
However, \scheme outperforms Goten in the WAN/GPU setting when running on a more powerful \LTH (\LTH-SoC), which has high local communication bandwidth and runs at a higher clock frequency.\footnote{\change{We believe {\scheme} outperforms in this case because Goten requires more communication for its secure multiplication. As the communication cost analysis for Goten is not available, we estimate the costs using analytical results in their paper (Table 1); and as an example, Goten's communication for VGG16 is estimated to be 273MB compared to {\scheme}'s 188 MB.}}
These results confirm the main intuition behind the \scheme design, that small high-security hardware can be sufficient when primarily used to perform non-linear operations.
}


\begin{table*}[ht]
\small 
\renewcommand\arraystretch{0.9}
    \centering
    \caption{\change{Latency (s) of running the entire batch of size 128 in a malicious setting.}}\vspace{-.9em}
    \begin{tabular}{x{1.8cm}|x{0.86cm}x{0.86cm}x{0.86cm}x{0.86cm}|x{0.86cm}x{0.86cm}x{0.86cm}x{0.86cm}|x{0.86cm}x{0.86cm}x{0.86cm}x{0.86cm}}
    \hline
    \hline
    
        &  \multicolumn{4}{c|}{Network-A} &\multicolumn{4}{c|}{Network-B}
        &\multicolumn{4}{c}{Network-C}\\
        &  LAN & LAN &WAN &WAN & LAN & LAN &WAN &WAN & LAN & LAN &WAN &WAN \\
        Framework&  GPU & CPU &GPU &CPU & GPU & CPU &GPU &CPU & GPU & CPU &GPU &CPU \\\hline
        
        Falcon+ &0.1921&0.3293&3.359&3.3701&0.6279&0.6416&3.504&3.0743&5.639&6.758&23.26&20.8988
\\\hline
        \scheme-chip    &\textbf{0.0913}&\textbf{0.2567}&\textbf{0.5150}&\textbf{0.7307}&\textbf{0.1594}&\textbf{0.2860}&\textbf{0.5803}&\textbf{0.7246}&\textbf{1.157}&\textbf{3.661}&\textbf{2.196}&\textbf{4.258}\\

        speed-up    &2.10$\times$&1.51$\times$&6.52$\times$&4.61$\times$&3.94$\times$&2.82$\times$&6.04$\times$&4.24$\times$&4.87$\times$&2.12$\times$&10.6$\times$&4.91$\times$
\\
        \scheme-SoC &0.0200	&0.1854	&0.4437	&0.6594  &0.0357	&0.1623	&0.4566	&0.6010 &0.0325	&2.536	&1.071	&3.134\\
        speed-up   &9.59$\times$	&2.09$\times$	&7.57$\times$	&5.11$\times$	&    17.5$\times$	&4.98$\times$	&7.67$\times$	&5.12$\times$	&    173$\times$	&3.06$\times$	&21.7$\times$	&6.67$\times$	
        \\\hline
    \hline
        &\multicolumn{4}{c|}{LeNet}&\multicolumn{4}{c|}{AlexNet}&\multicolumn{4}{c}{Transformer}\\
        &  LAN & LAN &WAN &WAN & LAN & LAN &WAN &WAN & LAN & LAN &WAN &WAN \\
        Framework&  GPU & CPU &GPU &CPU & GPU & CPU &GPU &CPU & GPU & CPU &GPU &CPU \\\hline
        Falcon+ &7.492&15.38&28.82&32.57&13.89&41.88&100.9&123.2&10.99&56.10&777.9&821.3\\\hline
        \scheme-chip    &\textbf{2.106}&\textbf{10.64}&\textbf{2.929}&\textbf{10.80}&\textbf{3.653}&\textbf{36.80}&\textbf{5.538}&\textbf{36.10}&\textbf{2.073}&\textbf{47.37}&\textbf{13.03}&\textbf{59.30}\\
        speed-up    &3.56$\times$&1.55$\times$&9.84$\times$&3.01$\times$&3.80$\times$&1.27$\times$&18.2$\times$&3.41$\times$&5.30$\times$&-&59.6$\times$&-\\
        \scheme-SoC    &0.5136	&9.052	&1.337	&9.213  &0.7738	&33.92	&2.659	&33.22    &1.720	&47.02	&12.68	&58.95\\
        speed-up   &14.5$\times$	&1.82$\times$	&21.5$\times$	&3.54$\times$	&    17.9$\times$	&1.37$\times$	&37.9$\times$	&3.71$\times$	&    6.39$\times$	&1.19$\times$	&61.3$\times$	&13.9$\times$
        \\\hline\hline

        &\multicolumn{4}{c|}{VGG16}&\multicolumn{4}{c|}{ResNet18}&\multicolumn{4}{c}{\change{Word2Vec}}\\
        &  LAN & LAN &WAN &WAN & LAN & LAN &WAN &WAN & \change{LAN }&\change{ LAN }&\change{WAN }&\change{WAN} \\
        Framework&  GPU & CPU &GPU &CPU & GPU & CPU &GPU &CPU& \change{GPU }&\change{ CPU }&\change{GPU }&\change{CPU }\\\hline
        Falcon+ &136.9&-&407.8&-&1550&-&4993&-&\change{3.985}&\change{32.80}&\change{202.5}&\change{231.2}
        \\\hline
        \scheme-chip    &\textbf{51.54}&-&\textbf{68.09}&-&\textbf{639.3}&-&\textbf{772.2}&-&\change{\textbf{1.327}}&\change{\textbf{30.44}}&\change{\textbf{2.209}}&\change{\textbf{31.44}}
        \\
        speed-up    &2.66$\times$&-&5.99$\times$&-&2.43$\times$&-&6.47$\times$&-&\change{3.00$\times$}&\change{1.08$\times$}&\change{91.6$\times$}&\change{7.35$\times$}
\\
        \scheme-SoC &20.06	&-	&36.62	&-  &166.2	&-	&299.1	&-&\change{
        1.258}&\change{30.37}&\change{2.140}&\change{31.37}
        \\
        speed-up   &6.82$\times$	&-	&11.1$\times$	&-    &9.33$\times$	&-	&16.6$\times$	&-   &\change{3.17$\times$}&\change{1.08$\times$}&\change{ 94.6$\times$}&\change{7.37$\times$}
        \\\hline
    \hline
    \end{tabular}

    \label{tab:time_mal}

\end{table*}

\begin{table*}[ht]
\small 
    \centering
    \caption{\change{Communication (MB) for the entire batch of size 128 in a semi-honest setting. Brackets indicate an altered batch size.}}\vspace{-.9em}
    \begin{tabular}{cc|ccccccccc}
    \hline
    
        \multicolumn{2}{c|}{Framework} &  Network-A    &Network-B  &Network-C  &LeNet  &AlexNet    &Transformer    &VGG16 &ResNet18 &\change{Word2Vec}\\\hline
        
        Falcon+ &Inter-party    &1.536  &6.272 &64.87  &95.33    &173.5    &72.21    &1730 &22933&\change{12.61}\\\hline

        AriaNN  &Inter-party      &2.816  &  - &38.54      &55.04     &121.6      &-      &1161    &18944&\change{-}\\\hline
        CryptGPU &Inter-party    &3.012  &9.911&33.35&122.7&99.05&-&1714(32)&2729(8)&\change{95.46}\\\hline
        \multirow{2}{*}{\scheme}    &Inter-party    &0.2058	&0.8371	&5.328	&7.931	&12.31	    &19.21 &187.7	&2106	&\change{0.4624}\\\cline{2-11}
        &\LTH-CPU &0.4585	    &0.7958	    &7.235&10.24	&18.53	&2.270   &202.5	&3044	&\change{0.412}\\\hline
    
    \end{tabular}

    \label{tab:comm_sem}
    \vspace{-2mm}
\end{table*}
\begin{table*}[ht]
\small 
\renewcommand\arraystretch{1}
    \centering
    \caption{\change{Inference communication (MB) of the entire batch of size 128 in a malicious setting.}}\vspace{-.9em}
    \begin{tabular}{cc|ccccccccc}
    \hline
    
        \multicolumn{2}{c|}{Framework} &  Network-A    &Network-B  &Network-C  &LeNet  &AlexNet    &Transformer    &VGG16 &ResNet18 &\change{Word2Vec}\\\hline
        
        Falcon+     &Inter-party    &10.51     &41.33     &423.4&620.1    &1135    &340.0    &11543&139287 &\change{99.49}\\\hline
        \multirow{2}{*}{\scheme}    &Inter-party        &0.8443     &2.0704	    &21.02&28.80    &48.68	&131.2     &838.6	&7500 &\change{25.08}	\\
    \cline{2-11}
        &\LTH-CPU &1.070	&1.856	&16.88	&23.90	&43.23&5.296	&472.5	&7103 &\change{0.826}
        \\\hline
    
    \end{tabular}

    \label{tab:comm_mal}
    \vspace{-2mm}
\end{table*}
\begin{table*}[ht]
\small 
\setlength\tabcolsep{3pt}
    \centering
    \caption{\change{Inference time (s) breakdown of a batch of size 128 in a semi-honest WAN/GPU setting, comparing with Falcon+.}}\vspace{-.9em}
    \begin{tabular}{x{1.2cm}x{1.4cm}|x{0.50cm}x{0.50cm}x{0.50cm}x{0.50cm}x{0.50cm}x{0.50cm}x{0.50cm}x{0.50cm}x{0.50cm}x{0.50cm}x{0.50cm}x{0.50cm}x{0.50cm}x{0.50cm}x{0.50cm}x{0.50cm}x{0.50cm}x{0.50cm}}
    \hline
        \multirow{2}{*}{Framework}&  \multirow{2}{*}{Component}&  \multicolumn{2}{c}{Network-A}    &\multicolumn{2}{c}{Network-B} &\multicolumn{2}{c}{Network-C}  &\multicolumn{2}{c}{LeNet} &\multicolumn{2}{c}{AlexNet}   &\multicolumn{2}{c}{Transformer}    &\multicolumn{2}{c}{VGG16}     &\multicolumn{2}{c}{ResNet18}  &\multicolumn{2}{c}{\change{Word2Vec}}\\
        && Time &Ratio& Time &Ratio& Time &Ratio& Time &Ratio& Time &Ratio& Time &Ratio& Time &Ratio& Time &Ratio& \change{Time}&\change{Ratio}\\\hline
        \multirow{3}{*}{\scheme} &CPU/GPU &0.04  & 17\%  & 0.06  & 23\%  & 0.19  & 19\%  & 0.28  & 21\%  & 0.29  & 12\%  & 0.30  & 6\%  & 12.8  & 43\%  & 98.0 & 28\%  & \change{0.47}&\change{27\%}\\ 
       & Comm. &0.18  & 70\%  & 0.16  & 58\%  & 0.35  & 34\%  & 0.35  & 27\%  & 0.94  & 38\%  & 4.56  & 91\%  & 3.34  & 11\%  & 44.0 & 13\%  & \change{1.22}&\change{71\%}\\ 
       & LTH &0.03  & 13\%  & 0.05  & 19\%  & 0.48  & 47\%  & 0.68  & 52\%  & 1.24  & 50\%  & 0.16  & 3\%  & 13.8  & 46\%  & 207 & 59\%  & \change{0.03}&\change{2\%}\\ \hline
        \multirow{2}{*}{Falcon+} &CPU/GPU &0.08  & 6\%  & 0.20  & 14\%  & 1.80  & 25\%  & 2.50  & 28\%  & 4.09  & 11\%  & 3.95  & 1\%  & 47.6  & 39\%  & 523 & 36\%  & \change{1.55}&\change{2\%}\\ 
        &Comm. &1.40  & 94\%  & 1.27  & 86\%  & 5.52  & 75\%  & 6.37  & 72\%  & 34.4  & 89\%  & 317  & 99\%  & 74.4  & 61\%  & 916 & 64\%& \change{89.3}&\change{99\%}\\\hline
    \end{tabular}

    \label{tab:breakdown}
\end{table*}

\de{To study the effect of {\scheme} on execution time breakdown, i}In \autoref{fig:chartlinear}, we show the time breakdown of semi-honest inference over AlexNet in the WAN / GPU setting. In both {\scheme} and Falcon+, linear layers\de{, including convolution layers and fully-connected layers with almost the same protocols,} take a similar amount of time. However, they contribute only 5.1\% of the execution time in Falcon+, and over 36\% in {\scheme}, because the non-linear layers' runtime is significantly reduced from about 94.9\% to 18.6\% (63.6\% if we roughly consider all operations on {\LTH} are related to non-linear layers. There is some overhead, such as the local transmission in step 4 of \autoref{p:matmulrelu}, which cannot be assigned to be only linear or non-linear operations). \autoref{tab:breakdown} shows another breakdown of the execution time in the WAN/GPU setting: CPU/GPU, communication, and {\LTH}. 
Note that the CPU/GPU execution time is also reduced\de{ compared with Falcon+} because major parts of most non-linear computations are moved to {\LTH}.

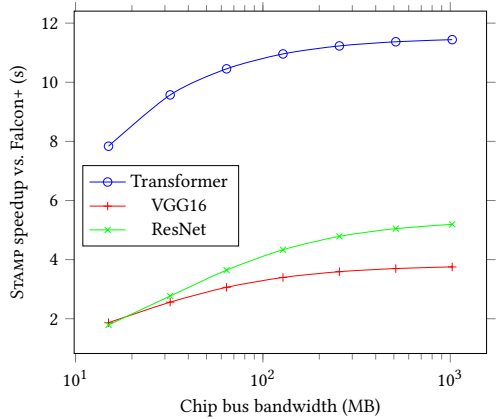
\begin{figure}[htp] 
\centering

\scalebox{0.8}{\begin{tikzpicture} 

\begin{axis}[
    xlabel=Chip bus bandwidth (MB), 
    ylabel=\scheme speedup vs. Falcon+ (s), 
    xmode = log,
    tick align=outside, 
    legend style={at={(0.02,0.55)},anchor=north west}
    ]

\addplot[smooth,mark=o,blue] plot coordinates { 
    (15,7.8363633 )  
    (32,9.57233077)
    (64,10.45380405)
    (128,10.95835719)
    (256,11.22934962)
    (512,11.36993485)
    (1024,11.44155582)
};

\addlegendentry{Transformer}

\addplot[smooth,mark=+,red] plot coordinates { 
    (15,1.86581376 )  
    (32,2.5598407)
    (64,3.06239306)
    (128,3.39571985)
    (256, 3.59116013)
    (512,3.69756665)
    (1024,3.7531701)
};
\addlegendentry{VGG16}
\addplot[smooth,mark=x,green] plot coordinates { 
    (15,1.78917723)  
    (32,2.76678852)
    (64,3.64559595)
    (128,4.33387423)
    (256,  4.78563108)
    (512,5.04876933)
    (1024,5.19149673)
};
\addlegendentry{ResNet}

\end{axis}
\end{tikzpicture}}
\caption{The speedup with different bus bandwidth in the semi-honest setting under LAN/GPU.}
\label{fig:bandwidth}
\end{figure}

\change{
\textbf{Discussion.} The acceleration achieved by {\scheme} varies significantly with the architectural design of the model. Convolutional neural networks (CNNs), such as AlexNet and VGG16, exhibit less pronounced speed improvements compared to language models like the Transformer and Word2Vec. This discrepancy aligns with the observation that language models employ computationally intensive non-linear operations more frequently, notably \textsf{Softmax} in our case. For instance, the Transformer model applies \textsf{Softmax} within each of its multiple attention heads. Word2Vec, despite its simplicity and consisting of only two linear layers, incurs a high computation cost for non-linear operations due to the inclusion of \textsf{Softmax} (we also keep the \textsf{Softmax} in the final layer of Word2Vec to demonstrate its impact). Consequently, \scheme tends to offer greater benefits for models that extensively leverage more complex non-linear operations. In contrast, models such as ResNet18 or VGG16, which are computationally heavy for linear operations but relying on simpler non-linear activation functions like \textsf{ReLU}, do not exhibit as significant speed-ups. 
This observation explains the higher speed-up numbers in the language models compared to the CNNs.
}

\scheme trades off the inter-party communication with the local communication between {\LTH} and the untrusted CPU/GPU. However, as shown in \autoref{tab:comm_sem} and \autoref{tab:comm_mal}, for {\LTH-chip}, the low (15 MBps) bus bandwidth becomes a bottleneck of our performance with large networks, especially in the LAN setting. In \autoref{tab:comm_sem}, almost 3GB of data is transmitted through the {\LTH}-CPU bus for the ResNet18 reference, causing more than 200 seconds of communication time, which is about 60\% of the total execution time of our scheme.
{\LTH-SoC} provides a much higher LTH-CPU bandwidth and significantly alleviates this bottleneck.

\autoref{fig:bandwidth} shows how the speedup over Falcon+ can change if we use a higher-bandwidth interconnect for the {\LTH}. In this figure, we choose two computation-heavy networks, VGG16 and ResNet18, and a communication-heavy Transformer network (due to frequently used \textsf{Softmax}) as examples. 
We can observe a considerable boost in performance with a higher {\LTH} bandwidth. 

\subsection{Accuracy}

The precision of the inference using the plaintext computation and model weights is shown in \autoref{tab:acc}, where the model weights are from the plaintext training. 
For Network-A, B, C and LeNet, which are measured by Falcon \cite{wagh2020falcon}, {\scheme} has the same accuracy that Falcon achieved. {\scheme} optimizes overhead, but does not change computation precision with the same $L$ and $\textsf{fp}$\de{, and we use the same size $L$ of a finite field and the same fixed-point precision}. In most schemes, for each batch of 128 elements, only one more sample would be classified incorrectly compared with the plaintext results, mainly due to the quantization when converting data from floating-point precision to fixed-point precision. 
However\de{, this simple transfer from a plaintext model to a fixed-point scheme may not work well in certain circumstances. I} if some weights and activations are outside of the fix-point representation range,
the accuracy may degrade more significantly. Carefully capping the values during training can potentially help avoid this issue. 


\begin{table}[htp]
\vspace{-1.5mm}
\small 
    \centering
    \caption{Accuracy on different networks. The weights of ResNet18 are from Torchvision \cite{marcel2010torchvision}.}\vspace{-.9em}
    \begin{tabular}{ccc}
   
    \hline
        Network& Plaintext Accuracy& \scheme Accuracy\\\hline
        
        Network-A  &98.18\% &97.42\%         \\
        Network-B  &98.93\% &97.81\%         \\
        Network-C  &99.16\% &98.64\%         \\
        LeNet      &99.76\% &99.15\%     \\
        ResNet18@1 &84.76\% &84.37\%             \\
        ResNet18@5 &95.80\% &95.50\%              \\
    
    \hline
    \end{tabular}
    
    \label{tab:acc}
\end{table}

   
        


\subsection{Hardware Overhead of LTH}\label{sec:area}


The LTH in \scheme consists of two parts:

    (1) The core microcontroller with the same capability as the entire TPM. We refer to the design of ST33TPM12SPI~\cite{TPMSPI} as a baseline with 0.40$mm^2$ area for the ARM SecurCore SC300. 
    The microcontroller has
    a peak power consumption of 12$mW$.
    
    (2) An AES engine performing pseudo-random number generation. A previous study \cite{dong201945nm} reports a cost of 0.13$mm^2$ and 56$mW$ in area and peak power consumption. The AES engine serves as the PRF $F$ in $\Pi_{\textsf{LTH.GenMask}}$ and $\Pi_{\textsf{LTH.GenMaskShare}}$.

The combined overhead of 0.53$mm^2$
and 68$mW$ is quite small, suggesting that {\LTH}\de{ represents a lightweight solution which} is cheaper and easier to deploy compared to adding a TEE to a high-performance processor. 
{\LTH} may even be implemented as a simple extension of the existing TPM hardware or the on-chip SoC security subsystem. Furthermore\de{, because {\LTH}-chip can be built independently of the main compute engines such as CPUs, GPUs, and accelerators}, our protocol can be deployed with existing or future hardware platforms without integrating new TEE features directly into them.
Note that the {\LTH} overhead here does not represent the full power consumption of {\scheme}, which also runs an MPC protocol on an untrusted CPU/GPU.

\subsection{Trusted Computing Base (TCB)}\label{sec:loc}

As the security of a system is difficult to quantify, the TCB size is often used as a proxy when comparing system designs. Our estimates suggest that {\LTH} has a much smaller TCB compared to a high-performance TEE.
For the hardware TCB, open-source microcontrollers whose complexity is comparable to {\LTH} that we use have $<$20k Lines-of-Code (LoC) (OpenRISC: ~16k LoC + AES: ~1k LoC). 
While the LoC for commercial TEE hardware is not publicly available, the area of high-performance processors 
(Intel Skylake: 
$322mm^2\sim698mm^2$, 
Intel Sapphire Rapids: 
$\sim400mm^2$) is much larger than the size of {\LTH} (0.53$mm^2$). 

For the software TCB, our {\LTH} software implementation has $\sim$13k LoC. On the other hand, the software TCB for the Intel SGX experiment includes Gramine ($\sim$50k LoC) and PyTorch ($\sim$166k LoC) inside a TEE. For, virtual machine (VM) based TEEs such as Intel TDX and AMD SEV 
the software TCB can be much larger as the entire operating system (OS), drivers, and ML software stack (PyTorch) all need to run inside a TEE (millions of LoC for Linux).

%
%




\section{Related Work}


\noindent \textbf{Encrypted computation (MPC/HE) for machine learning.} 
Cryptographic techniques such as garbled circuits \cite{chandran2017ezpc, riazi2019xonn}, secret sharing \cite{ryffel2020ariann, mohassel2018aby3, wagh2019securenn}, homomorphic encryption \cite{mohassel2017secureml, zheng2019helen} have been applied for privacy-preserving inference. 
Gazelle and Delphi \cite{juvekar2018gazelle,mishra2020delphi} combine homomorphic encryption and garbled circuits for their advantages in linear and non-linear operations, respectively. 
Falcon \cite{wagh2020falcon} implements a 3-party malicious secure protocol, combining techniques from SecureNN \cite{wagh2019securenn} and ABY3 \cite{mohassel2018aby3}. 
Blaze \cite{patra2020blaze} achieves not only 3-party malicious security but also fairness in an honest majority setting. 
AriaNN \cite{ryffel2020ariann} leverages function secret sharing to reduce communication rounds for specific functions, but at the cost of increasing the total amount of communication data in some cases. 
CrypTen \cite{crypten2020} provides a general software framework that makes secure MPC primitives more easily used by integrating them into a popular ML framework, PyTorch. 
%
GForce~\cite{ng2021gforce} proposed fusing layers in MPC, and more specifically combined dequantization and quantization layers into a truncation before and after ReLU and MaxPooling. 
Our protocol also applies layer fusing when applicable, but in the context of reducing overhead for non-linear operations in \LTH.
Our work leverages the recent developments in MPC for PPML, but shows that a simple security processor can significantly reduce the high overhead of today's MPC-based PPML methods.

\noindent \textbf{Combination of trusted hardware and crypto-based secure computation.} 
Recent studies explored multiple approaches to improve MPC/HE for machine learning using trusted hardware.
However, the previous work typically assumes a high-performance TEE such as Intel SGX and relies on the TEE to perform a significant amount of computation, which will be too slow on a small security processor.
To the best of our knowledge, our work is the first to show that even a small low-performance security processor can significantly improve the performance of MPC if the protocol can be carefully designed for lightweight trusted hardware.

For performance improvements, the previous studied proposed using a TEE (Intel SGX) to accelerate bootstrapping \cite{katz2007universally,lu2021correlated}, perform faster functional encryption \cite{fisch2017iron}, and simplify certain protocols \cite{choi2019hybrid, katz2007universally,felsen2019secure}.  
The previous work also investigated splitting the work between a TEE (Intel SGX) and MPC.
For example, Gupta et al. \cite{gupta2016using} propose splitting secure computation between garbled circuits and Intel SGX. 
Zhou et al. \cite{zhou2022ppmlac} introduce a two-party TEE-aided MPC scheme that focuses on improving multiplication overhead by moving part of the linear operations to a TEE. 
HYBRTC~\cite{wuhybrid} decides where the computation should be run based on whether or not the parties trust a TEE; a hybrid protocol moves the computation to the TEE or just performs an MPC protocol. 
While the high-level approach of offloading computation from MPC to trusted hardware is similar, the previous work offloaded heavy computation to a high-performance TEE while our work studies how to leverage a low-performance security processor.

Slalom \cite{tramer2018slalom} and Darknight \cite{hashemi2020darknight} propose to run a private machine learning computation on an untrusted GPU by securely outsourcing linear operations from the CPU TEE (SGX) to the GPU using secret sharing, and later Goten \cite{ng2021goten} proposed an improved scheme compared with Slalom by introducing ``dynamic quantization" for training. While the use of pseudorandom masks is similar to our protocol in Slalom, Slalom uses masking only for outsourcing linear operations, as the other two papers. As a result, non-linear operations cannot be offloaded, and the CPU TEE still needs to perform as many linear operations as a GPU in an offline phase.  
These approaches require a high-performance TEE, and the TEE performance limits the overall secure computation performance.
{\scheme}, on the other hand, only requires small low-performance trusted hardware for non-linear operations by performing linear operations on untrusted CPUs/GPUs using MPC. Also, Slalom utilizes its pseudorandom masks with the pure additive linear homomorphism of functions. Our approach of computing \textsf{Softmax} has a similar idea but involves multiplicative homomorphism as shown in \autoref{sec:softmax}.

Trusted hardware can also be used to improve the security of an MPC protocol.
For example, CryptFlow \cite{kumar2020cryptflow} runs MPC protocols on Intel SGX and leverages SGX's integrity protection to achieve malicious security.
Another work \cite{broadnax2021fortified} uses Intel SGX to protect the data of parties in MPC even if they are remotely compromised.





\noindent \textbf{Trusted hardware-based privacy-preserving machine learning.} 
The previous work investigated performing and optimizing machine learning computation inside a CPU TEE (Intel SGX) \cite{kim2020vessels}, and providing stronger side-channel protection through data-oblivious computation \cite{ohrimenko2016oblivious}.
The performance of a TEE can be further improved by introducing the TEE capabilities to GPUs \cite{volos2018graviton,jang2019heterogeneous} and domain-specific accelerators \cite{hua2020guardnn}.
While TEE on a high-performance CPU/GPU/accelerator is capable of providing much higher performance compared to MPC-based machine learning computation, the approach comes with the challenges in securing complex high-performance hardware as well as the cost of developing and deploying new hardware and software.
{\scheme} is the first to combine a small security processor with MPC for privacy-preserving machine learning, introducing a new security and performance trade-off.



\section{Conclusion and Future Work}

This paper introduces a new PPML system which significantly reduces the overhead of MPC with the assistance of an LTH. 
{\scheme} can guarantee security against malicious parties in an honest-majority 3-party setting. 
Theoretical analysis and experimental results show that {\scheme} achieves significantly higher performance over state-of-the-art MPC protocols in various environments, even with an LTH whose performance is comparable to a TPM. 

\change{
While {\scheme} provides significant speed-ups, we believe that this work represents the first step in exploring a broad design space of combining cryptographic protection and small high-security hardware to unlock a better security-efficiency trade-off, opening up interesting future directions.
From the system's point of view, while today's SoC security subsystems such as Apple Secure Enclave is closed and its software is tightly controlled by the SoC vendors, it will be valuable if we can integrate \scheme into today's SoCs to more fully understand the performance, security, and functionality of today's \LTH. 
It will also be interesting to broaden the applicability \scheme to a wider array of modern ML models in practice, including Large Language Models (LLMs) and diffusion models. In particular, previous MPC studies found that the polynomial approximation of softmax can cause serious accuracy challenges in large Transformers. We leave a study on practical MPC-based private LLM inference with sufficient performance and accuracy for future work. 
The experiments in this paper, while showing promising speed-ups, also show the challenges from the limited performance of \LTH.
In that sense, further optimizations of the protocol to reduce the overhead, extending the protocols to other types of MPC protocols, and the use of \LTH is other types of operations beyond non-linear layers will be all promising future directions.   
} 

\begin{acks}
This material is based upon work supported by the U.S. National Science Foundation under award No. CCF-2118709. Any opinions, findings, and conclusions or recommendations expressed in this material are those of the author(s) and do not necessarily reflect the views of the National Science Foundation. Thang Hoang was supported by an unrestricted gift from Robert Bosch, 4-VA, and the Commonwealth Cyber Initiative (CCI), an investment in the advancement of cyber R\&D, innovation, and workforce development. For more information about CCI, visit \url{www.cyberinitiative.org}.
\end{acks}

\newpage
\bibliographystyle{ACM-Reference-Format}
\bibliography{BIBTEEMPC}

\appendix




\section{Basic MPC Protocols}\label{sec:MPCbasic}

\textbf{Correlated Randomness.} A large number of random shares have to be obtained by the parties during the offline phase to reduce the communication cost during the offline phase.
The 3-out-of-3 randomness is defined as each $P_i$ holding a share of 0: $\alpha=\llb \alpha\rrb^L_1+\llb \alpha\rrb^L_2+\llb \alpha\rrb^L_3$ where $\alpha=0$ and $P_i$ holds $\llb\alpha\rrb^L_i$. They can be efficiently generated locally by a pseudo-random function (PRF). The security of the PRF function indicates that the output of a PRF is computationally indistinguishable (indistinguishable by a computationally bounded adversary) from the output of a truly random function. 
Given $k_i$ as the key that $P_i$ and $P_{i+1}$ share through a key exchange protocol for each $i$ and $\hat F$ as the PRF that is public to all parties, each $P_i$ can generate shares $\llb \alpha\rrb^L_i$ as $\llb \alpha\rrb^L_i=\hat F_{k_i}(\textsf{ctr})-\hat F_{k_i}(\textsf{ctr})$ with increments of \textsf{ctr} each time this process is invoked. 

\textbf{Input Phase.} To construct $\llb x\rrb^L$ from the generated 3-out-of-3 randomness and $\bar x$ which is provided by $P_i$, $P_i$ compute $\llb x\rrb^L_i = x + \llb \alpha \rrb^L_i$ and share it with $P_{i-1}$. $P_{i-1}$ will send $\llb \alpha \rrb^L_{i-1}$ to $P_{i+1}$ and $P_{i+1}$ will send $\llb \alpha \rrb^L_{i+1}$ to $P_i$. In an honest majority malicious setting, additionally $P_{i+1}$ should compute $ \llb \alpha \rrb^L_{i+1}+\llb \alpha \rrb^L_{i-1}$ and send to $P_i$, then $P_i$ confirm $\llb \alpha \rrb^L_{i+1}+\llb \alpha \rrb^L_{i-1} = -\llb \alpha \rrb^L_i$, therefore $P_{i_1}$ behaves honestly; $P_{i}$ should compute $ \llb \alpha \rrb^L_{i+1}+\llb \alpha \rrb^L_{i}$ and send to $P_{i-1}$, then $P_{i-1}$ confirm by $\llb \alpha \rrb^L_{i+1}+\llb \alpha \rrb^L_{i} = -\llb \alpha \rrb^L_{i-1}$, therefore $P_{i+1}$ behaves honestly. Since $P_i$ is allowed to send an arbitrary input $\bar x$, we do not need to check the share it sends. 

\textbf{Linear Operations.} For RSS shared secrets, most linear operations can be performed locally. For shares $\llbracket x\rrbracket^L$, $\llbracket y\rrbracket^L$, and the public scalar $c$, we have the following:
\begin{itemize}
	\item $\llbracket x\rrbracket^L+ c$  = $(\llbracket x\rrbracket^L_1+c,\llbracket x\rrbracket^L_2,\llbracket x\rrbracket^L_3)$
	\item $c\cdot\llbracket x\rrbracket^L$ = $(c\cdot\llbracket x\rrbracket^L_1,c\cdot\llbracket x\rrbracket^L_2,c\cdot\llbracket x\rrbracket^L_3)$
	\item $\llbracket x\rrbracket^L+\llbracket y\rrbracket^L$ = $(\llbracket x\rrbracket^L_1+\llbracket y\rrbracket^L_1,\llbracket x\rrbracket^L_2+\llbracket y\rrbracket^L_2,\llbracket x\rrbracket^L_3+\llbracket y\rrbracket^L_3)$
\end{itemize}
That can be done without any communication. However, multiplication between shared secrets cannot be done locally.

\textbf{Reconstruction.} $x\gets\Pi_{\textsf{Reconst}}(\llb x\rrb^L)$: To reconstruct the plaintext $x$ from the shares $\llb x\rrb^L$, each party $P_i$ will send $\llb x\rrb^L_i$ to $P_{i+1}$ in the semi-honest setting. After completion, all parties have all 3 shares to reconstruct $x$. In a malicious setting, $P_i$ will also send $\llb x\rrb^L_{i+1}$ to $P_{i-1}$. Then, under the honest majority assumption, at most one of the two copies of the share they receive is altered. A party can compare the values received from the other two and abort if an inconsistency occurs.

\section{Detailed STAMP Protocols for ReLU}\label{sec:reluapp}

In this section, we introduce $\Pi_{\textsf{ReLU}}$, the protocol to offload \textsf{ReLU} operations under MPC to trusted hardware. 
The steps we take are as follows: 
First, $P_i$ invokes $\Pi_{\textsf{LTH.GenMask}}$ to get the pseudo-random masks, and sends $\llb x \rrb_i$ to $P_{i+1}$ after adding them with the masks; 
Second, $P_{i+1}$ adds the received value with the two shares it holds, then sends the results to the {\LTH}; 
Third, $H_{i+1}$ recovers the plaintext value by invoking $\Pi_{\textsf{LTH.GenMask}}$ using the same key and the counter with the same recorded number, and computes \textsf{ReLU} in plaintext, re-masks the result with $\Pi_{\textsf{LTH.GenMaskShare}}$, and sends them back to $P_{i+1}$. 
The other two parties will also generate their common share in the meantime; 
Fourth, $P_{i+1}$ re-shares the received values to complete the construction of the RSS of the outputs.

\begin{algorithm*}[ht]
\small
\caption{$\llb\textsf{ReLU}(\bmx)\rrb^L\gets\Pi_{\textsf{ReLU}}(\llb\bmx\rrb^L)$ Do $\textsf{ReLU}$ on shares of vector $\bmx$}
\label{p:reluapp}
\setstretch{0}
\raggedright
\textbf{Input. }Each $\{P_i\}$ owns $\llb\bmx\rrb^L$.
\sbline
\textbf{Output. }Each $\{P_i\}$ gets $\llbracket\bmz\rrbracket^L=\llbracket\textsf{ReLU}(\bmx)\rrbracket^L$.
\begin{enumerate}[topsep=0.5pt,itemsep=0.5ex,partopsep=0ex,parsep=0ex]
    \item $P_i$ calls $H_i$ to execute
    $\Pi_{\textsf{LTH.GenMask}}(n,L,{i-1})$ to obtain the masks $\bmm_{i-1}\in\mathbf{Z}_L^n$ then compute $\bmx_{i-1}'=\llb\bmx\rrb^L_{i-1}+\bmm_{i-1}$. 
    
    \textcolor{blue}{\underline{\textit{Malicious:}} $P_{i-1}$ generates $\bmm_{i-1}=\Pi_{\textsf{LTH.GenMask}}(n,L,{i-1})$ and $\bmx_{i-1}'=\llb\bmx\rrb^L_{i-1}+\bmm_{i-1}$, and $P_{i-1}$, $P_{i+1}$ also generates $\bmm_{i+1}=\Pi_{\textsf{LTH.GenMask}}(n,L,i+1)$, $\bmx_{i+1}'=\llb\bmx\rrb^L_{i+1}+\bmm_{i+1}$}.

    \item $P_i$ send $\bmx_{i-1}'$ to $P_{i+1}$. 
    
    \textcolor{blue}{\underline{\textit{Malicious:}} $P_{i-1}$ send $\bmx_{i-1}'$ to $P_{i+1}$;  $P_{i-1}$ and $P_{i+1}$ send $\bmx_{i+1}'$ to $P_i$.}
    \item $P_{i+1}$, after receiving $\bmx_{i-1}'$ add it to $\llb\bmx\rrb^L_{i},\llb\bmx\rrb^L_{i+1}$ and pass it to $H_{i+1}$.
    
    \textcolor{blue}{\underline{\textit{Malicious:}} $P_{i}$, $P_{i+1}$ check the two copies received and abort if any inconsistency is found. $P_{i}$ execute as above with index $i-1$ replacing index $i$.}
    \item \textbf{{\LTH} Only} 
    : $H_{i+1}$ recovers the plaintext $\bmx=(\bmx_{i-1}'+\llb\bmx\rrb^L_{i}+\llb\bmx\rrb^L_{i+1})-\bmm_{i-1}$ 
    where $\bmm_{i-1}$ is generated with $\Pi_{\textsf{LTH.GenMask}}(n,L,i-1)$. 
    Set $\bmb=(\bmx>\mathbf{0})$.
    Then $H_{i+1}$ invokes $\Pi_{\textsf{LTH.GenMaskShare}}(n,L)$
    to get 
    $\llb \bmz^*\rrb^L_{i},\llb \bmz^*\rrb^L_{i+1}\in\mathbb{Z}_L^n$, 
    and compute: $(\llb z_j\rrb^L _i,\llb z_j\rrb^L _{i+1})=((b_j?x_j:0)+\llb z^*_j\rrb^L _i,\llb z^*_j\rrb^L _{i+1})$.
    Return their values to $P_{i+1}$.
    
    $H_i$ and $H_{i-1}$ invokes $\Pi_{\textsf{LTH.GenMaskShare}}(n,L)$ to get $\llb \bmz\rrb^L_{i-1}\in\mathbb{Z}_L^n$, and invokes $\Pi_{\textsf{LTH.GenMaskShare}}(n,2)$ to get $\llb \bmb\rrb^L_{i-1}\in\mathbb{Z}_2^n$. Return them respectively to $P_i$ and $P_{i-1}$.
    
    \textcolor{blue}{\underline{\textit{Malicious:}} $H_{i}$ perform the same step as above with index $i-1$ replacing index $i$, replacing $\Pi_{\textsf{LTH.GenMaskShare}}$ with $\Pi_{\textsf{LTH.GenMaskShare}}'$, while also masking the results with the masks of index $i$ (i.e., $(\llb z_j\rrb^L _{i-1},\llb z_j\rrb^L _{i})=(\llb z^*_j\rrb^L _{i-1},(b_j?x_j:0)+\llb z^*_j\rrb^L _{i})$). $H_{i-1}$ and $H_{i+1}$ generate the remaining share $\llb z_j\rrb^L _{i+1}$ for $P_{i-1}$ and $P_{i+1}$ respectively.}
    
    \item $P_{i+1}$ send $\llb \bmz\rrb^L_i$ to $P_i$, send $\llb \bmz\rrb^L_{i+1}$ to $P_{i-1}$. Now $\llb \bmz\rrb^L$ and $\llb \bmb\rrb^L$ are calculated and shared to each party.
    
    \textcolor{blue}{\underline{\textit{Malicious:}} $P_{i}$ shares to $P_{i-1}$ and $P_{i+1}$ respectively. Each party checks the copies they receive and aborts if an inconsistency is found.}
\end{enumerate}

\end{algorithm*}

After the protocol, $\{\mathsf{ctr}^i_{1,2},\mathsf{ctr}^i_{2,3},\mathsf{ctr}^i_{3,1},\mathsf{ctr}^i_s\}$ for $i=1,2,3$ all increase by $n$, and $\{\hat {\mathsf{ctr}}^i_{1,2},\hat {\mathsf{ctr}}^i_{2,3},\hat {\mathsf{ctr}}^i_{3,1},\hat {\mathsf{ctr}^i_s}\}$ for $i=1,2,3$ all increase by $n$ too in a malicious setting. 
The synchronization of the counters together with shared keys guarantees the correlated randomness among the {\LTH}s.

\section{Security Analysis}\label{sec:proof}

\begin{figure}[ht]
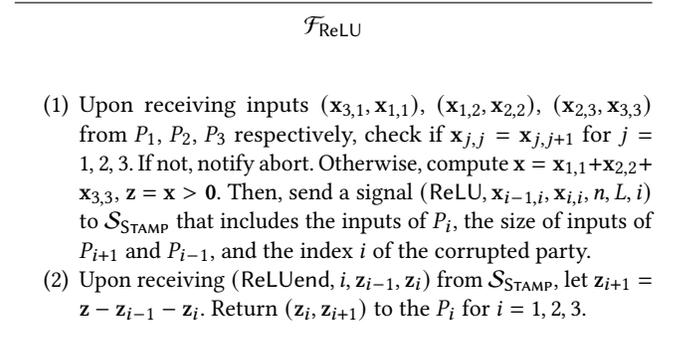

    \begin{func}{$\mathcal{F}_{\textsf{ReLU}}$}
        
        \begin{enumerate}
        \item Upon receiving inputs $( \bmx_{3,1}, \bmx_{1,1})$, $( \bmx_{1,2}, \bmx_{2,2})$, $( \bmx_{2,3}, \bmx_{3,3})$ from $\party{P}_1$, $\party{P}_2$, $\party{P}_3$ respectively, check if $ \bmx_{j,j}= \bmx_{j,j+1}$ for $j=1,2,3$. If not, notify abort. Otherwise, compute $\bmx= \bmx_{1,1}+ \bmx_{2,2}+ \bmx_{3,3}$, $\bmz =\bmx >\mathbf{0}$. Then, send a signal $(\textsf{ReLU}, \bmx_{i-1,i},\bmx_{i,i},n,L,i)$ to $\Sim$ that includes the inputs of $P_i$, the size of inputs of $P_{i+1}$ and $P_{i-1}$, and the index $i$ of the corrupted party. 
        
        \item Upon receiving $(\textsf{ReLUend},i,\bmz_{i-1},\bmz_i)$ from $\Sim$, let $\bmz_{i+1}= \bmz-\bmz_{i-1}-\bmz_i$. Return $( \bmz_i, \bmz_{i+1})$ to the $\party{P}_i$ for $i=1,2,3$.  
        \end{enumerate}
    
    \end{func}
    \caption{Ideal functionality for $\Pi_{\textsf{ReLU}}$.}
    \label{f:relu}
\end{figure}

We claim the following three theorems hold:
\begin{Theorem}\label{t:1}
	Under the assumption of secure PRF and LTH,
	$\Pi_{\textsf{ReLU}}$ (in \autoref{p:reluapp}) securely realizes the ideal functionality $\mathcal{F}_{\textsf{ReLU}}$ (in \autoref{f:relu}) against any non-uniform PPT malicious adversary that can corrupt up to 1 out of 3 parties with static corruption.
\end{Theorem}
\begin{Theorem}\label{t:2}
	Under the assumption of secure PRF and LTH,
	$\Pi_{\textsf{MatMalReLU}}$ (in \autoref{p:matmulrelu}) securely realizes the ideal functionality $\mathcal{F}_{\textsf{MatMulReLU}}$ \de{(in \autoref{f:matmulrelu}) }against any non-uniform PPT malicious adversary that can corrupt up to 1 out of 3 parties with static corruption.
\end{Theorem}
\begin{Theorem}\label{t:3}
	Under the assumption of secure PRF and LTH,
	$\Pi_{\textsf{Softmax}}$ (in \autoref{p:softmax}) securely realizes the ideal functionality $\mathcal{F}_{\textsf{Softmax}}$ \de{(in \autoref{f:softmax}) }against any non-uniform PPT malicious adversary that can corrupt up to 1 out of 3 parties with static corruption.
\end{Theorem}

In this section, we analyze the security of our proposed techniques. The extra or different steps done for a malicious setting than in a semi-honest setting are marked in blue.

\begin{figure}[h]
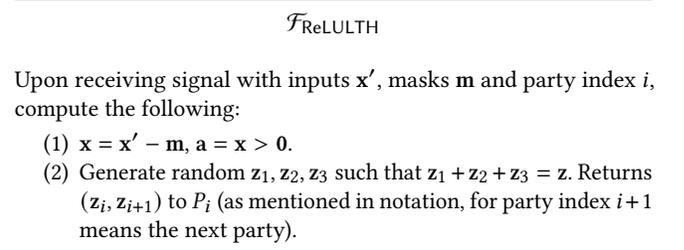

    \begin{func}{$\mathcal{F}_{\textsf{ReLULTH}}$}
        Upon receiving signal with inputs $\bmx'$, masks $\bmm$ and party index $i$, compute the following:
        \begin{enumerate}
            \item $\bmx=\bmx'-\bmm$, $\bma =\bmx >\mathbf{0}$.
            \item Generate random $ \bmz_1, \bmz_2, \bmz_3$ such that $ \bmz_1+ \bmz_2+ \bmz_3=\bmz$. Returns $( \bmz_{i}, \bmz_{i+1})$ to $\party{P}_i$ (as mentioned in notation, for party index $i+1$ means the next party).
        \end{enumerate}
    
    \end{func}
    \caption{Ideal functionality for the \LTH part of $\Pi_{\textsf{ReLU}}.$}
    \label{f:relulth}
\end{figure}

\begin{proof}[Proof for \autoref{t:1}]

We first prove \autoref{t:1} of $\Pi_{\textsf{ReLU}}$ by constructing a simulator \Sim~such that no non-uniform PPT environment $\Env$ can distinguish between: (i) the execution of the real protocol $\textsc{exec}_{\Pi_{\textsf{ReLU}},\mathcal{A},\Env}$ where parties $\party{P}_1,\party{P}_2,\party{P}_3$ run $\Pi_{\textsf{ReLU}}$ and the corrupted parties are controlled by a dummy adversary $\mathcal{A}$ who simply forward messages from/to \Env, and (ii) the ideal execution $\textsc{exec}_{\idealFuncReLU,\Sim,\Env}$ where parties interact with $\idealFuncReLU$, while the simulator \Sim~has the control over the corrupted party. 
Compared to the semi-honest scheme, the changed actions are written in \blue{blue}.\\\vspace{-.7em}


\noindent \textbf{The Environment $\Env$.} 
The environment \Env~provides inputs $( \bmx_{3,1}, \bmx_{1,1})$ to $\party{P}_1$, $( \bmx_{1,2}, \bmx_{2,2})$ to $\party{P}_2$, $( \bmx_{2,3}, \bmx_{3,3})$ to $\party{P}_3$, which are forwarded to the ideal functionality $\mathcal{F}_{\textsf{ReLU}}$ in \autoref{f:relu}.
The environment $\Env$ also indicates which party is corrupted to the ideal functionality.\\\vspace{-.7em}
%
%

%

%

\noindent \textbf{Case 1:} $\party{P}_1$ is corrupted ($i=1$) and $\party{P}_2$, $\party{P}_3$ are honest. \\\vspace{-.7em}

\noindent \textbf{The Simulator.} 
%
\Sim~simulates the following interactions on receiving the signal from $\mathcal{F}_{\textsf{ReLU}}$:

\begin{itemize}
	\item Upon receiving $(\textsf{ReLU},1, \bmx_{3,1},\bmx_{1,1},n,L)$ from $\mathcal{F}_{\textsf{ReLU}}$, \Sim~
 
 generates a random $\hat\bmx_2$, and use $(\bmx_{1,1},\hat\bmx_2)$ and $(\hat\bmx_2, \bmx_{3,1})$ as dummy inputs for $\party{P}_2$ and $\party{P}_3$, respectively.
 
	\item \Sim~acts as $\idealFuncGenMask$ to generate random masks $\bmm_3$ for $\party{P}_1$ and $\party{P}_3$, then computes $\bmx_3'=\bmx_{3,1}+\bmm_3$ and sends ($\bmm_3,\bmx_3'$) to $\party{P}_1$, computes \blue{$\bmx_3''=\bmx_{3,1}+\bmm_3$ and sends ($\bmm_3,\bmx_3''$) to $\party{P}_3$} then also sends $\tilde \bmx_3'$ (as adversary input for $\party{P}_1$) \blue{and $\bmx_3''$} to $\party{P}_2$ on behalf of $\party{P}_1$ and $\party{P}_3$, respectively. 
 
    \blue{\Sim~also acts as $\idealFuncGenMask$ to generate random masks $\bmm_2$ for $\party{P}_3$ and $\party{P}_2$, then computes $\bmx_2'=\hat\bmx_2+\bmm_2$ and sends ($\bmm_2,\bmx_2'$) to $\party{P}_3$, computes $\bmx_2''=\hat\bmx_2+\bmm_2$ and sends ($\bmm_2,\bmx_2''$) to $\party{P}_2$ then also sends $\bmx_2'$ and $\bmx_2''$ to $\party{P}_1$ on behalf of $\party{P}_3$ and $\party{P}_2$, respectively.}
    \item \blue{\Sim~check $\bmx_3''=\tilde \bmx_3'$, $\bmx_2''=\bmx_2'$ on behalf of $\party{P}_2$ and $\party{P}_1$ respectively, signal abort to \idealFuncReLU if inconsistency found.}
	\item \Sim~acts as~$\idealFuncLTHReLU$ with $\party{P}_2$'s input $\hat \bmx =\bmx_3'+\hat \bmx_2+\bmx_{1,1}$ to (re)generate $\bmm_3$, randoms $\bmz^*_1,\bmz^*_2,\bmz_3$ such that $\bmz^*_1+\bmz^*_2+\bmz_3=0$, and then compute
    $ (\bmz_{1}, \bmz_{2})=(\textsf{ReLU}(\hat \bmx - \bmm_3)+\bmz^*_1, \bmz^*_2)$. \Sim~sends $ (\bmz_{1}, \bmz_{2})$ to $\party{P}_2$. \Sim~sends $\tilde \bmz_{1}$ to $\party{P}_1$ and sends $\bmz_{2}$ to $\party{P}_3$ (on behalf $\party{P}_2$).

    \blue{\Sim~acts as~$\idealFuncLTHReLU$ with $\party{P}_1$'s input $\hat \bmx' =\bmx_2'+\bmx_{3,1}+\bmx_{1,1}$ to (re)generate $\bmm_2$, randoms $\bmz^*_1,\bmz^*_2,\bmz_3$ such that $\bmz^*_1+\bmz^*_2+\bmz_3=0$, and then compute
    $ (\bmz_{3}, \bmz'_{1})=(\bmz_3,\textsf{ReLU}(\hat \bmx' - \bmm_3)+\bmz^*_1)$. \Sim~sends $ (\bmz_{3}, \bmz'_{1})$ to $\party{P}_1$. \Sim~ sends $\tilde \bmz_{3}$ to $\party{P}_3$ and sends $\tilde \bmz'_{1}$ to $\party{P}_2$ (as adversary inputs on behalf $\party{P}_1$).}


    \item \Sim~acts as $\idealFuncGenMaskShr$ for $\party{P}_1$ to (re)generate random $\bmz_3$ and sends it to $\party{P}_1$. Similarly, \Sim~acts as $\idealFuncGenMaskShr$ for $\party{P}_3$ to (re)generate random as $\bmz_3$ and sends it to $\party{P}_3$. \blue{\Sim~also generate $ \bmz^*_2$ for $\party{P}_2$ and $\party{P}_3$}.

    \item \blue{\Sim~checks if $\bmz'_{1}=\bmz_{1}$ received by $\party{P}_1$, also $\party{P}_2$; if same $\bmz^*_{2}$ received by $\party{P}_2$, also $\party{P}_3$; if same $\bmz_{3}$ received by $\party{P}_3$, also $\party{P}_1$. Signal abort to \idealFuncReLU if an inconsistency is found. If no abort is signaled,} \Sim~signals $(\textsf{ReLUend},1,\bmz_3,\bmz_1)$ to $\mathcal{F}_{\textsf{ReLU}}$.
		
\end{itemize}

\noindent \textbf{Indistinguishability.} We prove the indistinguishability argument by constructing a sequence of hybrid games as follows.

\noindent \textbf{Hybrid $\Hybrid_0$}:
This is the real protocol execution.

\noindent \textbf{Hybrid $\Hybrid_1$}:
$\Hybrid_1$ is the same as $\Hybrid_0$, except that $\Pi_{\textsf{LTH.GenMask}}$ is replaced with simulated $\idealFuncGenMask$ that outputs random $ \hat \bmm_3,\hat \bmm_2$ for both step 1) and 4). 

We claim that $\Hybrid_0$ and $\Hybrid_1$ are computationally indistinguishable.
This is because $\Pi_{\textsf{LTH.GenMask}}$ generates pseudo-random $\bmm_3, \bmm_2$ to $\party{P}_1$ using \LTH. 
Due to the secure hardware assumption, there exists a simulator that is indistinguishable from the real hardware protocol execution.
Moreover, due to the security of PRF used in LTH, 
the random $ \bmm_3,\bmm_2$ produced by LTH is computationally indistinguishable from the random $ \hat \bmm_3, \hat \bmm_2$ generated by the simulator.
Therefore,
$\Hybrid_0$ and $\Hybrid_1$ are computationally indistinguishable.

\noindent \textbf{Hybrid $\Hybrid_2$}:
$\Hybrid_2$ is the same as $\Hybrid_1$, except that we replace step 4) with the simulated $\idealFuncLTHReLU$.

We claim that $\Hybrid_1$ and $\Hybrid_2$ are computationally indistinguishable using the same argument on the trusted hardware and PRF security as in $\Hybrid_1$. 
%
%
Specifically, the random vectors generated by $\Pi_{\textsf{LTH.GenMaskShare}}$ in the LTH are based on PRF and therefore, they are computationally indistinguishable from the random vectors generated by the simulator. Therefore,
$\Hybrid_1$ and $\Hybrid_2$ are computationally indistinguishable.

\noindent \textbf{Hybrid $\Hybrid_3$}:
$\Hybrid_3$ is the same as $\Hybrid_2$, except that $\party{P}_2, \party{P}_3$  use dummy inputs for interaction, instead of the ones provided by the environment. 
In this hybrid, we introduce an ideal functionality $\mathcal{F}_{\textsf{ReLU}}$ that takes the environments' actual inputs and returns the corresponding outputs. 

We claim that $\Hybrid_2$ and $\Hybrid_3$ are indistinguishable.
Since the corrupted party is $\party{P}_1$, \Sim~knows $\bmx_{3,1}=\bmx_{3,3}, \bmx_{1,1}=\bmx_{1,2}$. The  dummy inputs would be $ \bmx_{2,2}= \bmx_{2,3}$ (represented by $\hat \bmx_2$ in \Sim). The distribution of the computation result, $\hat \bmx =\bmx_3'+\hat \bmx_2+\bmx_{1,1}=(\bmx_{3,1}+\bmm_3)+\hat \bmx_2+\bmx_{1,1}$ and $\hat \bmx' =\bmx_2'+\bmx_{3,1}+\bmx_{1,1} = (\hat \bmx_2+\bmm_2)+\bmx_{3,1}+\bmx_{1,1}$ are uniformly random since $\bmm_3, \bmm_2$ are random. Therefore, $\Hybrid_2$ and $\Hybrid_3$ are indistinguishable.

The adversary's view of $\Hybrid_3$ is identical to $\textsf{EXEC}_{\idealFunc,\Sim,\Env}$. Therefore, in \textbf{Case 1} the view of $\mathcal{A}$ and \Env are indistinguishable in the real and the simulated world.

Putting it all together, we have that
$\Hybrid_0 \approx \Hybrid_1 \approx \Hybrid_2 \approx \Hybrid_3 = \Sim$.\\\vspace{-.7em}

\noindent \textbf{Case 2:} $\party{P}_2$ is corrupted ($i=2$) and $\party{P}_1$, $\party{P}_3$ are honest.   \\\vspace{-.7em}

\noindent \textbf{The Simulator.}
%
\Sim~simulates the following interactions on receiving the signal from $\mathcal{F}_{\textsf{ReLU}}$:

\begin{itemize}
	\item Upon receiving $(\textsf{ReLU},2, \bmx_{1,2},\bmx_{2,2},n,L)$ from $\mathcal{F}_{\textsf{ReLU}}$, \Sim~
 
 generates a random $\hat\bmx_3$, and use $(\bmx_{2,2},\hat\bmx_3)$ and $(\hat\bmx_3, \bmx_{1,2})$ as dummy inputs for $\party{P}_3$ and $\party{P}_1$, respectively.
 
	\item \Sim~acts as $\idealFuncGenMask$ to generate random masks $\bmm_3$ for $\party{P}_1$ and $\party{P}_3$, then computes $\bmx_3'=\hat\bmx_3+\bmm_3$ and sends ($\bmm_3,\bmx_3'$) to $\party{P}_1$, computes \blue{$\bmx_3''=\hat\bmx_3+\bmm_3$ and sends ($\bmm_3,\bmx_3''$) to $\party{P}_3$} then also sends $\bmx_3'$ \blue{and $\bmx_3''$} to $\party{P}_2$ on behalf of $\party{P}_1$ and $\party{P}_3$, respectively. 
 
    \blue{\Sim~also acts as $\idealFuncGenMask$ to generate random masks $\bmm_2$ for $\party{P}_3$ and $\party{P}_2$, then computes $\bmx_2'=\bmx_{2,2}+\bmm_2$ and sends ($\bmm_2,\bmx_2'$) to $\party{P}_3$, computes $\bmx_2''=\bmx_{2,2}+\bmm_2$ and sends ($\bmm_2,\bmx_2''$) to $\party{P}_2$ then also sends $\bmx_2'$ and $\tilde\bmx_2''$  (as adversary input for $\party{P}_3$)  to $\party{P}_1$ on behalf of $\party{P}_3$ and $\party{P}_2$, respectively.}
    \item \blue{\Sim~check $\bmx_3''=\bmx_3'$, $\tilde\bmx_2''=\bmx_2'$ on behalf of $\party{P}_2$ and $\party{P}_1$ respectively, signal abort to \idealFuncReLU if inconsistency found.}
	\item \Sim~acts as~$\idealFuncLTHReLU$ with $\party{P}_2$'s input $\hat \bmx =\bmx_3'+\bmx_{2,2}+\bmx_{1,2}$ to (re)generate $\bmm_3$, randoms $\bmz^*_1,\bmz_2,\bmz^*_3$ such that $\bmz^*_1+\bmz_2+\bmz^*_3=0$, and then compute
    $ (\bmz_{1}, \bmz_{2})=(\textsf{ReLU}(\hat \bmx - \bmm_3)+\bmz^*_1, \bmz_2)$. \Sim~sends $ (\bmz_{1}, \bmz_{2})$ to $\party{P}_2$. \Sim~sends $\tilde\bmz_{1}$ to $\party{P}_1$ and sends $\tilde\bmz_{2}$ to $\party{P}_3$  (as adversary inputs for $\party{P}_2$) .

    \blue{\Sim~acts as~$\idealFuncLTHReLU$ with $\party{P}_1$'s input $\hat \bmx' =\bmx_2'+\hat\bmx_3+\bmx_{1,2}$, (re)generate $\bmm_2$, generate randoms $\bmz^*_1,\bmz_2,\bmz^*_3$ such that $\bmz^*_1+\bmz_2+\bmz^*_3=0$, and then compute
    $ (\bmz_{3}, \bmz'_{1})=(\bmz^*_3,\textsf{ReLU}(\hat \bmx' - \bmm_3)+\bmz^*_1)$. \Sim~sends $ (\bmz_{3}, \bmz'_{1})$ to $\party{P}_1$. 
    \Sim~sends $\bmz_{3}$ to $\party{P}_3$ 
    and sends $\bmz'_{1}$ to $\party{P}_2$
    (on behalf of 
    $\party{P}_1$
    ). }


    \item \Sim~acts as $\idealFuncGenMaskShr$ for $\party{P}_1$ to (re)generate random $ \bmz^*_3$ and sends it to $\party{P}_1$. Similarly, \Sim~acts as $\idealFuncGenMaskShr$ for $\party{P}_3$ to (re)generate random $ \bmz^*_3$ and sends it to $\party{P}_3$. \blue{\Sim~also generate $ \bmz_2$ for $\party{P}_2$ and $\party{P}_3$}.

    \item \blue{\Sim~checks if $\bmz'_{1}=\tilde\bmz_{1}$ received by $\party{P}_1$, also $\party{P}_2$; if same $\bmz_{2}$ received by $\party{P}_2$, also $\party{P}_3$; if same $\bmz^*_{3}$ received by $\party{P}_3$, also $\party{P}_1$. Signal abort to \idealFuncReLU if an inconsistency is found. If no abort is signaled,} \Sim~signals $(\textsf{ReLUend},2,\bmz_1,\bmz_2)$ to $\mathcal{F}_{\textsf{ReLU}}$.
		
\end{itemize}

\noindent \textbf{Indistinguishability.} We prove the indistinguishability argument by constructing a sequence of hybrid games as follows. Notice that the first 3 games, \textbf{Hybrid $\Hybrid_0$}, \textbf{Hybrid $\Hybrid_1$} and \textbf{Hybrid $\Hybrid_2$} are identical as \textbf{Case 1}'s. The proofs between \textbf{Hybrid $\Hybrid_0$} and \textbf{Hybrid $\Hybrid_1$}, \textbf{Hybrid $\Hybrid_1$} and \textbf{Hybrid $\Hybrid_2$} are exactly the same.






\noindent \textbf{Hybrid $\Hybrid_3$}:
$\Hybrid_3$ is the same as $\Hybrid_2$, except that $\party{P}_1, \party{P}_3$  use dummy inputs for interaction, instead of the ones provided by the environment. 
In this hybrid, we introduce an ideal functionality $\mathcal{F}_{\textsf{ReLU}}$ that takes the environments' actual inputs and returns the corresponding outputs. 

We claim that $\Hybrid_2$ and $\Hybrid_3$ are indistinguishable.
Since the corrupted party is $\party{P}_2$, \Sim~knows $\bmx_{2,2}=\bmx_{2,3}, \bmx_{1,2}=\bmx_{1,1}$. The  dummy inputs would be $ \bmx_{3,3}= \bmx_{1,3}$ (represented by $\hat \bmx_3$ in \Sim). The distribution of the computation result, $\hat \bmx =\bmx_3'+\bmx_{2,2}+\bmx_{1,2}=(\hat\bmx_3+\bmm_3)+\bmx_{2,2}+\bmx_{1,2}$ and $\hat \bmx' =\bmx_2'+\hat\bmx_3+\bmx_{1,2} = (\hat \bmx_{2,2}+\bmm_3)+\hat\bmx_3+\bmx_{1,2}$ are uniformly random since $\bmm_3,\bmm_2$ are random. Therefore, $\Hybrid_2$ and $\Hybrid_3$ are indistinguishable.

The adversary's view of $\Hybrid_3$ is identical to $\textsf{EXEC}_{\idealFunc,\Sim,\Env}$. Therefore, in \textbf{Case 2} the view of $\mathcal{A}$ and \Env are indistinguishable in the real and the simulated world.

Putting it all together, we have that
$\Hybrid_0 \approx \Hybrid_1 \approx \Hybrid_2 \approx \Hybrid_3 = \Sim$.\\\vspace{-.7em}

\noindent \textbf{Case 3:} $\party{P}_3$ is corrupted ($i=3$) and $\party{P}_1$, $\party{P}_2$ are honest. \\\vspace{-.7em}

\noindent \textbf{The Simulator.}
%
\Sim~simulates the following interactions on receiving the signal from $\mathcal{F}_{\textsf{ReLU}}$:

\begin{itemize}
	\item Upon receiving $(\textsf{ReLU},3, \bmx_{2,3},\bmx_{3,3},n,L)$ from $\mathcal{F}_{\textsf{ReLU}}$, \Sim~
 
 generates a random $\hat\bmx_1$, and use $(\bmx_{3,3},\hat\bmx_1)$ and $(\hat\bmx_1, \bmx_{2,3})$ as dummy inputs for $\party{P}_1$ and $\party{P}_2$, respectively.
 
	\item \Sim~acts as $\idealFuncGenMask$ to generate random masks $\bmm_3$ for $\party{P}_1$ and $\party{P}_3$, then computes $\bmx_3'=\bmx_{3,3}+\bmm_3$ and sends ($\bmm_3,\bmx_3'$) to $\party{P}_1$, computes \blue{$\bmx_3''=\bmx_{3,3}+\bmm_3$ and sends ($\bmm_3,\bmx_3''$) to $\party{P}_3$} then also sends $\bmx_3'$ \blue{and arbitrary $\tilde \bmx_3''$} to $\party{P}_2$ on behalf of $\party{P}_1$ and $\party{P}_3$, respectively. 
 
    \blue{\Sim~also acts as $\idealFuncGenMask$ to generate random masks $\bmm_2$ for $\party{P}_3$ and $\party{P}_2$, then computes $\bmx_2'=\bmx_{2,3}+\bmm_2$ and sends ($\bmm_2,\bmx_2'$) to $\party{P}_3$, computes $\bmx_2''=\bmx_{2,3}+\bmm_2$ and sends ($\bmm_2,\bmx_2''$) to $\party{P}_2$ then also sends $\tilde\bmx_2'$  (as adversary input for $\party{P}_3$)  and $\bmx_2''$ to $\party{P}_1$ on behalf of $\party{P}_3$ and $\party{P}_2$, respectively.}
    \item \blue{\Sim~check $\tilde\bmx_3''=\bmx_3'$, $\bmx_2''=\tilde\bmx_2'$ on behalf of $\party{P}_2$ and $\party{P}_1$ respectively, signal abort to \idealFuncReLU if inconsistency found.}
	\item \Sim~acts as~$\idealFuncLTHReLU$ with $\party{P}_2$'s input $\hat \bmx =\bmx_3'+\bmx_{2,3}+\hat \bmx_1$ to (re)generate $\bmm_3$, randoms $\bmz^*_1,\bmz_2,\bmz_3$ such that $\bmz^*_1+\bmz_2+\bmz_3=0$, and then compute
    $ (\bmz^{**}_{1}, \bmz_{2})=(\textsf{ReLU}(\hat \bmx - \bmm_3)+\bmz^*_1, \bmz_2)$. \Sim~sends $ (\bmz^{**}_{1}, \bmz_{2})$ to $\party{P}_2$. \Sim~sends $\bmz^{**}_{1}$ to $\party{P}_1$ and sends $\bmz_{2}$ to $\party{P}_3$ (on behalf $\party{P}_2$).

    \blue{\Sim~acts as~$\idealFuncLTHReLU$ with $\party{P}_1$'s input $\hat \bmx' =\bmx_2'+\hat\bmx_1+\bmx_{3,3}$, (re)generate $\bmm_2$, generate randoms $\bmz^*_1,\bmz_2,\bmz_3$ such that $\bmz^*_1+\bmz^*_2+\bmz_3=0$, and then compute
    $ (\bmz_{3}, \bmz'_{1})=(\bmz_3,\textsf{ReLU}(\hat \bmx' - \bmm_3)+\bmz^*_1)$. \Sim~sends $ (\bmz_{3}, \bmz'_{1})$ to $\party{P}_1$. \Sim~sends $\bmz_{3}$ to $\party{P}_3$ and sends $\bmz'_{1}$ to $\party{P}_2$ (on behalf $\party{P}_1$).}


    \item \Sim~acts as $\idealFuncGenMaskShr$ for $\party{P}_1$ to (re)generate random $\bmz_3$ and sends it to $\party{P}_1$. Similarly, \Sim~acts as $\idealFuncGenMaskShr$ for $\party{P}_3$ to (re)generate random $\bmz_3$ and sends it to $\party{P}_3$. \blue{\Sim~also generate $ \bmz_2$ for $\party{P}_2$ and $\party{P}_3$}.

    \item \blue{\Sim~checks if $\bmz'_{1}=\bmz^{**}_{1}$ received by $\party{P}_1$, also $\party{P}_2$; if same $\bmz_{2}$ received by $\party{P}_2$, also $\party{P}_3$; if same $\bmz_{3}$ received by $\party{P}_3$, also $\party{P}_1$. Signal abort to \idealFuncReLU if an inconsistency is found. If no abort is signaled,} \Sim~signals $(\textsf{ReLUend},3,\bmz_2,\bmz_3)$ to $\mathcal{F}_{\textsf{ReLU}}$.
		
\end{itemize}

\noindent \textbf{Indistinguishability.} We prove the indistinguishability argument by constructing a sequence of hybrid games as follows. Notice that the first 3 games, \textbf{Hybrid $\Hybrid_0$}, \textbf{Hybrid $\Hybrid_1$} and \textbf{Hybrid $\Hybrid_2$} are identical as \textbf{Case 1}'s. The proofs between \textbf{Hybrid $\Hybrid_0$} and \textbf{Hybrid $\Hybrid_1$}, \textbf{Hybrid $\Hybrid_1$} and \textbf{Hybrid $\Hybrid_2$} are exactly the same.






\noindent \textbf{Hybrid $\Hybrid_3$}:
$\Hybrid_3$ is the same as $\Hybrid_2$, except that $\party{P}_1, \party{P}_3$  use dummy inputs for interaction, instead of the ones provided by the environment. 
In this hybrid, we introduce an ideal functionality $\mathcal{F}_{\textsf{ReLU}}$ that takes the environments' actual inputs and returns the corresponding outputs. 

We claim that $\Hybrid_2$ and $\Hybrid_3$ are indistinguishable.
Since the corrupted party is $\party{P}_2$, \Sim~knows $\bmx_{2,3}=\bmx_{2,2}, \bmx_{3,3}=\bmx_{3,1}$. The  dummy inputs would be $ \bmx_{1,1}= \bmx_{1,2}$ (represented by $\hat \bmx_1$ in \Sim). The distribution of the computation result, $\hat \bmx =\bmx_3'+\bmx_{2,3}+\hat \bmx_1=(\bmx_{3,3}+\bmm_3)+\bmx_{2,3}+\hat \bmx_1$ and $\hat \bmx' =\bmx_2'+\hat\bmx_1+\bmx_{3,3} = (\hat \bmx_{2,3}+\bmm_3)+\hat\bmx_1+\bmx_{3,3}$ are uniformly random since $\bmm_3,\bmm_2$ are random. Therefore, $\Hybrid_2$ and $\Hybrid_3$ are indistinguishable.

The adversary's view of $\Hybrid_3$ is identical to $\textsf{EXEC}_{\idealFunc,\Sim,\Env}$. Therefore, in \textbf{Case 1} the view of $\mathcal{A}$ and \Env are indistinguishable in the real and the simulated world.

Putting it all together, we have that
$\Hybrid_0 \approx \Hybrid_1 \approx \Hybrid_2 \approx \Hybrid_3 = \Sim$ and this completes the proof.
\end{proof}

Notice that as discussed in \autoref{sec:expanse}, the above proof works identically for \textsf{MaxPooling} and \textsf{BatchNorm} since only the plaintext computations
after subtracting the masks are different.

Next, we will prove the \autoref{t:2}. We provide the complete proof for case 1 (where $\party{P}_1$ is corrupted) due to the space limit, and the proof of the other two cases are similar as provided in the proof for $\Pi_{\textsf{ReLU}}$.

\begin{figure}[h]
    \begin{func}{$\mathcal{F}_{\textsf{MatMulReLU}}$}
        
        \begin{enumerate}
        \item Upon receiving inputs $(\bA_{3,1}, \bA_{1,1},\bB_{3,1}, \bB_{1,1})$, $( \bA_{1,2}, \bA_{2,2}, \bB_{1,2}, \bB_{2,2})$, $( \bA_{2,3}, \bA_{3,3},\bB_{2,3}, \bB_{3,3})$ from $\party{P}_1$, $\party{P}_2$, $\party{P}_3$ respectively, check if $ \bA_{j,j}= \bA_{j,j+1},\bB_{j,j}= \bB_{j,j+1}$ for $j=1,2,3$. If not, notify abort. Otherwise, compute $\bC'= (\bA_{1,1}+\bA_{2,2}+\bB_{3,3})\times(\bB_{1,1}+\bB_{2,2}+\bB_{3,3})\gg\textsf{fp}$, $\bC =\bC' >\mathbf{0}$. Then, send a signal $(\textsf{MatMalReLU},i, \bA_{i-1,i},\bA_{i,i},\bB_{i-1,i},\bB_{i,i},a,b,c,L)$ to $\Sim$ that includes the inputs of $P_i$, the size of inputs of $P_{i+1}$ and $P_{i-1}$, and the index $i$ of the corrupted party. 
        
        \item Upon receiving $(\textsf{MatMalReLUend},i,\bC_{i-1},\bC_i)$ from $\Sim$, let $\bC_{i+1}= \bmz-\bC_{i-1}-\bC_i$. Return $( \bC_i, \bC_{i+1})$ to the $\party{P}_i$ for $i=1,2,3$.  
        \end{enumerate}
    
    \end{func}
    \caption{Ideal functionality for $\Pi_{\textsf{MatMulReLU}}$.}
    \label{f:matmulrelu}
\end{figure}

\begin{figure}[h]
    \begin{func}{$\mathcal{F}_{\textsf{MatMulReLULTH}}$}
        Upon receiving signal with inputs $\hat\bC'$, masks $\bM$ and party index $i$, compute the following:
        \begin{enumerate}
            \item $\bD=\hat\bC'-\bM$, $\bE =\bD >\mathbf{0}$.
            \item Generate random $ \bC_1, \bC_2, \bC_3$ such that $ \bC_1+ \bC_2+ \bC_3=\bC$. Returns $( \bC_{i}, \bC_{i+1})$ to $\party{P}_i$ (as mentioned in notation, for party index $i+1$ means the next party).
        \end{enumerate}
    
    \end{func}
    \caption{Ideal functionality for the \LTH part of $\Pi_{\textsf{ReLU}}.$}
    \label{f:matmulrelulth}
\end{figure}

\begin{proof}[Proof for \autoref{t:2}]




We prove \autoref{t:2} by constructing a simulator and a series of hybrid games similar to the Proof of \autoref{t:1}, with \Env~providing inputs to parties.

\noindent \textbf{Case 1:} $\party{P}_1$ is corrupted ($i=1$) and $\party{P}_2$, $\party{P}_3$ are honest. \\\vspace{-.7em}

\noindent \textbf{The Simulator.} 
%
\Sim~simulates the following interactions on receiving the signal from $\mathcal{F}_{\textsf{MatMalReLU}}$:

\begin{itemize}
	\item Upon receiving $(\textsf{MatMalReLU},1, \bA_{3,1},\bA_{1,1},\bB_{3,1},\bB_{1,1},$ $a,b,c,L)$ from $\mathcal{F}_{\textsf{MatMalReLU}}$, \Sim~generates random $\hat\bA_2,\hat\bB_2$, and use $(\bA_{1,1},\hat\bA_2,\bB_{1,1},\hat\bB_2)$ and $(\hat\bA_2, \bA_{3,1},\hat\bB_2, \bB_{3,1})$ as dummy inputs for $\party{P}_2$ and $\party{P}_3$, respectively.

    \item \blue{\Sim~invokes the simulator of the secure multiplication protocol $\Pi_{\textsf{mal-arith-mult}}$ of \cite{mohassel2018aby3} without the truncation. In the end, $(\hat\bC_{3,1},\hat\bC{1,1}),(\hat\bC_{1,2},\hat\bC{2,2}),(\hat\bC_{2,3},\hat\bC{3,3})$ are distributed accordingly and \Sim~will signal abort to \idealFuncMMR if inconsistency was found in the distributed shares.}
    
	\item \Sim~acts as $\idealFuncGenMask$ to generate random masks $\bM_2$ for $\party{P}_2$ \blue{and $\party{P}_3$}, $\bM_3$ for $\party{P}_3$ \blue{and $\party{P}_1$} so that $\bM_1+\bM_2+\bM_3=\mathbf{0}$. Then \Sim~computes $\hat\bC_{2,2}'=\hat\bC_{2,2}+\bM_2$ as $\party{P}_2$ \blue{and $\hat\bC_{2,3}'=\hat\bC_{2,3}+\bM_2$ as $\party{P}_3$}; $\hat\bC_{3,3}'=\hat\bC_{3,3}+\bM_3$ as $\party{P}_3$ \blue{and $\hat\bC_{3,1}'=\hat\bC_{3,1}+\bM_3$ as $\party{P}_1$}. \blue{\Sim~sends $\hat\bC_{2,2}',\hat\bC_{2,3}'$ to $\party{P}_1$ as $\party{P}_2$ and $\party{P}_3$, sends $\hat\bC_{3,3}',\hat\bC_{3,1}'$ to $\party{P}_2$ as $\party{P}_3$ and $\party{P}_1$. \Sim~aborts if any inconsistency is found on the pairs.}. 
 
    \item \Sim~computes $\hat \bC_1'=\hat \bC_{2,2}'+\hat \bC^L_{3,1}+\hat \bC^L_{1,1}$ as $\party{P}_1$, \blue{$\hat \bC_2'=\hat \bC_{3,3}'+\hat \bC^L_{1,2}+\hat \bC^L_{2,2}$ as $\party{P}_2$. }

	\item \Sim~acts as~$\idealFuncLTHReLU$ ($H_1$) with $\party{P}_1$'s input $\hat \bC_1'$ to (re)generate $\bM_2$, computes $\hat\bC=(\hat \bC_1'- \bM_2)\gg\textsf{fp}$. \Sim~computes $\bD=\hat\bC>\mathbf{0}$, generates $\bZ'_1+\bZ'_2+\bZ'_3=\mathbf{0}$ (as $\idealFuncGenMaskShr$), then obtain $(\bZ'_3,\bZ'_1+\bD)$ as $(\bZ_{3,1},\bZ_{1,1})$ for $\party{P}_1$. \blue{\Sim~acts as~$\idealFuncLTHReLU$ ($H_2$) with $\party{P}_2$'s input $\hat \bC_2'$ to (re)generate $\bM_3$, computes $\hat\bC=(\hat \bC_2'- \bM_3)\gg\textsf{fp}$. \Sim~computes $\bD=\hat\bC>\mathbf{0}$ (as $\idealFuncGenMaskShr$), generates $\bZ'_1+\bZ'_2+\bZ'_3=\mathbf{0}$, then obtain $(\bZ'_1+\bD,\bZ'_2)$ as $(\bZ_{1,2},\bZ_{2,2})$ for $\party{P}_1$.}


    \item \Sim~acts as $\idealFuncGenMaskShr$ for $\party{P}_3$ to (re)generate $(\bZ_{2,3},\bZ_{3,3})=(\bZ'_2,\bZ'_3)$ and sends it to $\party{P}_3$. \Sim~compare $\bZ_{1,1}=\bZ_{1,2}$ as $\party{P}_1$ and $\party{P}_2$; $\bZ_{2,2}=\bZ_{2,3}$, $\bZ_{3,3}=\bZ_{3,1}$ as $\party{P}_3$. Signal abort to \idealFuncMMR if inconsistency was found.
		
\end{itemize}

\noindent \textbf{Indistinguishability.} We prove the indistinguishability argument by constructing a sequence of hybrid games as follows.

\noindent \textbf{Hybrid $\Hybrid_0$}:
This is the real protocol execution.

\noindent \textbf{Hybrid $\Hybrid_1$}:
$\Hybrid_1$ is the same as $\Hybrid_0$, except that $\Pi_{\textsf{LTH.GenMask}}$ is replaced with simulated $\idealFuncGenMask$ that outputs random $\bM_2,\bM_3$ for both step 2) and 5). 

%

\noindent \textbf{Hybrid $\Hybrid_2$}:
$\Hybrid_2$ is the same as $\Hybrid_1$, except that we replace step 5) with the simulated $\idealFuncLTHMMR$.


Proofs of $\Hybrid_0$ and $\Hybrid_1$, $\Hybrid_1$ and $\Hybrid_2$ being computationally indistinguishable are similar to the proof of \autoref{t:1}, with the actual random masks repalced by $\bM_2,\bM_3$.



\noindent \textbf{Hybrid $\Hybrid_3$}:
$\Hybrid_3$ is the same as $\Hybrid_2$, except that $\party{P}_2, \party{P}_3$  use dummy inputs for interaction, instead of the ones provided by the environment. Also $\Pi_{\textsf{mal-arith-mult}}$ of \cite{mohassel2018aby3} is replaced by its corresponding simulator.
In this hybrid, we introduce an ideal functionality $\mathcal{F}_{\textsf{ReLU}}$ that takes the environments' actual inputs and returns the corresponding outputs. 

We claim that $\Hybrid_2$ and $\Hybrid_3$ are indistinguishable. 
$\Hybrid_2$ and $\Hybrid_3$ are only different in inputs and step 1), and their indistinguishability  directly comes from the security of the simulator of $\Pi_{\textsf{mal-arith-mult}}$. Again we refer the reader to \cite{mohassel2018aby3} for more details. Since the view of $\party{P}_1$ does not change after step 1), it remains the same for the whole protocol since later steps remain the same in both hybrids. Therefore, $\Hybrid_2$ and $\Hybrid_3$ are indistinguishable.

The adversary's view of $\Hybrid_3$ is identical to $\textsf{EXEC}_{\idealFunc,\Sim,\Env}$. Therefore, in \textbf{Case 1} the view of $\mathcal{A}$ and \Env are indistinguishable in the real and the simulated world.

Putting it all together, we have that
$\Hybrid_0 \approx \Hybrid_1 \approx \Hybrid_2 \approx \Hybrid_3 = \Sim$.\\\vspace{-.7em}
\end{proof}


	
 
	


Notice that as discussed in \autoref{sec:expanse}, the above proof works identically for \textsf{MatMulBatchNormReLU} and \textsf{MatMulMaxPoolReLU} since only the plaintext computations
after subtracting the masks are different.

Next, we will prove the \autoref{t:3}. We provide the complete proof for case 1 (where $\party{P}_1$ is corrupted) due to the space limit, and the proof of the other two cases is similar to what we did previously.

\begin{figure}[h]
    \begin{func}{$\mathcal{F}_{\textsf{Softmax}}$}
        
        \begin{enumerate}
        \item Upon receiving inputs $( \bmx_{3,1}, \bmx_{1,1})$, $( \bmx_{1,2}, \bmx_{2,2})$, $( \bmx_{2,3}, \bmx_{3,3})$ from $\party{P}_1$, $\party{P}_2$, $\party{P}_3$ respectively, check if $ \bmx_{j,j}= \bmx_{j,j+1}$ for $j=1,2,3$. If not, notify abort. Otherwise, compute $\bmx= \bmx_{1,1}+ \bmx_{2,2}+ \bmx_{3,3}$, $\bmz =\lfloor\exp(\bmx\gg\textsf{fp})\ll\textsf{fp}\rfloor$. Then, send a signal $(\textsf{Softmax}, \bmx_{i-1,i},\bmx_{i,i},n,L,i)$ to $\Sim$ that includes the inputs of $P_i$, the size of inputs of $P_{i+1}$ and $P_{i-1}$, and the index $i$ of the corrupted party. 
        
        \item Upon receiving $(\textsf{Softmaxend},i,\bmz_{i-1},\bmz_i)$ from $\Sim$, let $\bmz_{i+1}= \bmz-\bmz_{i-1}-\bmz_i$. Return $( \bmz_i, \bmz_{i+1})$ to the $\party{P}_i$ for $i=1,2,3$.  
        \end{enumerate}
    
    \end{func}
    \caption{Ideal functionality for $\Pi_{\textsf{Softmax}}$.}
    \label{f:softmax}
\end{figure}

\begin{figure}[h]
    \begin{func}{$\mathcal{F}_{\textsf{SoftmaxLTH}}$}
        Upon receiving signal with inputs $\bmx'$, masks $\bmm$ and party index $i$, compute the following:
        \begin{enumerate}
            \item $\bmx=\bmx'-\bmm$, $\bma =\lfloor\exp(\bmx\gg\textsf{fp})\ll\textsf{fp}\rfloor$.
            \item Generate random $ \bmz_1, \bmz_2, \bmz_3$ such that $ \bmz_1+ \bmz_2+ \bmz_3=\bmz$. Returns $( \bmz_{i}, \bmz_{i+1})$ to $\party{P}_i$ (as mentioned in notation, for party index $i+1$ means the next party).
        \end{enumerate}
    
    \end{func}
    \caption{Ideal functionality for the \LTH part of $\Pi_{\textsf{Softmax}}.$}
    \label{f:softmaxlth}
\end{figure}

\begin{proof}[Proof for \autoref{t:3}]



We prove \autoref{t:3} by constructing a simulator and a series of hybrid games similar to the Proof of \autoref{t:1}, with \Env~providing inputs to parties.

\noindent \textbf{Case 1:} $\party{P}_1$ is corrupted ($i=1$) and $\party{P}_2$, $\party{P}_3$ are honest. \\\vspace{-.7em}

\noindent \textbf{The Simulator.} 
%
\Sim~simulates the following interactions on receiving the signal from $\mathcal{F}_{\textsf{ReLU}}$:

\begin{itemize}
	\item Upon receiving $(\textsf{Softmax},1, \bmx_{3,1},\bmx_{1,1},n,L)$ from $\mathcal{F}_{\textsf{Softmax}}$, 
 
 \Sim~generates a random $\hat\bmx_2$, and use $(\bmx_{1,1},\hat\bmx_2)$ and $(\hat\bmx_2, \bmx_{3,1})$ as dummy inputs for $\party{P}_2$ and $\party{P}_3$, respectively.
 
	\item \Sim~acts as $\idealFuncGenMask$ to generate random masks $\alpha_j\in \mathbb{Z}_{2^{52}}$ and $\beta_j\in\mathbb{Z}_{2^{32}}$ for $j=1,...,n$ for $\party{P}_1$, then computes $\bar r = \overline{ \exp((\bmx_{3,1})_j\gg\textsf{fp})}$ for $j=1,...,n$. Let $\bar r = \overline{ \exp( (\bmx_{3,1})_j\gg\textsf{fp}) }= 2^{q_j}\cdot (m_j\gg 52)$ as noted. \Sim~computes $\{m^*_j= (m_j+\alpha_j)_{2^{52}}, q^{*}_j=(q_j+\beta_j)_{2^{32}}\}$ for $i=1,...,n$ as $P_1$ sending to $P_2$.
    \blue{\Sim~repeat above for $\party{P}_3$, replacing $\bmx_{3,1}$ with $\bmx_{3,3}$; \Sim}
    
    \blue{again repeat above as $\party{P}_2$ and $\party{P}_3$, each replacing $\bmx_{3,1}$ with $\bmx_{2,2}$ and $\bmx_{2,3}$, with $\Pi_{\textsf{LTH.}}'$ generating $(\bm\alpha' ,\bm\beta')$, and get $\{'m^*_j, 'q^{*}_j\}$ respectively. $\{m^*_j, q^{*}_j\}$ are for $P_2$ provided by $P_1$ and $P_2$, and $\{'m^*_j, 'q^{*}_j\}$ are for $P_1$ provided by $P_2$ and $P_3$}
 
    \item \blue{\Sim~compare the received copies as $P_2$ and $P_1$, and signal abort to \idealFuncSoftmax if an inconsistency is found}. \Sim~computes as $P_2$: $ 2^{\hat (q_2)_j}\cdot \hat (m_2)_j:=\overline{ \exp(( (\bmx_{2,2})_j+(\bmx_{1,2})_j)\gg\textsf{fp}) }$ for $j=1,...,n$. \blue{\Sim~computes as $P_1$: $ 2^{\hat (q_1)_j}\cdot \hat (m_1)_j:=\overline{ \exp(( (\bmx_{1,1})_j+(\bmx_{3,1})_j)\gg\textsf{fp}) }$ for $j=1,...,n$.}
	\item \Sim~acts as~$\idealFuncLTHSoftmax$ with $\party{P}_2$'s input $\{\bmq^*+\hat \bmq_2,\bmm^*,\hat \bmm_2\}$ to $H_1$, to (re)generate $(\bm\alpha ,\bm\beta)$, then locally compute $\bmy$ as in step 4) 
    \Sim~then generate $\bmm_1+\bmm_2+\bmm_3=\mathbf{0}$ and $(\bmm_1+\bmy,\bmm_2)$ as $(\bmy_{1,2},\bmy_{1,2})$ for $P_2$. \blue{\Sim~acts as~$\idealFuncLTHSoftmax$ with $\party{P}_1$'s input $\{'\bmq^*+\hat \bmq_1,'\bmm^*,\hat \bmm_1\}$ to $H_1$, to (re)generate $(\bm\alpha' ,\bm\beta')$, then locally compute $\bmy$ as in step 4). \Sim~then generate $\bmm_1+\bmm_2+\bmm_3=\mathbf{0}$ and $(\bmm_3,\bmy+\bmm_1)$ as $(\bmy_{3,1},\bmy_{1,1})$ for $P_1$. \Sim~send }

    \item \blue{\Sim~checks \Sim~compare $\bmy_{1,1}=\bmy_{1,2}$ as $\party{P}_1$ and $\party{P}_2$; $\bmy_{2,2}=\bmy_{2,3}$, $\bmy_{3,3}=\bmy_{3,1}$ as $\party{P}_3$. Signal abort to \idealFuncSoftmax if an inconsistency is found. If no abort is signaled,} \Sim~signals $(\textsf{Softmaxend},1,\bmy_{3,1},\bmy_{1,1})$ to $\mathcal{F}_{\textsf{Softmax}}$.
		
\end{itemize}

\begin{table*}[ht]

\centering
\caption{Analytical cost analysis of the network communication rounds and amount (in Bytes) under the semi-honest setting, and the local bus communication with the LEE. Here, $n=m\times m$ is the input size, $s$ is the stride, $w\times w$ is the filter size, and $e$ is the precision parameter for the exponent ($\exp(x)\approx(1+\frac{x}{2^e})^{\frac{x}{2^e}}$). \change{We did not include CryptGPU and Goten because they did not focus on optimization of their non-linear layer protocols and provide no analytical cost analysis.}
}
\begin{tabular}{c|c|cx{5cm}ccc}

\hline
\textsf{Communication Type}&\textsf{Framework}& \textsf{ReLU} & \textsf{MaxPool}   &\textsf{BatchNorm}& \change{\textsf{Softmax} }
\\\hline
\multirow{3}{*}{Network comm. rounds} 
&Falcon+ &$10$           & $12(w^2-1)$            &335                &\change{$12n+p+317$}
\\
&AriaNN  &2              &3                      &9                  &\change{-}
\\  
&{\scheme} &2              &2                      &2                  &\change{2}
\\\hline
\multirow{3}{*}{Network comm. data}  
&Falcon+ &$16n$          & $(20+w^2)(\frac{m}{s}-1)^2$           &$~224n$                &\change{\change{$(\frac{m}{s}-1)^2(n+20)+(110+e)n$}}
\\
&AriaNN  &$12n$          &$(\frac{m}{s}+1)^2(w^4+1)$ &$72n$             \change{-}     &\change{-}
\\
&{\scheme} &$5n$           &$\frac{2}{3}(2n+2w^2+5(\frac{m}{s}+1)^2)$ &$4n$  
&\change{$\frac{20}{3}n$   }
\\\hline
\LTH comm.  &{\scheme} &$\frac{25}{3}n$     &$\frac{1}{3}(8m^2+8w^2+15(\frac{m}{s}+1)^2)$    &$\frac{16}{3}n$ &\change{\change{$\frac{44}{3}n$}}\\

\hline
\end{tabular}


\label{tab:cost}
\end{table*}

\noindent \textbf{Indistinguishability.} We prove the indistinguishability argument by constructing a sequence of hybrid games as follows.

\noindent \textbf{Hybrid $\Hybrid_0$}:
This is the real protocol execution.

\noindent \textbf{Hybrid $\Hybrid_1$}:
$\Hybrid_1$ is the same as $\Hybrid_0$, except that$\Pi_{\textsf{LTH.GenMask}}$ is replaced with simulated $\idealFuncGenMask$ that outputs random $ (\bm\alpha,\bm\beta)$ and $ (\bm\alpha',\bm\beta')$ for both step 1) and 3). 

%

\noindent \textbf{Hybrid $\Hybrid_2$}:
$\Hybrid_2$ is the same as $\Hybrid_1$, except that we replace step 4) with the simulated $\idealFuncLTHSoftmax$.


Proofs of $\Hybrid_0$ and $\Hybrid_1$, $\Hybrid_1$ and $\Hybrid_2$ being computationally indistinguishable are similar to the proof of \autoref{t:1}, with the actual random masks repalced by $ (\bm\alpha,\bm\beta)$ and $ (\bm\alpha',\bm\beta')$.

\noindent \textbf{Hybrid $\Hybrid_3$}:
$\Hybrid_3$ is the same as $\Hybrid_2$, except that $\party{P}_2, \party{P}_3$  use dummy inputs for interaction, instead of the ones provided by the environment. 
In this hybrid, we introduce an ideal functionality $\mathcal{F}_{\textsf{Softmax}}$ that takes the environments' actual inputs and returns the corresponding outputs. 

We claim that $\Hybrid_2$ and $\Hybrid_3$ are indistinguishable.
Since the corrupted party is $\party{P}_1$, \Sim~knows $\bmx_{3,1}=\bmx_{3,3}, \bmx_{1,1}=\bmx_{1,2}$. The  dummy inputs would be $ \bmx_{2,2}= \bmx_{2,3}$ (represented by $\hat \bmx_2$ in \Sim). The computation result sent to $P_1$ by $P_3$ in step 1) used the dummy inputs, and $\{'m^*_j= ('m_j+\alpha_j')_{2^{52}}, 'q^{*}_j=('q_j+\beta_j')_{2^{32}}\}$ for $i=1,...,n$ are uniformly random since $(\bm\alpha',\bm\beta')$ are random. Therefore the views of adversary in step 1) are not distinguishable in both hybrids, and its views of step 3) and 4) are also not distinguishable in both hybrids due the uniformly masked output. Therefore, $\Hybrid_2$ and $\Hybrid_3$ are indistinguishable.

The adversary's view of $\Hybrid_3$ is identical to $\textsf{EXEC}_{\idealFunc,\Sim,\Env}$. Therefore, in \textbf{Case 1} the view of $\mathcal{A}$ and \Env are indistinguishable in the real and the simulated world.

Putting it all together, we have that
$\Hybrid_0 \approx \Hybrid_1 \approx \Hybrid_2 \approx \Hybrid_3 = \Sim$.\\\vspace{-.7em}

\end{proof}

\section{Analytical cost Analysis}\label{sec:analytical}

The cost analysis of our protocol is shown in \autoref{tab:cost}, compared with baselines with analytical results provided in their work. The byte size of the finite field is chosen to be 4 and we count the exponent and the mantissa part in \autoref{p:softmax} as 4 and 8 bytes.
We see significant improvements in inter-party communication rounds compared to Falcon, and significant theoretical reduction in the amount of communication data compared to both Falcon and AriaNN.



The actual speedup of a particular neural network depends on its structure including the ratio between linear and non-linear operations, the order of linear/non-linear operations/layers (which determines if protocols like \autoref{p:matmulrelu} can be applied), the input dimensions, etc. The communication setting and the computational power also matter. We discuss the performance in \autoref{sec:experiment}.

\section{\change{Memory Usage Analysis}}
\label{sec:memory}

\change{
As \LTH typically does not support off-chip DRAM, \scheme needs to be able to run using a small on-chip SRAM. Here, we analyze the \LTH memory usage of \scheme and show that the small on-chip SRAM is sufficient even for large ML models. 
For our experiments, our implementation runs on a Arduino Due microcontroller, as discussed in \autoref{sec:setup}. In this prototype, the code occupies 23 KB of flash memory, which is less than 4\% of the total capacity (512 KB) of Arduino Due's flash memory. The \LTH code uses up to 43 KB of on-chip SRAM during the execution, including the buffers for variables, space used by Arduino's libraries and middleware (e.g., SerialUSB functions), and other usage like the function call stack. 
}


\change{In order to more fully understand the \LTH memory requirement for larger models that were not run in our experiments, we provide analytical memory usage numbers of different non-linear operations.
We list all the SRAM memory usage of the non-linear operations in \autoref{tab:SRAM}.} 
\change{Note that the memory usage is constant for \textsf{ReLU} and \textsf{BatchNorm}. This is because they are scalar-wise operations during the inference phase. Therefore, each individual scalar of an input vector can be processed independently. The \textsf{MaxPool} operation has a dependency on the window size, which is rather small in all the models used in practice. All of the above three operations need no more than 0.5 KB of SRAM and, therefore, will not pose any memory usage issue even for larger models. As a reference point, our prototype has 96 KB SRAM for \LTH.
}


\begin{table}[ht]
\vspace{-2mm}
\small

\centering
\caption{\change{Analytical analysis of the minimum SRAM usage for each operations. Here, $n$ is the plaintext input vector size, $w\times h$ is the maxpool window size, and $l$ is the normal variable size (in our case 4 bytes). Dynamic buffer reuse is considered.}}\vspace{-.9em} 
\begin{tabular}{c|c|c}

\hline
\textsf{Operations} & \textsf{Parameters} & \LTH \textsf{Least SRAM Usage} \\\hline
\textsf{ReLU} & - & $2l$\\\hline
\textsf{MaxPool}  & $\{w,h\}$ & $2whl$\\\hline
\textsf{BatchNorm} & - & $2l$\\\hline
\textsf{LayerNorm} & $n$ & $(2n+1)l$\\\hline
\textsf{Softmax} & $n$ & $6nl$\\\hline
\end{tabular}

\label{tab:SRAM}
\end{table}

\change{However, the minimum memory usage of the \textsf{Softmax} and \textsf{LayerNorm} operation depends on the input vector length, which can be large in some model structure. In our prototype, \LTH cannot host all the variables needed within its SRAM with $n$ greater than 3925. For the models in our experiments, the largest vector size is 200. 
Even modern Transformer models such as GPT-3 2.7B have the maximum $n$ not larger than 2560 in its model structure. 
In that sense, \LTH will be able to support many modern ML models even with a relatively small SRAM capacity without protocol changes. 
}

\change{
In a case when an input vector size is too large to fit into the \LTH SRAM, the \scheme protocol can be slightly modified to break down the input vector into multiple smaller chunks, and perform non-linear operations in multiple rounds. 
Here, we show this approach using \textsf{Softmax}. The party first cuts the input of step (2) of \autoref{p:softmax} into chunks in the host CPU, and sends them to the \LTH; The \LTH recovers and computes the exponent of each input chunk as in step (3) of \autoref{p:softmax} until the $\bullet$, accumulates the exponents locally and then sends the exponent results back to the host CPU with temporal generated masks calling \autoref{p:LTH.GenMask}; After all chunks are summed, the party sends again the masked exponent results and \LTH will continue step (3) of \autoref{p:softmax}. 
If such a multi-round \textsf{Softmax} is used, the communication cost of \LTH will increase from $\frac{20}{3}n$ to $\frac{68}{3}n$. 
Similar steps can be taken for \textsf{ LayerNorm}.}

\change{In summary, \scheme should be able to handle most models with our current \LTH setting, and can be modified to handle even larger models with some increase in the \LTH local communication cost.}

\end{document}